\documentclass[acmsmall, screen, nonacm]{acmart}

\AtBeginDocument{%
  }

\setcopyright{cc}
\copyrightyear{2018}
\setcctype{by}
\acmDOI{10.1145/3729296}
\acmYear{2025}
\acmJournal{PACMPL}
\acmNumber{PLDI}
\received[accepted]{2025-03-06}

\usepackage{array}
\usepackage{cleveref}
\usepackage [autostyle, english = american]{csquotes}
\usepackage{extarrows}
\usepackage[shortcuts]{extdash} 
\usepackage{graphicx}
\usepackage{mathtools}
\usepackage{mdframed}
\usepackage[frozencache=true,cachedir=minted-cache]{minted}
\usepackage{newunicodechar}
\usepackage{pifont}
\usepackage{semantic}
\usepackage{tabularray}
\usepackage{textgreek}
\usepackage{xspace}
\usepackage{yfonts}
\usepackage{enumitem}

\newunicodechar{⇒}{\ensuremath{\Rightarrow}}
\newunicodechar{→}{\ensuremath{\rightarrow}}
\newunicodechar{λ}{\ifmmode\lambda\else\textlambda\fi}
\newunicodechar{Θ}{\ifmmode\Theta\else\textTheta\fi}
\newunicodechar{Γ}{\ifmmode\Gamma\else\textGamma\fi}
\newunicodechar{μ}{\ifmmode\mu\else\textmu\fi}
\newunicodechar{η}{\ifmmode\eta\else\texteta\fi}
\newunicodechar{Π}{\ifmmode\Pi\else\textPi\fi}
\newunicodechar{𐤄}{\ifmmode{\rotatebox[origin=c]{15}{\textphnc{e}}\hspace{-1pt}}\else{\textphnc{e}}\fi}
\newunicodechar{⊥}{\ensuremath{\bot}}
\newunicodechar{×}{\ensuremath{\times}}
\newunicodechar{⊳}{\ensuremath{\rhd}}
\newunicodechar{∩}{\ensuremath{\cap}}
\newunicodechar{∪}{\ensuremath{\cup}}
\newunicodechar{≠}{\ensuremath{\neq}}
\newunicodechar{∅}{\ensuremath{\varnothing}}
\newunicodechar{⊢}{\ensuremath{\vdash}}
\newunicodechar{△}{\ensuremath{\bigtriangleup}}
\newunicodechar{≡}{\equiv}
\newunicodechar{≢}{\not\equiv}
\newunicodechar{≤}{\ensuremath\leq}
\newunicodechar{⊕}{\oplus}
\newunicodechar{⊖}{\ominus}
\newunicodechar{⟦}{\lBrack}
\newunicodechar{⟧}{\rBrack}
\newunicodechar{∖}{\setminus}

\newcommand{\BNF}{\quad\operatorname{::=}\quad}
\newcommand{\BNFOR}{\quad\operatorname{\big{|}}\quad}
\newcommand{\CASE}[6]{\langword{case}_{#1}#2\langword{of}\INL #3 ⇒ #4 ; \INR #5 ⇒ #6}
\newcommand{\CASEm}[6]{\begin{aligned}[t]\langword{case}_{#1}#2\langword{of}&\INL #3 ⇒ #4 ;\\& \INR #5 ⇒ #6\end{aligned}}
\newcommand{\COMM}[2]{\langword{com}_{#1;#2}}
\newcommand{\cmark}{\ding{51}}%
\newcommand{\DEF}{{\quad\operatorname{\triangleq}\quad}}
\newcommand{\DEFCASE}{\quad\operatorname{\xRightarrow{△}}\quad}
\newcommand{\DOT}{\langword{.}}
\newcommand{\eg}{\textit{e.g.}\xspace}
\newcommand{\FLR}[1]{\left\lfloor{#1}\right\rfloor}
\newcommand{\FST}[1]{\langword{fst}_{#1}}

\newcommand{\ie}{\textit{i.e.}\xspace}
\newcommand{\INL}{\langword{Inl}}
\newcommand{\inlinecode}[2][haskell]{\mintinline[breaklines]{#1}{#2}}
\newcommand{\INR}{\langword{Inr}}
\newcommand{\langword}[1]{\operatorname{\mathsf{#1}}}
\newcommand{\LOOKUP}[2]{\langword{lookup}^{#1}_{#2}}
\newcommand{\mask}{\ifmmode{\operatorname{⊳}}\else{⊳}\fi\xspace}
\newcommand{\myference}[3]{\inference[\textsc{#1}]{#2}{#3}}
\newcommand{\netstep}[2]{\xlongrightarrow{ #1 \ifthenelse{\equal{#1}{}}{}{:} #2 }}
\newcommand{\nonempty}[1]{{#1^{+}}}
\newcommand{\noop}[2]{\mathtt{noop}^{\mask #1}\!\!(#2)}
\newcommand{\PAIR}{\langword{Pair}}
\newcommand{\prcstep}[2]{\xlongrightarrow{ ⊕#1 ; ⊖#2 }}
\newcommand{\RECV}[1]{\langword{recv}_{#1}}
\newcommand{\roles}[1]{\mathtt{roles}(#1)}
\newcommand{\SEND}[1]{\langword{send}_{#1}}
\newcommand{\set}[1]{\left\{#1\right\}}
\newcommand{\SND}[1]{\langword{snd}_{#1}}
\newcommand{\step}{\operatorname{\longrightarrow}}
\newcommand{\stepname}[1]{$\mathfrak{#1}$}
\newcommand{\vdbl}{\\[8pt]}
\newcommand{\xmark}{{\large $\times$}}%

\newtheorem{theorem}{Theorem}
\newtheorem{lemma}{Lemma}
\newtheorem{corollary}{Corollary}
\newtheoremstyle{ourdef}
	{0.25em}
	{0.25em}
	{\hangindent=2em}
	{0.0em}
	{}
	{\,:}
	{0.8em}
	{\textsc{\thmname{#1}}\quad\textit{\thmnote{#3}}}
\theoremstyle{ourdef}
\newtheorem*{definition}{definition}

\newcommand{\ChoreographyTS}{ChoreoTS\xspace}
\newcommand{\chorLambda}{Chor$\lambda$\xspace}
\newcommand{\Chorus}{ChoRus\xspace}
\newcommand{\HasChor}{Has\-Chor\xspace}
\newcommand{\HLSCentral}{$\boldsymbol{\lambda}_{\boldsymbol{C}}$\xspace}
\newcommand{\HLSLocal}{$\boldsymbol{\lambda}_{\boldsymbol{L}}$\xspace}
\newcommand{\HLSNet}{$\boldsymbol{\lambda}_{\boldsymbol{N}}$\xspace}
\newcommand{\MultiChor}{Multi\-Chor\xspace}

\MakeOuterQuote{"}

\newcolumntype{L}[1]{>{\raggedright\let\newline\\\arraybackslash\hspace{0pt}}m{#1}}
\newcolumntype{C}[1]{>{\centering\let\newline\\\arraybackslash\hspace{0pt}}m{#1}}
\newcolumntype{R}[1]{>{\raggedleft\let\newline\\\arraybackslash\hspace{0pt}}m{#1}}

\begin{document}

\title{Efficient, Portable, Census-Polymorphic Choreographic Programming}

\author{Mako Bates}
\email{mako.bates@uvm.edu}
\orcid{0009-0001-9933-1728}
\affiliation{%
 \institution{University of Vermont}
 \city{Burlington}
 \state{Vermont}
 \country{USA}}

\author{Shun Kashiwa}
\orcid{0009-0001-3665-0182}
\affiliation{%
  \institution{University of California, San Diego}
  \country{USA}
  \city{La Jolla}}
\email{skashiwa@ucsd.edu}

\author{Syed Jafri}
\email{sajafri@uvm.edu}
\orcid{0009-0009-4372-4715}
\affiliation{%
 \institution{University of Vermont}
 \city{Burlington}
 \state{Vermont}
 \country{USA}}

\author{Gan Shen}
\orcid{0009-0006-0947-9531}
\affiliation{%
  \institution{University of California, Santa Cruz}
  \country{USA}
  \city{Santa Cruz}}
\email{gshen42@ucsc.edu}

\author{Lindsey Kuper}
\orcid{0000-0002-1374-7715}
\affiliation{%
  \institution{University of California, Santa Cruz}
  \country{USA}
  \city{Santa Cruz}}
\email{lkuper@ucsc.edu}

\author{Joseph P. Near}
\email{jnear@uvm.edu}
\orcid{0000-0002-3203-3742}
\affiliation{%
 \institution{University of Vermont}
 \city{Burlington}
 \state{Vermont}
 \country{USA}}

\begin{abstract}
  Choreographic programming (CP) is a paradigm for implementing
  distributed systems that uses a single global program to define the
  actions and interactions of all participants.
  Library-level CP implementations, like \HasChor, integrate well with
  mainstream programming languages but have several limitations:
  Their conditionals require extra communication;
  they require specific host-language features (\eg, monads);
  and they lack support for programming patterns that are essential for implementing
  realistic distributed applications.

  We make three contributions to library-level CP to specifically
  address these challenges. First, we propose and formalize
  \emph{conclaves} and \emph{multiply-located values}, which enable
  efficient conditionals in library-level CP without redundant
  communication.
  Second, we propose
  \emph{census polymorphism}, a technique for abstracting over the
  number of participants in a choreography.
  Third, we introduce a design pattern for library-level CP in host languages without support for monads.
  We demonstrate these
  contributions via implementations in Haskell, Rust, and TypeScript.
\end{abstract}




\maketitle

\section{Introduction}
\label{sec:introduction}

Choreographic programming~\citep{montesi-carbone-dfbd,montesi-dissertation,montesi_book} (CP)
is a paradigm for implementing distributed systems in which the programmer writes one unified program, called a choreography,
that describes from a third-party perspective how the endpoints of the system interact.
A choreography is then translated to a collection of executable programs for each endpoint via a compilation step called endpoint projection (EPP).
The CP approach has benefits both for understandability of distributed system implementations,
and for strong static guarantees about the deadlock-freedom of the resulting executable code~\citep{montesi-carbone-dfbd}.

Today, after over a decade of work on its theoretical underpinnings,
CP is poised to enter the mainstream in the form of frameworks and language extensions for widely used programming languages.
For instance,
Choral~\citep{giallorenzo-choral} is a choreographic language that compiles to Java and uses standard object-oriented abstractions to express CP concepts,
and
\HasChor~\citep{shen-haschor} takes a "just-a-library" approach to choreographic programming in Haskell,
enabling seamless use of the host language's ecosystem.

The integration of CP into mainstream languages via libraries is essential for its widespread adoption.
The library-level approach meets programmers where they are:
in their general-purpose language of choice, with access to that language's ecosystem and tools.
Library-level CP also enables integration of choreographic components into larger non-choreographic systems.
However, a purely library-level CP implementation approach, as exemplified by \HasChor, has several drawbacks that limit its applicability.
First, \HasChor has an inefficient implementation of choreographic conditionals:
it broadcasts the value of the scrutinee of a conditional expression to all parties in a choreography, regardless of where that information is actually needed.
Second, \HasChor and all other prior CP systems lack support for programming patterns that are essential
for implementing realistic distributed applications.
In particular, they have required exact (though possibly implicit) enumeration of the participants in a choreography.
This limitation makes it impossible to implement algorithms that abstract over their \emph{number} of participants.
Finally, \HasChor relies on specific host-language features, \eg, monads, that most would-be host languages lack.

\paragraph{Efficient conditionals.}
To address the first challenge, we present and formalize two new choreographic language features,
\emph{conclaves} (\Cref{sec:conclaves}) and \emph{multiply-located values} (MLVs) (\Cref{sec:multiply-located-values}).
Together these features provide improved communication efficiency compared to previous library-based approaches.
Prior approaches for optimal communication have not been amenable to library-level implementation
because unprojectable choreographies are detected by the EPP mechanism, not by the type system.
In a purely library-level approach, however, EPP happens at run time;
\HasChor's inefficient solution for conditional expressions was chosen to avoid run-time errors.
In \Cref{sec:main-ideas} we show how conclaves and MLVs can be used together to implement communication-efficient conditional expressions
whose safety is enforced at the type level, and formalize this property.
In \Cref{sec:multichor} we show that this strategy is compatible with the fundamental engineering approach of \HasChor
by presenting \MultiChor, a more efficient and more expressive CP library for Haskell.

\paragraph{Abstraction over number of participants.}
To address the second challenge, we present \emph{census polymorphism} (\Cref{sec:census-poly}),
the ability to safely express and type a choreography that is parametric over the \emph{number} of participating parties.
Many concurrent protocols are designed to work with varying numbers of participants,
either because there are a variable number of parties who might want to contribute,
or because they have features (security, fault tolerance) that can be scaled with the number of participants.
Some recent CP implementations allow choreographies that are polymorphic over
the \emph{identities} of the parties who fulfill each \emph{role},
but still require exact enumeration of the roles (including quantities) \citep{hirsch2021pirouette,chor-lambda}.
We demonstrate census polymorphism by implementing three protocols not previously expressible due to their parametricity over the number of participants:
a multi-server key-value store,
the low-trust randomization mechanism from DPrio~\cite{dprio2023},
and
the GMW~\cite{goldreich2019play} secure computation protocol (\Cref{sec:case-studies}).

\paragraph{Language-agnostic implementation pattern.}
To address the third challenge, we introduce \emph{endpoint projection as dependency injection} (\Cref{sec:epp-as-di}),
a technique for implementing run-time EPP using only higher-order functions and related language features
(instead of dynamic interpretation of free monads, the approach HasChor uses).
We use this pattern, in combination with conclaves and MLVs and the host languages' subtyping systems,
to implement CP libraries in two additional languages: Rust and TypeScript.
Together, these techniques enable the implementation of library-level choreographic programming in a wide variety of host languages with diverse features.

We demonstrate the feasibility and generality of our contributions by implementing the above concepts
in three libraries implemented in different host languages using a variety of language features: \MultiChor\footnote{
    \MultiChor:  \url{https://github.com/ShapeOfMatter/MultiChor}
}, implemented in Haskell;
\Chorus\footnote{
    \Chorus:  \url{https://github.com/lsd-ucsc/ChoRus}
}, implemented in Rust;
and \ChoreographyTS\footnote{
    \ChoreographyTS:  \url{https://github.com/shumbo/choreography-ts}
}, implemented in TypeScript.
All three libraries are open source and come with extensive case studies demonstrating their use~\citep{ourArtifact}.

\paragraph{Contributions.}
\label{sec:contributions}
In summary, we make the following contributions:
\begin{itemize}
    \item We introduce the \emph{conclaves-\&-MLVs} choreographic programming paradigm (\Cref{sec:conclaves,sec:multiply-located-values}) and describe how the use of conclaves and MLVs enables efficient choreographic conditionals.  We present a formalism that implements the conclaves-\&-MLVs paradigm (\Cref{sec:formalism}) and prove soundness and completeness of endpoint projection, from which deadlock freedom follows as a corollary.
    \item We introduce \emph{census polymorphism} as a desirable feature for CP systems (\Cref{sec:census-poly}),
      and use it to implement a selection protocols (\Cref{sec:case-studies}).
    \item We demonstrate that library level CP is practical in languages
	    without the specific features on which HasChor relies (\eg monadic programming).~(\Cref{sec:epp-as-di})
\end{itemize}

\section{Background on Choreographic Programming}
\label{sec:background}

In this section, we give a brief overview of choreographic programming (CP).  For a comprehensive introduction to the topic, we refer the reader to~\citet{montesi_book}.

Choreographic programming is a paradigm that expresses a distributed system
as a single, global program describing the behavior and interactions of all parties.
The global view of the distributed system enables easier reasoning about the system's behavior---for example, choreographic languages can ensure \emph{deadlock freedom}~\citep{montesi-carbone-dfbd}---and also simplifies the modular development of complicated interactions between parties.

\begin{figure}[tbp]
\begin{mdframed}
\begin{tabular}{c c}
\begin{minipage}{8cm}
\begin{minted}{haskell}
handleRequest :: Request -> IORef State -> IO Response

kvs :: Request @ "client" -> (IORef State) @ "server" 
  -> Choreo IO (Response @ "client")
kvs request stateRef = do
  -- send the request to the server
  request' <- (client, request) ~> server
  -- server handles the request and creates a response
  response <- locally server \un -> 
    handleRequest (un server request') (un server stateRef)
  -- server sends the response back to the client
  (server, response) ~> client
\end{minted}
\end{minipage}
&
\begin{minipage}{3.5cm}
\includegraphics[width=3.5cm]{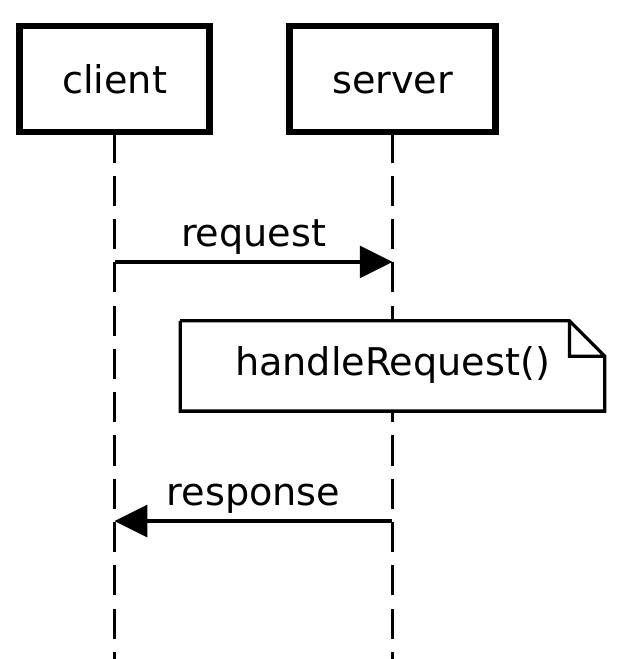}
\end{minipage}
\end{tabular}
\caption{A simple client-server interaction, written using the \HasChor~\cite{shen-haschor} Haskell library.}
\label{fig:simple-choreography}
    \Description{Twelve lines of Haskell code using the HasChor library, and a UML sequence diagram of that program.
	The code defines a choreography called "kvs".
	In the sequence diagram, first a party called "client" sends a message to a party called "server",
	then "server" calls a function called "handle Request",
	then "server sends "client" a message.}
\end{mdframed}
\end{figure}

As a first example, consider the protocol in Figure~\ref{fig:simple-choreography},
in which a client sends a \inlinecode{Request} to a key-value store application runing on a server, and the server sends back the appropriate response.
This simple example is written in Haskell using the \HasChor library~\cite{shen-haschor},
and demonstrates the main features of choreographic programming.
It mixes communication (using a monolithic \inlinecode{~>} operator to represent both transmission and reception of data)
with local computation, (\eg, a call to \inlinecode{handleRequest}, a function to handle requests) in a single global program.
In a choreographic program, known as a \emph{choreography}, each value has a \emph{location} indicating which party stores the value
(\eg, \inlinecode{request} is located at the client, while \inlinecode{request'} is located at the server).
\HasChor denotes such located values with the type
\inlinecode{t @ l}, where \inlinecode{t} is a type and \inlinecode{l} is a location
(\eg \inlinecode{Response @ "client"}).
In \Cref{fig:example-choreography} we will show that our library \MultiChor represents its equivalent type as \inlinecode{Located '[l] t},
and in \Cref{sec:epp-as-di} we will introduce \ChoreographyTS and \Chorus, which have their own respective syntaxes.

We elide the implementation of the \inlinecode{handleRequest} function, which runs locally on the server and is not a choreographic function, and show only its type signature (line 1).  In the type of \inlinecode{handleRequest}, the \inlinecode{Request} type represents a request to either get or update the value associated with a key, the \inlinecode{IORef State} type represents the mutable state of the key-value store, and \inlinecode{IO Response} denotes that the function runs in Haskell's \inlinecode{IO} monad, which enables updating the mutable store.

The \inlinecode{kvs} function demonstrates \HasChor's approach to implementing library-level choreographic programming in Haskell.
The \inlinecode{Choreo IO (Response @ "client")} in the type signature of the \inlinecode{kvs} function (lines 3-4) denotes that \inlinecode{kvs} is a choreography---the \inlinecode{Choreo} monad is defined by the \HasChor library, and \HasChor defines choreographies as computations in this monad. The \inlinecode{Choreo} monad uses a ``local'' monad (in this case, \inlinecode{IO})
and allows returning a value
(in this case, \inlinecode{Response @ "client"}.)

In \HasChor, the \inlinecode{~>} operator is used for communication\footnote{
    The \inlinecode{~>} syntax, sometimes pronounced ``comm'' (short for communication), is a popular choice in the CP literature.
    Other options include $\rightsquigarrow$, $\leftsquigarrow$, \inlinecode{comm}, and \inlinecode{com}.
} (lines 7 and 12). In line 7, the \inlinecode{kvs} choreography uses the \inlinecode{~>} operator to send the request from the client to the server, calling the result (now located at the server) \inlinecode{request'}. The \inlinecode{locally} function (line 9) is used to perform local computation: \inlinecode{locally l f} runs the function \inlinecode{f} \emph{locally} at location \inlinecode{l}---all other participants skip executing \inlinecode{f}. The function \inlinecode{f} is given an \emph{unwrapper} (by convention, often named \inlinecode{un}) with type \inlinecode{T @ l -> T} that can transform a located value (at location \inlinecode{l}) into a regular Haskell value.

\subsection{Endpoint Projection}\label{sec:background-epp}

Executing a choreography requires compiling it into separate local programs
for each of the parties to run---a process called \emph{endpoint projection} (EPP)~\citep{carbone-cdl-epp-esop,carbone-cdl-epp}.\footnote{
 Unlike in the literature on multiparty session types~\citep{honda-mpsts},
 in which endpoint projection refers to projecting a \emph{global type} to a collection of \emph{local types},
 in choreographic programming we are concerned with projecting a \emph{global program} (that is, a choreography)
 to a collection of \emph{local programs}.
 The dual of choreographic EPP, \emph{extraction} \citep{carbone2018mcc}, is not considered in this work.
}
EPP "projects" a choreography to a given "endpoint" (process, machine, location, \textit{etc})
in a sense analogous to geometric projection of a high-dimension object to its lower-dimensional shadow.
For example, EPP can transform the program in Figure~\ref{fig:simple-choreography}
into two separate programs, one for the server and one for the client,
such that each resulting program is a single-threaded (and correct) implementation of the respective party's behavior in the original choreography.
Each communication in the original choreography becomes a call to \inlinecode{send} for the original sender
and a \inlinecode{recv} for the original receiver.
Since the original choreography exactly specifies the sequence of communications,
as long as EPP is correct,
the network of projected programs will not contain deadlocks.

The \inlinecode{send} and \inlinecode{recv} functions can be implemented with traditional network primitives like blocking sockets;
some systems, including \HasChor and the systems we present in this paper, allow projection to multiple transport mechanisms.
Specifically, a single HasChor choreography can be executed as either a protocol in which machines communicate using HTTPS
\emph{or} as a protocol in which threads on a single machine communicate using sockets.
Furthermore, users can write their own adapters to use other transport mechanisms.
All three of our implementations have the same options.

\subsection{Knowledge of Choice}
\label{sec:koc}

Choreographies with conditionals---\inlinecode{if}-expressions or anything that could be used for conditional control-flow---introduce
a challenge for endpoint projection:
\emph{some parties might not know which branch to take!}
This challenge is referred to as the \emph{knowledge of choice}  (KoC)~\citep{castagna-knowledge-of-choice} problem.
All choreographic programming languages include a strategy for KoC
that ensures that relevant parties have enough information to play their part in the program.

Standalone CP languages typically ensure that
each branching operation is controlled by a single party,
and that they communicate their choice to other relevant parties using
a designated \inlinecode{select} operator~\citep{giallorenzo-choral, chor-lambda-2, chor-lambda, hirsch2021pirouette}.
Choreographies without correct KoC management are deemed \emph{unprojectable},
because they fail a step called \emph{merging} during EPP.
In this paper we refer to this KoC management strategy as \emph{select-\&-merge}.

The \inlinecode{select} operator is flexible enough to avoid unneeded communication between parties in a choreography,
so it is common in standalone languages.
Unfortunately, the \emph{select-\&-merge} approach is difficult to implement in library-based systems for choreographic programming
because determining whether KoC has been handled correctly
(\ie, whether the choreography is projectable) requires actually performing EPP for all endpoints.
For a select-\&-merge system to do this at run time would imply unpredictable run-time errors.
In some contexts it is possible to carry out EPP prior to run time by implementing it using a sufficiently powerful macro system 
(\eg, \citet{chorex-github}),
but for purely library-level CP in most languages, the only system available for static analysis is the host language's type system.
No technique to date encodes a projectability test for select-\&-merge choreographies in a type system.

\HasChor~\cite{shen-haschor} solves the KoC problem in what \citeauthor{shen-haschor} describe as an ``admittedly heavy-handed'' way: by broadcasting the chosen branch of each conditional to all parties.
This approach reduces programmer burden compared to select-\&-merge KoC
and allows Haskell's type checker to guarantee choreographies are projectable
(which enables safe run-time EPP).
Unfortunately, such \emph{broadcast-based} KoC management can result in unneeded additional communication.
For example, parties who do not participate in either branch of a conditional expression should be able to skip the whole expression;
sending the branch choice to these parties is a waste of communication.

\section{Main Ideas}
\label{sec:main-ideas}

This section describes three new concepts in choreographic programming.
The first two, \emph{conclaves} (\Cref{sec:conclaves}) and \emph{multiply-located values} (MLVs) (\Cref{sec:multiply-located-values}),
work together to avoid redundant communication in library-level implementations of CP.
The third, \emph{census polymorphism}, is a desirable feature for any user-facing CP system (\Cref{sec:census-poly}).
We use a case study, described in \Cref{sec:kvs-example}, as an example to introduce these concepts.
In \Cref{sec:formal-stuff} we show that these concepts are sound,
and in \Cref{sec:implementation} we discuss our implementations.

\subsection{Case Study: A Key-Value Store with Backup Servers}
\label{sec:kvs-example}

\Cref{fig:example-choreography} shows a choreographic function \inlinecode{kvs} in which a collection of servers
maintain replicas of a key-value store.
The function is polymorphic (parametric) over the number of servers that will actually be used,
but at least one, identified as \inlinecode{primary}, is guaranteed to exist and is the only one the client communicates with.
In the case of a \inlinecode{Get} request, the primary server can handle it unilaterally,
but they still multicast the request to the other servers so they can know to skip the work they'd do for a \inlinecode{Put}.
In the case of a \inlinecode{Put} request, the primary server will not respond to the client until they know
that the change to the key-value store has been replicated by all the servers,
but the way they do this actually maintains a polite fiction for the client:
Writes to the local states (stored in \inlinecode{IORefs}) can sometimes fail!
Checking for and correcting such mistakes is time-consuming,
so that process is deferred until after the client has gotten their response.
(A third request form, \inlinecode{Stop}, is used to shut the whole system down.)

The code in \Cref{fig:example-choreography} is modern Haskell
using officially-maintained GHC extensions (\eg \inlinecode{DataKinds}),
common packages (\eg \inlinecode{containers}),
and the \MultiChor library (one of our three implementations discussed in \Cref{sec:implementation}).

\begin{figure}[tbp]
\begin{mdframed}
\begin{tabular}{c c}
\begin{minipage}[b]{7.5cm}
\begin{minted}{haskell}
data Request = Put String String | Get String | Stop; data Response = Found String | NotFound | Stopped
type State = Map String String

newEmptyState :: IO (IORef State)  -- updateState returns the value previously stored under the key, or
                                   -- `NotFound`. Has a small chance of randomly saving the wrong value!
updateState :: IORef State -> String -> String -> IO Response
lookupState :: IORef State -> String -> IO Response
hashState :: IORef State -> IO Int

kvs :: (KnownSymbol client, KnownSymbol primary,
        KnownSymbols servers, KnownSymbols census) =>
       Member client census ->
       Member primary servers ->
       Subset servers census ->
       Faceted servers '[] (IORef State) ->
       Located '[client] Request ->
       Choreo census IO (Located '[client] Response)
kvs client primary servers stateRefs request = do
  let primary' = inSuper servers primary
  request' <- (client, request) ~> primary' @@ nobody
  requestShared <- (primary', request') ~> servers
  response' <- conclave servers do
    naked allOf requestShared >>= \case
      Put key val -> do
        responses <- parallel allOf ( \sr un -> updateState
                                  (viewFacet un sr stateRefs)
                                  key val )
        _ack <- fanIn (primary @@ nobody) \sr -> do
                  ack <- locally sr \_ -> pure ()
                  (sr, ack) ~> (primary @@ nobody)
        return $ localize primary responses
      Get key -> locally primary ( \un -> lookupState
                             (viewFacet un primary stateRefs)
                             key )
      Stop        -> locally primary \_ -> return Stopped
  response <- (primary',
               flatten (primary @@ nobody) refl response'
              ) ~> client @@ nobody
  _ <- conclave servers do
    naked allOf requestShared >>= \case
      Put _ _ -> do
        hash' <- parallel allOf \sr un ->
                   hashState $ viewFacet un sr stateRefs
        hashes <- gather allOf (primary @@ nobody) hash'
        let check = (1 <) . length . nub . toList
        needsReSynch <- locally primary \un ->
                          pure $ check $ un singleton hashes
        broadcast (primary, needsReSynch) >>= \case
          True -> resynch stateRefs  -- Could take a while!
          False -> return ()
      _ -> return ()
  return response
\end{minted}
\end{minipage}
&
\begin{minipage}[b]{5.0cm}
    \includegraphics[width=5.5cm]{figures/KVS8Paper.pdf}
\end{minipage}
\end{tabular}
\caption{A choreography called \inlinecode{kvs} written in Haskell using \MultiChor, with its corresponding sequence diagram.
	 An unspecified number of \inlinecode{servers} maintain copies of a key-value-store,
	 and a \inlinecode{client} makes a request against that store by communicating with the \inlinecode{primary} server.
	 The first three arguments are proof-witness objects that identify the parties and prove they have the needed relationships.
	 The forth argument \inlinecode{stateRefs} is a faceted value containing the servers' individual copies of the \inlinecode{State},
	 which could accidentally diverge during \inlinecode{updateState} operations.
	 The fifth argument is the client's \inlinecode{request}.
	 Lines~20-21: \inlinecode{primary} receives the request and forwards it to all the other servers.
	 Lines~22-35: The servers do work without the client;
	 they all examine the request and, if it's a \inlinecode{Put} request, update their local state.
	 Line~28: \inlinecode{primary} will not proceed until it's received \inlinecode{_ack} flags from all the other servers.
	 Line~36: \inlinecode{primary} sends the response to \inlinecode{client}, who can skip ahead to line~52 and exit.
	 Lines~39-51: If the store was updated by a \inlinecode{Put},
	 then the servers compare hashes of their copies of the store
	 (the actual comparison is done by \inlinecode{primary} on lines~45-47),
	 and if necessary they enter the expensive process of re-syncing their copies.
	}
\label{fig:example-choreography}
    \Description{A choreography written in Haskell using MultiChor.
	         The function "kvs" spans lines 18 through 52,
		 and has a accompanying UML sequence diagram.}
\end{mdframed}
\end{figure}

\subsection{Conclaves}
\label{sec:conclaves}
A \emph{conclave} is a sub-choreography with a sub-census.

\begin{definition}[Census]
	The set or list of parties or roles who are eligible to participate in a given choreographic expression.
	Usually tracked as part of the context during type-checking.
	Instructions in a choreography that involve parties not in the choreography's census are erroneous.
\end{definition}

\begin{definition}[Conclave]
	A section of a choreography that is itself a choreography and that has a census
	which is a subset of the census of the surrounding choreography.
	EPP to party not in the conclave's census should result in a "skip" expression instead of evaluating the conclave's body.
\end{definition}

Our systems are not the first to track the census, the set of parties "present", as part of the context in which expressions are type-checked,
but earlier systems (\eg, \citet{chor-lambda}) did not leverage this information as part of their KoC strategy,
so they did not need a word for it.
Because our systems use conclaves as part of the KoC strategy, users will be directly concerned with the censuses.
In \Cref{fig:example-choreography} (lines~10-17), the choreography \inlinecode{kvs}
has a polymorphic census denoted by the type-variable \inlinecode{census}.
The block on line~22 uses the \inlinecode{conclave} operator to reduce its census to just \inlinecode{servers},
a subset of \inlinecode{census}.

Censuses and conclaves improve "broadcasts" as a KoC strategy.
Because the \emph{select-\&-merge} paradigm is difficult to embed in a library,
the prior state of the art for library-level CP has been to broadcast the guard or scrutinee
of control-flow branches to all participants.
(In HasChor, the broadcast is implicit in the \inlinecode{cond} operator.)
Our systems take a similar approach, but do not suffer the same efficiency problem as HasChor
because \inlinecode{broadcast} is interpreted in the context of the local census.
For example, on line~48 \inlinecode{primary} broadcasts \inlinecode{needsReSynch}
so that the choreography can proceed with the \inlinecode{resynch} function (or not, as needed).
Because the \inlinecode{broadcast} happens inside a conclave to \inlinecode{servers} (line~39),
it does \emph{not} entail a message to \inlinecode{client}, who is not involved in the \inlinecode{resynch} process.
The same effect can also be accomplished using a "multicast" operation.
On line~21 \inlinecode{primary} sends \inlinecode{request'} to all of \inlinecode{servers};
since \inlinecode{servers} is exactly the census of the conclave on line~22,
they can unwrap \inlinecode{requestShared} using \inlinecode{naked} (line~23)
and all branch on it together.

\subsection{Multiply-Located Values}
\label{sec:multiply-located-values}
Previous choreographic languages have featured \emph{located values}, values owned by a particular participant
(\eg, \inlinecode{Request @ "client"} from line~3 of \Cref{fig:simple-choreography}).
Endpoint projection at the owner results in the value itself,
while endpoint projection at other parties results in a special "missing" value (\eg, $\bot$ or \inlinecode{Empty}).
\emph{Multiply located values} (MLVs) are exactly the same except they are annotated with a non-empty \emph{set} of parties.

\begin{definition}[Multiply located value (MLV)]
  A choreographic data type annotated with a list of owners.
  EPP to any of the owners will result in a normal value.
  Critically, all of the owners will arrive at the \emph{same} value.
  EPP to anyone else will result in a placeholder.
\end{definition}

In the \MultiChor example in \Cref{fig:example-choreography} line~16,
\inlinecode{Located '[client] Request} is the type of a \inlinecode{Request} located at the client,
similar to the \HasChor type shown in \Cref{fig:simple-choreography}.
In \MultiChor, unlike \HasChor, the type-level list allows encoding multiple locations for a located value.
For example, the multicast on line~21 returns an MLV, and \inlinecode{requestShared} has type
\inlinecode{Located servers Request} (where the type-variable \inlinecode{servers} has kind "list of parties").
Prior works have objects with multiple owners as emergent structures in a language (\eg, choreographic processes~\cite{giallorenzo-choral},
distributed choice types~\cite{chor-lambda-2}), but these project to each owner's distinct view of the structure.

Conclaves and MLVs enable re-use of KoC.
Conclaves allow KoC among a subset of the census \textit{via} broad- or multi-casting,
but without MLVs, KoC decisions made inside the conclave are lost when the conclave exits.
With MLVs, conclaves can return values reflecting KoC decisions made inside the conclave,
and these decisions can be re-used later in the choreography.
Alternately, an MLV resulting from a multicast can be unwrapped for use in sequential conclaves;
for example the \inlinecode{server}s branch on \inlinecode{requestShared} in the conclave started on line~22,
and then they branch on it again in the conclave on line~39.
No additional communication is needed for KoC in the second conditional!
We call this KoC management strategy \emph{conclaves-\&-MLVs}.

\subsection{Census Polymorphism}
\label{sec:census-poly}

It is common to describe protocols, even choreographies, without being specific about the number of participants,
but in prior CP systems it has not been possible to write actual code without enumerating the participants or the roles they will fulfil.
This is a serious limitation for writing choreographic software;
modern concurrent systems often use dozens to thousands of participants
and are defined parametrically over their number of participants~\cite{bigConcurrent1, corrigan2017prio, bigConcurrent3, bigConcurrent4, dprio2023}.
We assert that such parametric protocol declarations are a required feature for CP to find mainstream use;
our systems provide it in the form of \emph{census polymorphism}.

\begin{definition}[Census polymorphism]
  \emph{(of a choreographic expression)} Being polymorphic over the census,
  including not just the specific identities listed but also the quantity.\\
  \emph{(of a CP system or language)} The ability to express census polymorphic choreographies
  and related polymorphic types such as MLVs that are polymorphic over their ownership set.
\end{definition}

As already mentioned, the census of \inlinecode{kvs} in \Cref{fig:example-choreography} is a type variable.
The "proof witness" arguments (see \Cref{sec:membership}) \inlinecode{client}, \inlinecode{primary} and \inlinecode{servers}
tell us that those parties\footnote{
  The exact identities of \inlinecode{client} and \inlinecode{primary} are also polymorphic;
  this is "process polymorphism", which various CP systems to date have implemented.
}
are members of the census,
but \inlinecode{kvs} can be called with \emph{any} census satisfying the constraints.
At a shallow level, this suffices to call \inlinecode{kvs} census polymorphic,
but consider how members of \inlinecode{servers} \emph{other than} \inlinecode{primary}
might participate in the choreography using only the tools described so far:
They'll receive any broadcasts and participate in any active replication that applies to all of \inlinecode{servers},
but there's no way to specify them as the senders of messages, nor is there a way to specify that any of them should receive a message
\emph{except} by broadcasting it to \inlinecode{servers}.
\MultiChor and our other systems allow an unspecified quantity of parties to actively participate in the choreography.
For example, line~44 uses the \inlinecode{gather} operation,
in which a polymorphic list of participants (\inlinecode{allOf} the census, which is \inlinecode{servers})
each send a computed value to a common recipient (\inlinecode{primary}) who aggregates them.

Two things are needed to make census polymorphism useful: 
A way to loop over a polymorphic list of parties,
and the ability to express and use divergent data known by un-enumerated parties.
We discuss divergent located data first, and call our solution \emph{faceted values}\footnote{
    The word "faceted" is most commonly used in reference to a cut gemstone,
    but analogy to the facets of polyhedral playing dice might be more on-the-nose.
    Our faceted values are basically the same as the faceted values introduced in \citet{austin2012},
    except their public facet is always "$\bot$" and multiple parties have distinct private facets.
}.
For example, in \Cref{fig:example-choreography}, the type of the argument \inlinecode{stateRefs} is given on line~15
as \inlinecode{Faceted servers '[] (IORef State)},
because each server maintains their (possibly incorrect) copy of the \inlinecode{State} in a private \inlinecode{IORef}.

\begin{definition}[Faceted value]
  A choreographic data type annotated with a list of owners.
  EPP to any of the owners will result in a normal value specific to that party;
  there is no expectation for the owners to have the same value, or for them to know each other's values.
  EPP to a non-owner will result in a placeholder.
\end{definition}

Faceted values are necessary, among other use-cases, as the type of the argument of \inlinecode{gather}
and the return type of \inlinecode{scatter}.
Without them, each sender would need to generate its value to send "inside" the \inlinecode{gather} function,
and the only way for the sent values to be distinct would be by using private local state accessed by \inlinecode{locally}.
Even worse, \inlinecode{scatter}ed data would need to be aggregated into some non-faceted type before the \inlinecode{scatter} function could return,
and simply appending the scattered values to a list would erase their association with their recipients.
(Such a list conversion would be impossible anyway in the languages we use in this paper
because \inlinecode{Located} values with different owners have different types.)

Using faceted values, \inlinecode{scatter} and \inlinecode{gather} can be given sensible types.
A third operation, \inlinecode{parallel} (as used on line~42) generalizes the \inlinecode{locally} operator
(which is basically the same in \MultiChor as in HasChor) to represent divergent parallel local computation by a list of parties.
We find that these three operations can suffice for many uses cases.
For example, it would be possible to re-write the "acknowledgement" step on lines~28-30 using \inlinecode{parallel} and \inlinecode{gather}.
On the other hand, we do not advocate implementing those three functions as primitive operators;
they can each be implemented as a loop over a (type level) list of parties.
As we will see in \Cref{sec:implementation}, the \MultiChor library offers such a general-purpose loop operation,
but \Chorus and \ChoreographyTS use an intermediate strategy:
They implement two "loop" functions, \inlinecode{fanOut} and \inlinecode{fanIn} with somewhat restricted type signatures:
\inlinecode{fanOut}'s argument is any choreography that results in a \inlinecode{Located} value at the party identified by the loop variable;
it aggregates these results as a \inlinecode{Faceted}.
\inlinecode{fanIn} is almost the same, except that the owners of the resulting MLVs do not vary over the loop iterations,
and they are aggregated in a multiply-located \emph{quire} (see below) owned by that same list of recipients.
(In \MultiChor, these are derived helper functions.)
On line~28, acknowledgement that the \inlinecode{Put} request has been processed is expressed as a \inlinecode{fanIn};
the loop body sends a static "unit" flag~\inlinecode{()} from the loop party \inlinecode{sr} to \inlinecode{primary}.
\inlinecode{fanOut} and \inlinecode{fanIn} do not induce conclaves in their loop bodies;
the entire census may participate in every cycle of the loop.
If that is not desired, one can explicitly call \inlinecode{conclave} inside the loop body.

\begin{definition}[Quire]
  \emph{(in the context of CP)} A vector of values, all of the same type, indexed by the type level party with which each value is associated.
  Different from a length-indexed list only in that the indexing types have kind "party" instead of "bounded natural numbers".
  \\
  \emph{(in common parlance)} A stack of sheets of paper, all cut to the same size.\\
  \emph{(pronounced "\emph{choir}", rhymes with "\hspace{0.2em}\emph{buyer}\hspace{-0.1em}".)}
\end{definition}

Quires are useful in choreographies, \eg in the return type of \inlinecode{gather}.
That said, a quire is not a choreographic data type; EPP has no effect on it.
In \Cref{sec:implementation} we will see that a faceted value can be modeled as a mapping from parties to located values;
in this sense the sibling of a faceted value is specifically a \emph{multiply-located} quire.
On line~44, \inlinecode{hashes} has type \inlinecode{Located '[primary] (Quire servers Int)}.

In \Cref{sec:case-studies} we present additional examples of census polymorphic choreographies in each of our implementations.
In earlier CP systems it would be necessary to hard-code the number of participants when writing choreographies like these.
Our implementations each provide the needed features in different ways as discussed in \Cref{sec:implementation}.

\section{Formalizing Conclaves \& Multiply-Located Values}
\label{sec:formal-stuff}

The conclaves-\&-MLVs strategy for managing KoC is the most obvious unifying feature of the libraries we describe in Section~\ref{sec:implementation}.
It's more easily embedded in the type systems of general-purpose programming languages than select-\&-merge,
and doesn't induce spurious communication like \HasChor's auto-broadcast system.
In \Cref{sec:formalism} (and in greater detail in Appendix~\ref{sec:more-formalism})
we demonstrate its theoretical soundness by formalizing a small CP language and
proving the usual theorems of progress, preservation, and bisimulation.
Together, these theorems ensure deadlock freedom (\Cref{corr:deadlock}) for all well-typed programs.
Deadlock freedom is a classical result and one of the original motivations for the CP paradigm~\cite{montesi-carbone-dfbd}.
In \Cref{sec:select-n-merge-comparison} we compare the expresivity and efficiency of conclaves-\&-MLVs.

\paragraph{Regarding census polymorphism}
In all three of our implementations, census polymorphism can be resolved statically,
\ie, while one can write choreographies and choreographic functions that are census-polymorphic,
it is always possible in principal to unroll the top-level choreography
(that actually gets compiled)
into a monomorphic form.
As we show in \Cref{sec:census-poly-haskell},
Haskell's powerful type system lets us safely implement the types and functions needed for census polymorphism
as a transparent layer on top of a core API.
Therefore, we do not bother with a separate proof of the soundness of static census polymorphism.
We leave any clarity about the prospects of \emph{dynamic} census polymorphism for future work.

\begin{figure}[tbp]
    \begin{mdframed}
\small
      \begin{tblr}{colspec={r c},
	           hline{2-Y}={dotted,lightgray,endpos,l=-1,r=-1},
		   rowsep=0.5em
                  }
		    \textbf{(a)}&
$\begin{aligned}
M  \BNF   &  V
   \BNFOR    M M
   \BNFOR    \CASE{\nonempty{p}}{M}{x}{M}{x}{M}
					    \\[0.3em]
V  \BNF   &  x
   \BNFOR    (λ x:T \DOT M)@\nonempty{p}
   \BNFOR    ()@\nonempty{p}
   \BNFOR    \INL V
   \BNFOR    \INR V
   \BNFOR    \PAIR V V \\
   \BNFOR &  (V, \dots, V)
   \BNFOR    \FST{\nonempty{p}}
   \BNFOR    \SND{\nonempty{p}}
   \BNFOR    \LOOKUP{n}{\nonempty{p}}
   \BNFOR    \COMM{p}{\nonempty{p}}
                                            \\[0.3em]
T  \BNF   &  d@\nonempty{p}
   \BNFOR    (T → T)@\nonempty{p}
   \BNFOR    (T, \dots, T)
         \quad\quad\quad
d  \BNF      ()
   \BNFOR    d + d
   \BNFOR    d × d
\end{aligned}$\\
		    \textbf{(b)}&
$\begin{gathered}
\myference{TVar}
          {x : T \in Γ \quad T' = T \mask Θ}
          {Θ;Γ ⊢ x : T' }
          \quad
\myference{TLambda}
          {\nonempty{p};Γ,(x:T) ⊢ M : T' \quad
           \nonempty{p} \subseteq Θ \quad
           \noop{\nonempty{p}}{T}}
          {Θ;Γ ⊢ (λ x:T \DOT M)@\nonempty{p} : (T → T')@\nonempty{p}}
          \vdbl
\myference{TCase}
          {Θ;Γ ⊢ N : T_N \quad
           (d_l + d_r)@\nonempty{p} = T_N \mask \nonempty{p} \\
           \nonempty{p};Γ,(x_l:d_l@\nonempty{p}) ⊢ M_l : T \quad
           \nonempty{p};Γ,(x_r:d_r@\nonempty{p}) ⊢ M_r : T \quad
           \nonempty{p} \subseteq Θ}
          {Θ;Γ ⊢ \CASE{\nonempty{p}}{N}{x_l}{M_l}{x_r}{M_r} : T}
          \vdbl
\myference{TInl}
          {Θ;Γ ⊢ V : d@\nonempty{p}}
          {Θ;Γ ⊢ \INL V : (d + d')@\nonempty{p}}
          \quad
\myference{TCom}
          {s \in \nonempty{s} \quad
           \nonempty{s}\cup\nonempty{r} \subseteq Θ}
          {Θ;Γ ⊢ \COMM{s}{\nonempty{r}} : (d@\nonempty{s} → d@\nonempty{r})@(\set{s}\cup\nonempty{r})}
          \vdbl
\end{gathered}$\\[-0.5em]
		    \textbf{(c)}&
$\begin{gathered}
⟦\COMM{s}{\nonempty{r}}⟧_p \DEF
  \begin{cases}
      p = s, p \in \nonempty{r}      & \DEFCASE \SEND{\nonempty{r} ∖ \set{p}}^\ast \\
      p = s, p \not\in \nonempty{r}  & \DEFCASE \SEND{\nonempty{r}} \\
      p \not = s, p \in \nonempty{r} & \DEFCASE \RECV{s} \\
      \quad\text{else}                    & \DEFCASE  ⊥
  \end{cases}
\end{gathered}$\\
		    \textbf{(d)}&
$\begin{gathered}
\myference{LAbsApp}
          {}
          {(λ x \DOT B) L \prcstep{∅}{∅} \FLR{B[x:=L]}}
          \quad
\myference{LSend1}
          {}
          {\SEND{p^{\ast}} () \prcstep{\set{(p, ()) \mid p \in p^{\ast}}}{∅} ⊥}
          \vdbl
\myference{LRecv}
          {}
          {\RECV{p} L_0 \prcstep{∅}{\set{(p, L)}} L}
          \quad
\myference{NPro}
          {B \prcstep{μ}{∅} B'}
          {p[B] \netstep{p}{μ} p[B']}
          \vdbl
\myference{NCom}
          {\mathcal{N} \netstep{s}{μ∪\set{(r,L)}} \mathcal{N}'
           \quad B \prcstep{∅}{\set{(s, L)}} B'}
          {\mathcal{N} \mid r[B] \netstep{s}{μ} \mathcal{N}' \mid r[B']}
          \quad
\myference{NPar}
          {\mathcal{N} \netstep{}{∅} \mathcal{N}'}
          {\mathcal{N} \mid \mathcal{N}^{+} \netstep{}{∅} \mathcal{N}' \mid \mathcal{N}^{+}}
\end{gathered}$
\end{tblr}
    \caption{Selected rules and definitions from our formalism.
             \textbf{(a)}
	     The complete syntax of \HLSCentral terms ($M$, $N$) and types ($T$).
	     \textbf{(b)}
             \textsc{TVar}, \textsc{TLambda}, \textsc{TCase}, \textsc{TInl}, and \textsc{TCom}
             are representative of the thirteen total typing rules for \HLSCentral.
	     \textbf{(c)}
             $⟦\cdot⟧_p$ is EPP from \HLSCentral to \HLSLocal, parameterized by $p$;
             here we show only its definition for the communication operator.
	     \textbf{(d)}
             \HLSLocal is untyped; \textsc{LAbsApp}, \textsc{LSend1}, and \textsc{LRecv}
             are representative of its fifteen total semantics rules.
             \textsc{NPro}, \textsc{NCom}, and \textsc{NPar} are the complete semantics rules for \HLSNet.
             Appendix~\ref{sec:more-formalism} explains all of these systems and their relationships in detail.}
    \label{fig:typing-excerpt}
    \Description{Inference rules for the Lambda-C family of formal languages.}
    \end{mdframed}
\end{figure}

\subsection{A Conclaves-\&-MLVs Lambda Calculus}
\label{sec:formalism}

Our overall strategy (as in textbook CP models~\cite{montesi_book}) is to
\begin{itemize}[leftmargin=14pt]
    \item Define a syntax, semantics, and typing for a \emph{central choreographic language} \HLSCentral.
        We show that \HLSCentral has the usual $\lambda$-calculus properties of progress and preservation.
    \item Define EPP for \HLSCentral as a translation into a \emph{local language} \HLSLocal.
    \item Define semantics for \HLSNet, unordered collections of concurrent processes evaluating \HLSLocal asynchronously.
        We show the semantics of \HLSCentral are a sound and complete representation of \HLSNet.
\end{itemize}

\HLSCentral is a finite, monomorphic, higher-order choreographic lambda calculus
with the key features that unite our real implementations:
census tracking, conclaves, MLVs, and multicast communication.
The syntax of \HLSCentral is modeled on \chorLambda~\cite{chor-lambda},
but it uses more location annotations, and "data" (which can be communicated) is clearly distinguished from functions, \textit{etc.} (which cannot).

\Cref{fig:typing-excerpt}\textit{(a)} shows the syntax of \HLSCentral.
Meta-variables $M$ and $N$ represent expressions and $V$ is for values.
\Cref{fig:typing-excerpt}\textit{(b)} shows select typing rules for \HLSCentral (using name-scheme "\textsc{T...}").
The census is a typing context, typically denoted $\Theta$.
Because \HLSCentral is doesn't support subset-based subtyping,
the operator \mask is used in typing and semantics to limit the location annotations in an expression.
In \textsc{TVar} the expression $T \mask \Theta$ reads as "type $T$ masked to the local census $\Theta$"
or "$\Theta$'s view of $T$".
Many of the rules require that specified parties are in the census.
Both lambda expressions and case expressions conclave their body clauses;
a \inlinecode{conclave} operator can be accurately mimicked by simply wrapping any expression in a lambda-abstraction.
Functions are owned by their participants (and since they can't be communicated, this ownership doesn't typically change).
The precondition $\noop{\nonempty{p}}{T}$ in \textsc{TLambda} simply says that $T$ is already masked to $\nonempty{p}$.
\textsc{TCase} enforces KoC by conclaving to $\nonempty{p}$ and checking that all of $\nonempty{p}$ own the scrutinee $N$.
\HLSCentral supports products and co-products as multiply-located "data" (\eg \textsc{TInl}) as well as heterogeneous vectors at the expression level.
$\COMM{s}{\nonempty{r}}$ is a multicast operator from $s$ to $\nonempty{r}$.
Without subtyping, the recipient list must be precise;
we do not \emph{automatically} include any parties who already know the argument in the resulting ownership set,
but they can be explicitly included.
\HLSCentral's semantics are call-by-value, complicated by the need to manage the party-set annotations;
this management complexity is necessary to to support a monomorphic type-preservation theorem, which we prove in Appendix~\ref{sec:preservation-proof}.
In Appendix~\ref{sec:progress-proof} we prove the usual semantic-progress theorem.
Since \HLSCentral doesn't have recursive functions,
this suffices to show that well-typed \HLSCentral programs will always evaluate under \HLSCentral's own semantics,
but it remains for us to show that said semantics faithfully represent the "real" computational model we are concerned with.

\paragraph{EPP from \HLSCentral to Processes}
Our intuitive expectation of endpoint projection is that it should change a program very little for participants
and erase it (project it to a placeholder, in this case ⊥) for non-participants.
Accounting for the ⊥-normalizing operation "$\FLR{\cdot}$", that is exactly how EPP works for \HLSCentral.
We denote EPP to the party $p$ by $⟦\cdot⟧_p$.
The most interesting case of the definition, $⟦\COMM{s}{\nonempty{r}}⟧_p$, is excerpted in \Cref{fig:typing-excerpt}\textit{(c)}.
The target of \HLSCentral's EPP is the analogous process language \HLSLocal;
we use meta-variables $B$ and $L$ for \HLSLocal's expressions and values, respectively.
\HLSLocal has an operator $\RECV{s}$ denoting expectation and reception of a message from $s$.
($\RECV{s}$ takes an argument, which it ignores.)
It has two "send" operators, $\SEND{p^{\ast}}$ and $\SEND{p^{\ast}}^{\ast}$, which differ only in that,
after transmitting the argument, $\SEND{p^{\ast}}$ evaluates to ⊥ and $\SEND{p^{\ast}}^{\ast}$ to the unchanged argument.
\Cref{fig:typing-excerpt}\textit{(d)} includes rules \textsc{LAbsApp}, \textsc{LSend1}, and \textsc{LRecv} from \HLSLocal's semantics.
\HLSLocal-semantic steps are annotated by sets denoting sent ($⊕$) and received ($⊖$) messages,
\eg in \textsc{LRecv} the evaluating party can receive any value and the step-annotation reads \textit{"received $L$ from $p$"}.

\paragraph{Networks of Processes}
The semantics of \HLSLocal "lift" to a network setting \HLSNet by rules \textsc{NPro}, \textsc{NCom}, and \textsc{NPar},
shown in \Cref{fig:typing-excerpt}\textit{(d)}.
The meta-variable $\mathcal{N}$ denotes a \HLSNet network state: a table of parties and their respective \HLSLocal expressions.
\HLSNet-semantic steps are annotated only with messages that are sent and not received.
Since we are not concerned with dropped messages, the "ground truth" of our computational model is $∅$-annotated \HLSNet-semantic steps.
In other words, using \textsc{LSend1} in \HLSNet computation via \textsc{NPro} is only possible
if \textsc{NCom} is used sufficiently many times in the \emph{same step} to cancel out every $(p, ())$ in $\mu$.

\paragraph{Deadlock Freedom}
We prove by bi-simulation that the semantics of \HLSCentral are a sound and complete representation of
the behavior of the \HLSNet networks to which \HLSCentral choreographies project.
First, in Appendix~\ref{sec:soundness-proof} we show that
given some $M$ in \HLSCentral and its projection $\mathcal{N}$ in \HLSNet,
any network configuration that $\mathcal{N}$ can advance to
either will \emph{be} the projection of an expression $M'$ to which $M$ itself can advance,
or will \emph{be able to advance to} the projection of such an $M'$.
Second, in Appendix~\ref{sec:completeness-proof} we show that if a \HLSCentral choreography $M$ can advance to some $M'$,
then the projection of $M$ can advance to the projection of $M'$.
Since well-typed \HLSCentral programs never get stuck, it follows that their projections don't either, including by deadlocking.
\begin{corollary}[Deadlock Freedom]\label{corr:deadlock}
  If $Θ;∅ ⊢ M : T$ and $⟦M⟧ \netstep{}{∅}^{\ast} \mathcal{N}$,
    then either $\mathcal{N} \netstep{}{∅}^{\ast} \mathcal{N}'$
    or for every $p\in\roles{M}$, $\mathcal{N}(p)$ is a value.
\end{corollary}

\paragraph{Relation to the implementations}
\HLSCentral is a pure lambda calculus with a substitution-based semantics.
Our implementations are embedded within mainstream programming languages, and necessarily interact with an external world in impure ways.
Correctly implementing conclaves-\&-MLVs CP in an impure stateful language requires maintaining an invariant implicit in the \emph{completeness}
of \HLSCentral's centralized semantics:
that the copies of an MLV residing on its owning machines are always the same.
Haskell's type system allows clear delineation of pure \emph{vs} impure code,
and \MultiChor can enforce this invariant to exactly the extent that Haskell enforces said delineation.
The type systems of Rust and TypeScript are less concerned with side effects than Haskell’s is,
so users of our implementations in those languages
might use local state improperly, undermining the invariant and causing the choreography to go wrong.
(We have endeavored to design the APIs to discourage such accidents.)

Furthermore, the distributed semantics of \HLSNet do not account for the possibility of dropped messages.
In practice, the guarantees of CP only hold in the context of reliable communication and reliable participants who faithfully follow the choreography.
Cutting edge theoretical work such as \citet{graversen2025promisingfuture} may improve the situation regarding dropped messages.

\subsection{Expressivity of Conclaves-\&-MLVs}
\label{sec:select-n-merge-comparison}
In principle, the conclaves-\&-MLVs paradigm is equivalent to the select-\&-merge paradigm
in the computations and communication patterns it can express,
but this is not self-evident, is not without caveat, and depends greatly on the details of any given system.

\paragraph{Select-\&-merge is as expressive as conclaves-\&-MLVs}
While most select-\&-merge systems to date cannot represent MLVs
(and therefore cannot recycle KoC in sequential conditional expressions),
\citet{chor-lambda-2} showed that the \chorLambda language~\citep{chor-lambda}
can represent \emph{"distributed choice types"},
data structures equivalent in behavior to MLVs.
Therefore, assuming a sufficiently capable type system and ignoring implementation details,
any choreography that be be represented in a conclaves-\&-MLVs system (\eg, a well-typed \HLSCentral expression)
can also be represented as a (well-typed, projectable) choreography in certain select-\&-merge systems.

\paragraph{Conclaves-\&-MLVs is as expressive as select-\&-merge}
A feature of the select-\&-merge paradigm is that a party can participate in a conditional expression without having KoC,
provided its own behavior is the same in all branches.
Such a choreography can easily be transformed for a conclaves-\&-MLVs system by decomposing the conditional into
sequential conditionals (in conclaves) with the ignorant party's behavior interleaved.
When KoC must be communicated, select-\&-merge systems require that \inlinecode{select} statements to the party in question
align across all branches.
A conditional expression with \inlinecode{select} statements can be transformed for a conclaves-\&-MLVs system by
decomposing it into two sequential conclaved conditionals, a setup and a continuation.
Each branch of the setup will end where the \inlinecode{select} was, and return the selected flag.
In between the two conditionals the controlling party multicasts the chosen flag; the continuation branches on that multiply-located flag
and picks up where the setup left off.
In more complicated cases a party may need only \emph{partial} knowledge of choice,
\eg a party might need only binary knowledge of whether the first of three branches was chosen.
Transforming such a choreography for a conclaves-\&-MLVs system starts the same way but is more complicated:
Inside the continuation, the parties with more complete KoC will need to branch again in an inner conclave.
In all of our systems so far this process can involve \emph{"dead branches"},
clauses of \inlinecode{case} expressions that the type-checker or syntax expect to exist but which a human can observe will never be activated.

\section{Implementations}
\label{sec:implementation}

We demonstrate the practicality of the theory discussed in Section~\ref{sec:formal-stuff}
by presenting implementations in Haskell, Rust, and TypeScript.
One of our contributions in this work is to show that library-level CP is a practical technology to build and use,
so in this section we discuss our libraries in detail.
Any library-level implementation will inherit particular details from its host language that affect its capabilities and engineering.
Table~\ref{fig:implementations} notes the high-level differences between our three implementations.

\begin{table}[tbp]
    \begin{tabular}{
            R{6.6em}    | C{5.3em}    | C{5.2em}           | C{5.4em}           | C{5.5em}       | C{5.4em}  }
                        & \HasChor    & \HLSCentral        & \MultiChor         & \Chorus        & \ChoreographyTS            \\ \hline
 Multiply-located values \& multicast
                        & \xmark      & \cmark             & \cmark             & \cmark         & \cmark                     \\[1.3em]
 Censuses \& conclaves   & \xmark      & \cmark             & \cmark             & \cmark         & \cmark                     \\[1em]
 Membership constraints & n/a         & custom             & proof objects      & indexed traits & union types                \\[1em]
 Census polymorphism    & \xmark      & \xmark             & \cmark             & \cmark         & \cmark                     \\[1em]
 EPP strategy           & free-monad  & custom             & free-monad         & EPP-as-DI      & EPP-as-DI                  \\[1em]
 Host language          & Haskell     & n/a                & Haskell            & Rust           & TypeScript                 \\[0.3em]
    \end{tabular}\\[0.3em]
    \caption{Comparisons between \HasChor, \HLSCentral (our formal model), and our three libraries.
    Our formal model, which is not intended for use in real software, is entirety monomorphic.
    Our implementations all offer a superset of \HasChor's features.    }
    \label{fig:implementations}
\end{table}

\MultiChor, the Haskell implementation discussed in Section~\ref{sec:multichor},
uses the same free-monad approach as \HasChor to implement conclaves-\&-MLVs;
its core monad has minimalist foundational constructors supporting a derived ergonomic API.

The free-monad engineering approach is less popular outside of Haskell and related languages.
In Section~\ref{sec:epp-as-di} we demonstrate that dependency injection \textit{via} higher-order functions is sufficient to implement library-level CP.
We call this pattern \emph{endpoint projection as dependency injection}, or EPP-as-DI.
\ChoreographyTS, our TypeScript library, uses the EPP-as-DI pattern straightforwardly to implement conclaves-\&-MLVs CP.
Rust's type system places some challenging limits on higher-order and anonymous functions;
\Chorus, our Rust library, works around these limitations by using trait implementations for dependency injection, but the approach is still recognizably EPP-as-DI.

Correctly using both conclaves and MLVs involves checking if parties are members of relevant censuses and ownership lists;
we insist on having our host languages' type systems perform these checks to avoid run-time errors.
Haskell, Rust, and TypeScript have fundamentally different type systems;
in Section~\ref{sec:membership} we discuss the approach to membership constraints that we use in each of our three implementations.
\Cref{sec:census-poly-haskell,sec:census-poly-rust} discuss how our implementations support census polymorphism.

\subsection{Censuses, Conclaves, and MLVs in Haskell}
\label{sec:multichor}
The type of a choreography in \MultiChor is of the form \inlinecode{Choreo p m t},
where \inlinecode{p} is the census and \inlinecode{t} is the return type.
\MultiChor uses the same free-monad approach as \HasChor~\cite{shen-haschor} to implement choreographic programming, EPP,
and the final interpretation to a real communication mechanism.
Also like \HasChor, \MultiChor's \inlinecode{Choreo} monad is parameterised by a \emph{local monad}
(\inlinecode{m}) in which parties' local computations can be expressed.
A \MultiChor type \inlinecode{Located ps t} is a multiply-located \inlinecode{t} owned by the parties \inlinecode{ps}.
It is possible to write \MultiChor functions that look and work like each of \HasChor's three primitive operators,
but the derived API in which users write \MultiChor choreographies contains a clear analog of only one of \HasChor's primitives.
Haskell's monadic-\inlinecode{do} notation and purity-oriented type system make \MultiChor code concise and safe
(in the sense that users are unlikely to accidentally invalidate important invariants).

As explained in Section~\ref{sec:main-ideas},
our KoC strategy requires that the correctness (well-typed-ness) of choreographies be judged in the context of a census.
\MultiChor adds the census as a type parameter of the \inlinecode{Choreo} monad.
Its kind is \inlinecode{[Symbol]},
which is to say that the census is a type-level list of parties and parties are type-level strings.
\inlinecode{Choreo} is \emph{not} an \emph{indexed} monad (that is, executing a monadic operation doesn't change the census),
but monadic operations like \inlinecode{conclave} can take choreographies with smaller censuses as arguments.

The fundamental operations of \MultiChor's \inlinecode{Choreo} monad are
\inlinecode{conclave}, \inlinecode{broadcast'}, \inlinecode{locally'}, and \inlinecode{congruently'}.
Their type signatures are given in \Cref{fig:operators-multichor}.
Like in \HasChor, these are free-monad constructors; their behavior is implemented in interpreters
that carry out EPP or implement a centralized semantics.
Three of them have their names "primed" because the un-primed versions of these names are reserved for more ergonomic derived functions.
For example, \inlinecode{locally'} takes a single argument, a computation in the local monad, and requires that the census contains
\emph{a single party}, who will execute that computation.
The un-primed \inlinecode{locally} takes an additional argument that identifies a single party from a larger census;
it uses \inlinecode{conclave} to correctly call \inlinecode{locally'}.
\inlinecode{broadcast} shares a \inlinecode{Located} value with the entire census so the unwrapped value can be used;
by combining this with \inlinecode{conclave} we can implement point-to-point or multicast communication.
From the perspective of a centralized semantics, \inlinecode{conclave} doesn't do anything at all besides run
the sub-choreography,
but EPP to a party \emph{not} in the sub-census skips the sub-choreography and just returns \inlinecode{Empty}.

\inlinecode{congruently} lets us leverage MLVs to concisely write actively-replicated computations.
In contrast to \inlinecode{locally}, the computation is performed by multiple parties
and the result is multiply-located across all of them.\footnote{
    The entire census participates in the primed version, and its result is returned naked.
    The behavior of \inlinecode{conclave} and the more fundamental rules of monadic programming
    ensure the un-primed \inlinecode{congruently} behaves correctly.
}
For the execution of these actively-replicated computations to correctly return an MLV,
all the parties must be guaranteed to be doing a pure computation on the same data.
Haskell makes it easy to enforce such a guarantee to a practical (but not unbreakable) extent.
This is why \inlinecode{congruently} does not grant access to the local monad \inlinecode{m}.
It also requires that the computation not have access to the specific identity of the computing party,
unlike \inlinecode{locally} and the similar-looking function \inlinecode{parallel} mentioned in \Cref{sec:census-poly}.
Weakening (or subverting) these restrictions would allow a user to violate \MultiChor's invariant that MLVs (\inlinecode{Located} values)
have the same value across all their owners.

It is critical to the safety of \MultiChor that the projection of a choreography to any given party will not use
any other party's \inlinecode{Located} values.
We use the same basic strategy for this as \HasChor:
\inlinecode{Located}'s constructors, \inlinecode{Wrap} and \inlinecode{Empty}, are hidden inside the core module
and afforded only by dependency injection to \inlinecode{locally} and \inlinecode{congruently}.
The specific "unwrapper" functions afforded to \inlinecode{locally} and \inlinecode{congruently}
are known to user code only by their type signatures, which have respective aliases \inlinecode{Unwrap} and \inlinecode{Unwraps}.
\inlinecode{Located}'s constructors are also used by two less-critical functions, \inlinecode{flatten} and \inlinecode{othersForget}.
These are needed for shrinking ownership sets or un-nesting \inlinecode{Located} values;
for example in \Cref{fig:example-choreography} line~37,
\inlinecode{response'} has type
\inlinecode{Located servers (Located '[primary] Response)},
so it must be \inlinecode{flatten}ed before it can be used.
These two functions could be derived using \inlinecode{congruently},
but by implementing them in the core module where they can pattern-match \inlinecode{Located} values we are able to make them not-monadic,
and so more convenient.

\begin{figure*}[tbp]
  \begin{mdframed}
    \begin{minipage}[b]{13cm}
\begin{minted}{haskell}
locally' :: (KnownSymbol l) => (Unwrap l -> m a) -> Choreo '[l] m a

congruently' :: (KnownSymbols ls) => (Unwraps ls -> a) -> Choreo ls m a

broadcast' :: (Show a, Read a, KnownSymbol l) =>
              Member l ps -> (Member l ls, Located ls a) -> Choreo ps m a

conclave :: (KnownSymbols ls) => Subset ls ps -> Choreo ls m a -> Choreo ps m (Located ls a)
\end{minted}
    \end{minipage}
    \caption{
        The fundamental operators for writing expressions in \MultiChor's \inlinecode{Choreo} monad.
        Of these four operators, \inlinecode{conclave} is the only one users will usually call directly;
        the other three can combine with each other (and with \inlinecode{conclave}) to make more user-friendly alternatives.
    }
    \label{fig:operators-multichor}
    \Description{Four type signatures for Haskell functions using the MultiChor library.}
  \end{mdframed}
\end{figure*}

\subsection{Endpoint Projection as Dependency Injection}
\label{sec:epp-as-di}

Engineering patterns natural to Haskell, such as the free-monad pattern that \MultiChor uses in its implementation, do not always translate safely or ergonomically to other languages.
Therefore, we do not advocate for porting \MultiChor \emph{directly} into languages like Rust or TypeScript.
In this section we demonstrate an alternative, more portable, approach.

\emph{Endpoint projection as dependency injection} (EPP-as-DI) is a simple technique for implementing library-level choreographic programming.
EPP-as-DI asks little from the host language: support for \emph{higher-order functions} is all that is required.
The complete idea of EPP-as-DI is that we can implement CP by
representing a choreography as a host-language function that takes \emph{choreographic operators} as arguments.
Endpoint projection determines the semantics of the choreography by injecting implementations of the choreographic operators
that are specific to the location to which it is projected.
As in \HasChor and \MultiChor, EPP happens at runtime, but the \emph{ability} to project to all relevant parties is checked statically.

We present two choreographic programming libraries implemented with EPP-as-DI: \Chorus for Rust and \ChoreographyTS for TypeScript.
In each case, we take advantage of the native abstractions of the host language (\eg, TypeScript's interfaces; Rust's traits) to make the implementation idiomatic, but EPP-as-DI itself is a general technique and not specific to any particular host language.

\begin{figure*}[tbp]
  \begin{mdframed}
    \begin{minipage}[b]{13cm}
\begin{minted}{typescript}
// A choreographic operator for performing a local computation on a location
type Locally<L extends Location> =
  <L1 extends L, T>(location: L1, callback: (unwrap: Unwrap<L1>) => T) => Located<T, L1>
// A set of choreographic operators that can be used in a choreography
type Dependencies<L extends Location> = {
  locally: Locally<L>;
  comm: Comm<L>;
  ... // other operators
}
// A choreography is a function that's expressed in terms of the signature of the dependencies
type Choreography<L extends Location, A, R> = (deps: Dependencies<L>, args: A) => R;
// EPP is done by executing the choreography function with concrete implementations of the operators
function epp<L extends Location, T extends L, A, R>(
  target: T, choreography: Choreography<L, A, R>, args: A
): R {
  const locally: Locally<L> = async (location, callback) => {
    if(location === target) { return await callback(unwrap(target)); }
  };
  return choreography({ locally, /* other operators */ }, args);
}
\end{minted}
    \end{minipage}
    \caption{
        The definition of choreography in \ChoreographyTS.
        The choreographic operators are defined as aliases for function types,
        and these all get packaged together in a record type \inlinecode[typescript]{Dependencies}.
        A \inlinecode[typescript]{Choreography} is a function from a \inlinecode[typescript]{Dependencies}
        (and any other arguments)
        to some return value.
        In order to create a \inlinecode[typescript]{Dependencies}, it must have real implementations of each of the operator functions,
        most of which have additional type parameters.
        \inlinecode[typescript]{epp} builds a such an object implementing the behavior of each operation
        for the target party, and passes that object to the choreography function.
    }
    \label{fig:epp-as-di-ts}
    \Description{Twenty lines of TypeScript code using the ChoreTS library.
	  This code declares interfaces for "Locally", "Dependencies", and "Choreography",
	  and shows part of the definition of the "epp" function.}
  \end{mdframed}
\end{figure*}

\paragraph{\ChoreographyTS} Figure \ref{fig:epp-as-di-ts} shows the definition of the \inlinecode[typescript]{Choreography} type in \ChoreographyTS and a simplified version of the EPP function.
In \ChoreographyTS, a choreography is a function of type \inlinecode[typescript]{Choreography} (line 11), which takes an object of type \inlinecode{Dependencies}
and some arguments of type \inlinecode{A} and returns a result of type \inlinecode{R}.
The body of a choreography can use choreographic operators, such as \inlinecode{locally} and \inlinecode{comm},
afforded by the \inlinecode{Dependencies}.
Endpoint projection, implemented by the \inlinecode{epp} function (lines 13-20), is done by providing concrete implementations of the choreographic operators for the projection target.
For example, the \inlinecode{locally} operator (lines 16-18) compares the specified location and the projection target.
If they are the same, \inlinecode{locally} executes the provided callback; otherwise it does nothing.

\paragraph{\Chorus} Our Rust implementation, \Chorus, also uses the EPP-as-DI technique but differs from \ChoreographyTS by using traits
(Rust's main polymorphism construct)
for dependency injection instead of higher-order functions.
This is necessary because closures in Rust cannot be generic.
The consequence is that \Chorus choreographies are declared and composed at the type level,
and composition of higher-order choreographies can only happen statically.
It is still possible to represent choreographies that behave dynamically,
but the representation of higher-order choreographies gets increasingly verbose as the compositions get more complex.
\Cref{fig:chorus-implementation} shows the most important functions of the \inlinecode{ChoreoOp} trait in \Chorus,
and the full signature of \inlinecode{Choreography}.

\begin{figure*}[tbp]
  \begin{mdframed}
    \begin{minipage}[b]{13cm}
\begin{minted}{rust}
pub trait ChoreoOp<ChoreoLS: LocationSet> {
  fn locally<V, L1: ChoreographyLocation, Index>(
    &self, location: L1, computation: impl Fn(Unwrapper<L1>) -> V) -> MultiplyLocated<V, LocationSet!(L1)>
  where L1: Member<ChoreoLS, Index>;

  fn multicast<Sender: ChoreographyLocation, V: Portable, D: LocationSet, Index1, Index2>(
    &self, src: Sender, destination: D, data: &MultiplyLocated<V, LocationSet!(Sender)>
  ) -> MultiplyLocated<V, D>
  where Sender: Member<ChoreoLS, Index1>,
        D: Subset<ChoreoLS, Index2>;

  fn naked<S: LocationSet, V, Index>(&self, data: MultiplyLocated<V, S>) -> V
  where ChoreoLS: Subset<S, Index>;

  fn conclave<R, S: LocationSet, C: Choreography<R, L=S>, Index>(&self, choreo: C) -> MultiplyLocated<R, S>
  where S: Subset<ChoreoLS, Index>;

  fn fanout<V, QS: LocationSet, FOC: FanOutChoreography<V, L=ChoreoLS, QS=QS>, QSSubsetL, QSFoldable>(
    &self, locations: QS, c: FOC) -> Faceted<V, QS>
  where QS: Subset<ChoreoLS, QSSubsetL>,
        QS: LocationSetFoldable<ChoreoLS, QS, QSFoldable>;
}

pub trait Choreography<R=()> {
  type L: LocationSet;
  fn run(self, op: &impl ChoreoOp<Self::L>) -> R;
}
\end{minted}
    \end{minipage}
    \caption{
        The main API for writing choreographies in \Chorus.
        A number of additional functions are elided from the \inlinecode{ChoreoOp} trait here.
        \inlinecode{fanout} is discussed in \Cref{sec:census-poly}.
    }
    \label{fig:chorus-implementation}
    \Description{Twenty seven lines of Rust code using the ChoRus library.
	  This code shows part of the definition of the "ChoreoOp" trait
	  and the definition of the "Choreography" trait.}
  \end{mdframed}
\end{figure*}

Like most mainstream programming languages (and unlike Haskell), TypeScript and Rust do not distinguish between pure, stateful, or effectful code.
Indeed, this will be the case for most languages in which the EPP-as-DI approach will be more attractive than the free-monad approach.
It is therefore the user's responsibility to sequester impure code in \inlinecode{locally} clauses.
To simplify the distinction of where users should or should not use impure code,
\ChoreographyTS and \Chorus do not have the \inlinecode{congruently} operator that \MultiChor has.
Instead, a \inlinecode{naked} operator unwraps an MLV already known to the entire census.
The unwrapped value can then be used in pure computation unsequestered from the choreography;
such computation will be actively replicated across the entire census.

\subsection{Membership Constraints}
\label{sec:membership}
It is not sufficient to simply \emph{have} censuses in one's CP implementation;
the property that parties not in the census do not participate in a computation must be enforced.
Furthermore, observing that a particular participant owns a multiply-located value is non-trivial for the type-checker when the
ownership set is polymorphic.
Declaring membership and subset relations as class constraints can work in some cases,
but this strategy has serious limitations in most type systems.
For example, a rule as obvious as
$(p \in ps_1 \land ps1 \subseteq ps_2) \to p \in ps_2$,
represented in Haskell as
\inlinecode{instance ( IsMember p ps1, IsSubset ps1 ps2) => IsMember p ps2},
would be impossible to use because the compiler has no way of guessing which set \inlinecode{ps1}
it should be checking \inlinecode{p}'s membership in
(and even if it could \emph{guess}, it wouldn't backtrack and try a different guess if its first try didn't work).

\paragraph{\MultiChor.}
To work around such limitations, \MultiChor uses a strategy of \emph{term-level proof witnesses}
inspired by \citet{noonanGDP}.
We do not actually use the \inlinecode{gdp}\footnote{
    "Ghosts of Departed Proofs"~\cite{gdp_hackage}
} package; we found that writing our own purpose-specific system had a few advantages.
First, we were able to write everything we needed without hand-waving any foundations as \inlinecode{axiom}s.
Second, pattern matching against the constructors of \inlinecode{Member l ls} suffices to convince GHC that \inlinecode{ls} is not empty,
which is sometimes useful.
Finally, the implicit paradigm of "memberships as indices \& subsets as functions" was qualitatively easier to work with
when we were building the census-polymorphism tools described in Section~\ref{sec:census-poly}.

In \MultiChor, locations are identified by type-level strings, uninhabited types with "kind" \inlinecode{Symbol}.
A value of type \inlinecode{Member p ps} can be used both
as proof that \inlinecode{p} is eligible to take some action (because of their membership in \inlinecode{ps})
and as a term-level identifier for the party \inlinecode{p}.
It's actual form will be that of an index in the type-level list \inlinecode{ps} at which \inlinecode{p} appears.
Subset relations are expressed and used similarly.
A value of type \inlinecode{Subset ps qs} has the underlying form of a function
from \inlinecode{Member p ps} to \inlinecode{Member p qs},
\emph{universally quantified over \inlinecode{p}}.
Because these logical structures can be built from scratch inside Haskell's type system,
all of the machinery we use to do so can safely be exposed to end-users so that they can write their own proofs, as needed, inside choreographies.
In practice however, they will usually use higher-level operators.
For example in \Cref{fig:example-choreography} line~19,
we use \inlinecode{inSuper} to compose \inlinecode{primary} (proof that the type-level \inlinecode{primary} is a member of \inlinecode{servers})
with \inlinecode{servers} (proof that the type-level \inlinecode{servers} is a subset of \inlinecode{census})
to make \inlinecode{primary' :: Member primary census}.
On line~20,
 \inlinecode{primary' @@ nobody}, read as \emph{"\inlinecode{primary} and nobody else"},
makes a \inlinecode{Subset '[primary] census} out of a \inlinecode{Member primary census}.

\paragraph{\Chorus.}
\Chorus's strategy for enforcing membership and subset relations is more similar to the naïve approach of using class constraints;
\inlinecode[rust]{Member} and \inlinecode[rust]{Subset} are both traits
(parameterized by the containing list of locations)
that must be fulfilled by the member location or subset.
What makes these relations decidable for the typechecker is a second parameter of each trait that provides a concrete proof
(again in the form of indices) of the relation.
In other words, the type checker doesn't need to \emph{choose} from every possible way a membership relation might hold,
because the structure of the trait uniquely identifies \emph{which} member of a set the location in question is claiming to be.
Ironically, Rust's type checker can almost always infer these indices itself;
in none of our examples has it been necessary to write them out.

Because \Chorus doesn't need term level proofs,
and because Rust doesn't have type-level strings,
a \inlinecode[rust]{ChoreographyLocation} in \Chorus is an empty \inlinecode[rust]{struct} type
whose inhabitants can be used as term-level identifiers for the location.
Type-level lists are similarly inhabited, and their values are effectively lists of location values.
For example in the \inlinecode[rust]{multicast} operator (\Cref{fig:chorus-implementation} lines 6--10)
the type variable \inlinecode[rust]{Sender} is the sender of the message;
the function also takes an argument \inlinecode[rust]{src} of type \inlinecode[rust]{Sender},
which is used by the type checker to determine what \inlinecode[rust]{Sender} actually is.
We believe this combination of type-level membership relations and inhabited location types will be more intuitive for new uses.
The disadvantage is that \Chorus's \inlinecode[rust]{Member} and \inlinecode[rust]{Subset} traits are not manipulatable objects;
a rule like
$(p \in ps_1 \land ps_1 \subseteq ps_2) \to p \in ps_2$
is not representable in \Chorus.

\paragraph{\ChoreographyTS}
TypeScript's type system natively affords the features \ChoreographyTS needs to identify locations and enforce membership and subset relations.
First, TypeScript's string literal types naturally combine term-level and type-level identification of locations:
Each string literal has a corresponding type whose only inhabitant is that value.
For example, \inlinecode[typescript]{"alice"} has type \inlinecode[typescript]{"alice"}, and is the only value of that type.
Second, TypeScript's union types have subtyping behavior that matches the behavior of subset relations.
(There's no need to distinguish membership relations from subset relations, as a union of one type is not a distinct concept in TypeScript.)
A \ChoreographyTS MLV type like \inlinecode[typescript]{Located<number, "alice" | "bob">}
is a \inlinecode[typescript]{number} owned by \inlinecode[typescript]{"alice"} and \inlinecode[typescript]{"bob"};
it can be unwrapped by an \inlinecode[typescript]{Unwrap<L>}
provided \inlinecode[typescript]{L} is a sub-type of \inlinecode[typescript]{("alice" | "bob")},
\ie \inlinecode[typescript]{"alice"} or \inlinecode[typescript]{"bob"}.
One critical interaction between these features is the typing of list literals:
The type of \inlinecode[typescript]{["alice", "bob"]} is \inlinecode[typescript]{Array<"alice" | "bob">}.
\ChoreographyTS's API is designed so that it never needs to retrieve the list of strings from a union type,
so the above features are all that's needed for identifying parties and lists of parties and enforcing membership and subset relations.

\subsection{Census Polymorphism in \MultiChor}
\label{sec:census-poly-haskell}

We leverage the type system of modern Haskell to achieve useful census polymorphism in \MultiChor.
This behavior is implemented as a layer \textit{on top of} \MultiChor's central monad and data-types;
from a theory perspective \MultiChor gets census polymorphism "for free" because it's a Haskell library.

Key to \MultiChor's strategy is Haskell's ability to express quantified type variables.
For example, a \inlinecode{Faceted} value is (underneath a little boiler-plate) a function.
Its argument is a \inlinecode{Member} proof that some party is in the list of owners,
and it returns a \inlinecode{Located} value known to the party in question.
(In practice, this behavior is provided by the function \inlinecode{localize}.)
Notably, nothing about a type like \inlinecode{Faceted ps cmn x} indicates who the (type-level!) party indicated by the argument might be.
The second type parameter to \inlinecode{Faceted} (\ie \inlinecode{cmn}) represents parties who know \emph{all} the contained values.
It's frequently \inlinecode{'[]},
but in \Cref{fig:census-poly-haskell} we see that \inlinecode{scatter} returns a \inlinecode{Faceted recipients '[sender] a}
because the sender knows all the values they sent.
The ability to express such "common owners" of faceted values is unique to \MultiChor.

\inlinecode{Faceted ps cmn x} is actually a special case of a more general type,
\inlinecode{PIndexed ps f}, where \inlinecode{f} can be any \emph{type-level function} from a party to a concrete type.
A \inlinecode{PIndexed} is effectively a type-indexed vector,
except that the type of the value retrieved depends on the index.
The case where it does \emph{not} depend on the index, \ie when \inlinecode{f} is \inlinecode{Const},
is exactly \inlinecode{Quire}.
Because of its unusual \inlinecode{kind}, type classes that one would expect to apply to vectors generally do not apply to \inlinecode{PIndexed}.
What's actually needed for census polymorphism is the ability to \inlinecode{sequence} a \inlinecode{PIndexed} of choreographies.
Since \inlinecode{PIndexed} is not an instance of \inlinecode{Traversable},
we implement the needed function \inlinecode{sequenceP}, which is effectively just a \inlinecode{for}-loop
(in any monad) over type-level lists of parties.
These loops are not unrolled at compile time;
the type class \inlinecode{KnownSymbols} affords to the runtime environment sufficient knowledge of the type-level list.

\begin{figure*}[tbp]
  \begin{mdframed}
    \begin{minipage}[b]{13cm}
\begin{minted}{haskell}
sequenceP :: forall b (ls :: [Symbol]) m. (KnownSymbols ls, Monad m)
        => PIndexed ls (Compose m b) -> m (PIndexed ls b)

fanOut :: (KnownSymbols qs)
       => (forall q. (KnownSymbol q) => Member q qs
                                     -> Choreo ps m (Located (q ': common) a))
       -> Choreo ps m (Faceted qs common a)

scatter :: forall census sender recipients a m.
           (KnownSymbol sender, KnownSymbols recipients, Show a, Read a)
        => Member sender census
        -> Subset recipients census
        -> Located '[sender] (Quire recipients a)
        -> Choreo census m (Faceted recipients '[sender] a)
\end{minted}
    \end{minipage}
    \caption{
        Type signatures for \inlinecode{sequenceP}, \inlinecode{fanOut}, and \inlinecode{scatter}.
    }
    \label{fig:census-poly-haskell}
    \Description{Fourteen lines of Haskell code using the MulitChor library.}
  \end{mdframed}
\end{figure*}

Using \inlinecode{sequenceP}, we can implement the functions
\inlinecode{fanOut}, \inlinecode{fanIn}, \inlinecode{scatter}, \inlinecode{gather}, and \inlinecode{parallel}
described in \Cref{sec:census-poly}.
\Cref{fig:census-poly-haskell} shows the type signatures for \inlinecode{sequenceP}, \inlinecode{fanOut}, and \inlinecode{scatter}.
Keen readers may notice that the "\inlinecode{cmn}" parties' views of a \inlinecode{Faceted} are effectively just a \inlinecode{Quire},
and so wonder at the need for \inlinecode{fanIn}.
In fact, \inlinecode{fanIn} \emph{is} less often used than \inlinecode{fanOut},
but it's necessary for expressing choreographic loops that yield values which aren't known to the parties over whom the loop is defined.
For example, the GMW protocol (\Cref{sec:mpc}) cannot be written using only \inlinecode{fanOut}.

Modern Haskell language features, especially type-variable quantification,
enable \MultiChor's implementation of census polymorphism to be entirely type-safe and transparent to users.
This is a flexible system within which users can easily write their own novel and bespoke functions and data structures;
we will show in the following section that such flexibility is not entirely necessary for census polymorphism.

\subsection{Census Polymorphism in \Chorus and \ChoreographyTS}
\label{sec:census-poly-rust}

\Chorus and \ChoreographyTS provide fan-out and fan-in operators
and faceted values for census polymorphism.

All \inlinecode[rust]{LocationSet}s in \Chorus also implement the \inlinecode[rust]{LocationSetFoldable} trait,
which provides a method \inlinecode[rust]{fold}.
Since closures in Rust cannot be generic, the method takes another \inlinecode[rust]{struct} that implements \inlinecode[rust]{LocationSetFolder}.
Both \inlinecode[rust]{fanOut} and \inlinecode[rust]{fanIn} use this system to fold over the parties in a set of locations.
\inlinecode[rust]{fanOut} takes a \inlinecode[rust]{LocationSet} and a special choreography implementing \inlinecode[rust]{FanOutChoreography}
(\ie, given a location in the aforementioned set, it produces a value at that location).
\inlinecode[rust]{fanOut} returns a \inlinecode[rust]{Faceted} value that can be unwrapped by the locations in the looped-over set
the same way they would unwrap a \inlinecode[rust]{MultiplyLocated}.
\inlinecode[rust]{fanIn} is similar, but it consumes a \inlinecode[rust]{Faceted} instead of producing one
(the result is a \inlinecode[rust]{Quire}).
Although \inlinecode[rust]{LocationSetFoldable} and \inlinecode[rust]{LocationSetFolder} are safe to expose to end users,
the implementation of \inlinecode[rust]{Faceted} (and \inlinecode[rust]{Quire}) is not.
This may effect ergonomics in some edge cases.

TypeScript can infer the type of an array of string literals to be an array of a union of string literals.
We use this feature to implement \inlinecode{fanOut} and \inlinecode{fanIn} for \ChoreographyTS.
\inlinecode{fanOut} takes an array of locations and a function that, given a location in the array,
produces a choreography that results in a value at that location.
Since these are arrays of string literals, we can use the regular \inlinecode{for}-loop to iterate over them
and produce \inlinecode{Faceted} values, which maintain the same invariants as in \Chorus.
\inlinecode{fanIn} is similar.
(TypeScript has native syntax that amounts to \inlinecode{Quire}s; naming them as such in \ChoreographyTS would be pointless boilerplate.)
Similar to the situation in \Chorus, the correctness of \ChoreographyTS's implementation of \inlinecode{Faceted}
is not enforced by the type system.
For this reason we hide \inlinecode{Faceted}'s implementation from users,
meaning that they would not be able to implement \inlinecode{fanOut} themselves.

We conjecture that library-level CP implementations in many modern languages would be able to provide type-safe census polymorphism for their users
by packaging up careful implementations of \inlinecode{fanOut}, \inlinecode{fanIn}, and \inlinecode{Faceted} values.
\inlinecode{scatter} is trivial to implement using \inlinecode{fanOut}
and \inlinecode{gather} likewise uses \inlinecode{fanIn}.
\inlinecode{parallel} also uses \inlinecode{fanOut} to generalize \inlinecode{locally} to a list of parties.
In a language where safe versions of \inlinecode{fanOut} and \inlinecode{fanIn} are impractical,
implementing only the less-generalized functions
\inlinecode{scatter}, \inlinecode{gather}, and \inlinecode{parallel}
will suffice for many use-cases.

\section{Case Studies}
\label{sec:case-studies}

Each of our implementations is packaged with multiple example protocols\footnote{
  See the associated software artifact, or the respective Git repositories.
}
designed to leverage, test, and demonstrate the use of the various features described above,
and to extend examples used in prior literature.
One of them we have already featured: the choreography \inlinecode{kvs} in \Cref{fig:example-choreography}.
The actual source file is typical among the examples:
It contains relevant helper-functions
and an executable "\inlinecode{main} method" that detects which role to fulfil (via command-line argument),
performs EPP on the outer-most choreography, and executes the chosen party's behavior using HTTP for communication.

\paragraph{A key-value store in \Chorus}
A principal innovation of our systems is census polymorphism:
One can implement choreographic behavior (\eg replicating a request from a client across some backup servers)
without specifying how many participants there are (\eg the protocol is parametric on the number of backups).
In \Cref{sec:chorus-kvs} we show another key-value-store protocol written in Rust using \Chorus.
The choreography is very similar to the one in \Cref{fig:example-choreography},
but doesn't model corrupt stores or resynching.
Like the \MultiChor version, it uses \inlinecode[rust]{broadcast} inside a \inlinecode[rust]{conclave}
to achieve efficient KoC among the servers.
Writing census polymorphic or higher-order choreographies in \Chorus entails some boilerplate;
this case study showcases the implementation and use of a \inlinecode[rust]{gather} operation
as an instance of \inlinecode[rust]{FanInChoreography}.

\paragraph{A fair lottery in ChoreoTS}
DPrio~\cite{dprio2023} is a recent variation on a secure aggregation
protocol called Prio~\cite{corrigan2017prio}.
DPrio supports all of the same security guarantees as Prio and adds a layer of
differential privacy (DP)
so that client inputs cannot be inferred by the receiving analyst.
In \Cref{sec:lottery} we excerpt the novel part of DPrio as a stand-alone choreography written in TypeScript using \ChoreographyTS.
The key mechanism of DPrio is that every client generates noise for the DP layer, and
the servers randomly choose whose noise to actually use.
In effect, each client chooses a secret number (in the real DPrio protocol they should choose these randomly)
and the analyst receives one of those numbers at random and without knowing whose they got.
The intermediate servers---who each receive only a "share" of each secret, and so never learn the secrets themselves---
work together to choose which secret to forward to the analyst.
The process of choosing which secret, the "lottery", is such that as long as there's at least one server acting honestly,
the choice will still be random.
This many-to-many communication pattern relies heavily on \inlinecode[typescript]{fanin} and \inlinecode[typescript]{fanout};
the choreography is polymorphic over the quantities and identities of both the clients and the servers.

\paragraph{Secure multiparty computation in \MultiChor}
Secure multiparty computation~\cite{evans2018pragmatic} (MPC) is a family of techniques
that allow a group of parties to jointly compute an agreed-upon function of their distributed data
without revealing the data or any intermediate results to the other parties.
We consider a classic and general-purpose MPC protocol named Goldreich-Micali-Widgerson (GMW)~\cite{goldreich2019play} after its authors.
The GMW protocol requires the function to be computed to be specified as a binary circuit.
It's based on two important building blocks: \emph{additive secret sharing},
a method for encrypting distributed data that still allows computing on it,
and \emph{oblivious transfer} (OT)~\cite{naor2001efficient}, a building-block protocol in applied cryptography.
The GMW protocol starts by asking each party to secret-share its input values for the circuit.
Then, the parties iteratively evaluate the gates of the circuit while keeping the intermediate values secret-shared.
When evaluation finishes, the parties reveal their secret-shares of the output to each other to decrypt the final result.
Our implementation takes a data-structure representing the circuit as an argument,
and is parametric on the number of participants.
It represents secret-shares as \inlinecode{Faceted} values,
and does most of the computation in parallel.
To evaluate an AND gate, we must nest \inlinecode{FanOut}, \inlinecode{FanIn}, and \inlinecode{conclave}
to call the oblivious transfer sub-choreography (which has an explicit census of only two parties)
once for every ordered pair of distinct participants.
\Cref{sec:mpc} discusses this case-study in more detail;
our complete and working implementation of the GMW protocol is about three-hundred lines of code.

\section{Related Work}
\label{sec:related}

Early choreographies were specifications~\citep{w3c-cdl-primer} rather than executable programs.
The Chor language~\citep{montesi-carbone-dfbd, montesi-dissertation} pioneered the use
of endpoint projection~\citep{carbone-cdl-epp-esop,carbone-cdl-epp} as a way to make choreographies executable.
Much of the subsequent literature on CP focuses on developing its formal foundations~\citep{hirsch2021pirouette, chor-lambda, graversen2023polychor},
paving the way for practical implementations such as Choral~\citep{giallorenzo-choral}, a standalone choreographic language,
and HasChor~\citep{shen-haschor}, which takes a library-based approach in which endpoint projection is dynamic rather than static. Our work in this paper focuses on the latter, library-based approach.

Excitingly, CP libraries for a variety of languages have recently appeared on the scene. These include UniChorn~\citep{unichorn}, Chorex~\citep{chorex-github}, and Klor~\citep{klor-github}. All three are under development, so here we report on their current state.
UniChorn is a port of HasChor into the Unison programming language.
To implement EPP, it uses the Unison feature of \emph{abilities}, better known in the literature as algebraic effects~\citep{plotkin-2003,plotkin-2013}. This implementation approach, which was also recently proposed by~\citet{shen-alg-eff-cp}, can be thought of as a generalization of the free-monad approach.
As a direct port of HasChor, UniChorn does not support conclaves, MLVs, or census polymorphism.
Chorex is a CP system for Elixir, and Klor is a CP system for Clojure; both Chorex and Klor leverage the powerful macro systems of their respective host languages to carry out EPP. Chorex uses the select-\&-merge KoC strategy, and unprojectable choreographies can be detected at macro expansion time.
Klor, on the other hand, is effectively a conclaves-\&-MLVs system, but their API differs from the implementations we present here,
and the authors have not yet shown what safety guarantees it does or does not offer.
Neither Chorex nor Klor support census polymorphism.

Wysteria~\citep{wysteria} and Symphony~\citep{Sweet_2023} are domain-specific languages for \emph{secure multiparty computation}.
Programs in these languages can exhibit census polymorphism,
but they have homomorphic encryption baked into their
semantics for communication,
and they are not intended for general-purpose choreographic programming.
Wysteria has a \inlinecode{par} language construct, used for evaluating an expression at a set of locations, that is somewhat similar in spirit to our conclaves.  However, applying the conclave concept to choreographic programming, and to the choreographic knowledge-of-choice problem in particular, is to our knowledge a novel contribution of this paper.  Symphony does not support conditionals, and therefore KoC propagation is a moot point for them.

\citet{jongmans2022predicates} present a bespoke and highly efficient KoC strategy that differs from both traditional select-\&-merge
and from the conclaves-\&-MLVs approach we present here.
Their approach does not use MLVs and requires writing distinct guards for every participant in a conditional expression;
they show how to use predicate transformers to check that such distributed decisions are unanimous.
They also do not present an implementation of their approach.

\section{Conclusion}
\label{sec:conclusion}

This paper makes three contributions towards efficient, portable, and census-polymorphic library-based choreographic programming. 
To address library-specific challenges associated with efficient Knowledge of Choice, we introduce and formalize the concepts of \emph{conclaves} and \emph{multiply located values}. 
To address the need for choreographies to support a variable number of participants, we introduce the idea of \emph{census polymorphism}. 
To address the challenge of implementing library-based choreographic programming in host languages with varied features, we introduce the \emph{endpoint projection as dependency injection} approach that supports implementations in a variety of host languages. 
We demonstrate the generality of all three contributions by building implementations in Haskell, Rust, and TypeScript.

\begin{acks}
This material is based upon work supported by the National Science Foundation under Grants No. 2238442 and 2145367, and by a gift from the Stellar Development Foundation. Any opinions, findings and conclusions or recommendations expressed in this material are those of the author(s) and do not necessarily reflect the views of the sponsoring organizations.
\end{acks}

\bibliographystyle{ACM-Reference-Format}
\bibliography{refs}

\clearpage
\appendix

\section{The GMW Protocol in MultiChor}
\label{sec:mpc}

\emph{Secure multiparty computation}~\cite{evans2018pragmatic} (MPC) is a family of techniques that allow a group of parties to jointly compute an agreed-upon function of their distributed data without revealing the data or any intermediate results to the other parties. We consider an MPC protocol named Goldreich-Micali-Widgerson (GMW)~\cite{goldreich2019play} after its authors. The GMW protocol requires the function to be computed to be specified as a binary circuit, and each of the parties who participates in the protocol may provide zero or more inputs to the circuit. At the conclusion of the protocol, all participating parties learn the circuit's output.

The GMW protocol is based on two important building blocks: \emph{additive secret sharing}, a method for encrypting distributed data that still allows computing on it, and \emph{oblivious transfer} (OT)~\cite{naor2001efficient}, a building-block protocol in applied cryptography. The GMW protocol starts by asking each party to secret share its input values for the circuit. Then, the parties iteratively evaluate the gates of the circuit while keeping the intermediate values secret shared. Oblivious transfer is used to evaluate AND gates. When evaluation finishes, the parties reveal their secret shares of the output to decrypt the final result.

\begin{figure*}[tbp]
  \begin{mdframed}
    \begin{minipage}[b]{13cm}
\begin{minted}{Haskell}
gmw :: forall parties m. (KnownSymbols parties, MonadIO m, CRT.MonadRandom m)
    => Circuit parties -> Choreo parties (CLI m) (Faceted parties '[] Bool)
gmw circuit = case circuit of
  InputWire p -> do        -- process a secret input value from party p
    value :: Located '[p] Bool <- _locally p $ getInput "Enter a secret input value:"
    secretShare p value
  LitWire b -> do          -- process a publicly-known literal value
    let chooseShare :: forall p. (KnownSymbol p) =>
                       Member p parties -> Choreo parties (CLI m) (Located '[p] Bool)
        chooseShare p = congruently (p @@ nobody) $ \_ -> case p of First   -> b
                                                                    Later _ -> False
    fanOut chooseShare
  AndGate l r -> do        -- process an AND gate
    lResult <- gmw l; rResult <- gmw r;
    fAnd lResult rResult
  XorGate l r -> do        -- process an XOR gate
    lResult <- gmw l; rResult <- gmw r
    parallel (allOf @parties) \p un -> pure $ xor [viewFacet un p lResult, viewFacet un p rResult]

data Circuit :: [LocTy] -> Type where
  InputWire :: (KnownSymbol p) => Member p ps -> Circuit ps
  LitWire :: Bool -> Circuit ps
  AndGate :: Circuit ps -> Circuit ps -> Circuit ps
  XorGate :: Circuit ps -> Circuit ps -> Circuit ps

mpc :: forall parties m. (KnownSymbols parties, MonadIO m, CRT.MonadRandom m)
    => Circuit parties -> Choreo parties (CLI m) ()
mpc circuit = do
  outputWire <- gmw circuit
  result <- reveal outputWire
  void $ _parallel (allOf @parties ) $ putOutput "The resulting bit:" result
\end{minted}
    \end{minipage}
    \caption{A choreography for the GMW protocol. The choreography works for an arbitrary number of parties.
      Figure~\ref{fig:gmw-helpers-multichor-example} contains the \inlinecode{xor} function to compute the OR gate. In addition to the \inlinecode{secretShare} and \inlinecode{fAnd} choreographies to compute the results of the INPUT and AND gates respectively.
    \inlinecode{mpc} uses \inlinecode{gmw} protocol as well as \inlinecode{reveal} in Figure~\ref{fig:gmw-helpers-multichor-example} printing the resulting bit at each party.}
    \label{fig:gmw-multichor-example}
  \end{mdframed}
    \Description{Thirty one lines of Haskell code using the MultiChor library.
	This code defines a recursive choreographic function "gmw" and a top-level function "mpc".}
\end{figure*}

\begin{figure*}
  \begin{mdframed}
    \begin{minipage}[b]{13cm}
\begin{minted}{Haskell}
secretShare :: forall parties p m. (KnownSymbols parties, KnownSymbol p, MonadIO m)
            => Member p parties -> Located '[p] Bool -> Choreo parties m (Faceted parties '[] Bool)
secretShare p value = do
  shares <- locally p \un -> genShares p (un singleton value)
  PIndexed fs <- scatter p (allOf @parties) shares
  return $ PIndexed $ Facet . othersForget (First @@ nobody) . getFacet . fs

genShares :: forall ps p m. (MonadIO m, KnownSymbols ps) => Member p ps -> Bool -> m (Quire ps Bool)
genShares p x = quorum1 p gs'
  where gs' :: forall q qs. (KnownSymbol q, KnownSymbols qs) => m (Quire (q ': qs) Bool)
        gs' = do freeShares <- sequence $ pure $ liftIO randomIO -- generate n-1 random shares
                 return $ qCons (xor (qCons @q x freeShares)) freeShares

xor :: (Foldable f) => f Bool -> Bool
xor = foldr1 (/=)

fAnd :: forall parties m.
        (KnownSymbols parties, MonadIO m, CRT.MonadRandom m)
     => Faceted parties '[] Bool
     -> Faceted parties '[] Bool
     -> Choreo parties (CLI m) (Faceted parties '[] Bool)
fAnd uShares vShares = do
  let genBools = sequence $ pure randomIO
  a_j_s :: Faceted parties '[] (Quire parties Bool) <- _parallel (allOf @parties) genBools
  bs :: Faceted parties '[] Bool <- fanOut \p_j -> do
      let p_j_name = toLocTm p_j
      b_i_s <- fanIn (p_j @@ nobody) \p_i ->
        if toLocTm p_i == p_j_name
          then _locally p_j $ pure False
          else do
              bb <- locally p_i \un -> let a_ij = getLeaf (viewFacet un p_i a_j_s) p_j
                                           u_i = viewFacet un p_i uShares
                                         in pure (xor [u_i, a_ij], a_ij)
              conclaveTo (p_i @@ p_j @@ nobody) (listedSecond @@ nobody) (ot2 bb $ localize p_j vShares)
      locally p_j \un -> pure $ xor $ un singleton b_i_s
  parallel (allOf @parties) \p_i un ->
    let computeShare u v a_js b = xor $ [u && v, b] ++ toList (qModify p_i (const False) a_js)
    in pure $ computeShare (viewFacet un p_i uShares) (viewFacet un p_i vShares)
                           (viewFacet un p_i a_j_s)   (viewFacet un p_i bs)

ot2 :: (KnownSymbol sender, KnownSymbol receiver, MonadIO m, CRT.MonadRandom m) =>
  Located '[sender] (Bool, Bool) -> Located '[receiver] Bool
  -> Choreo '[sender, receiver] (CLI m) (Located '[receiver] Bool)
ot2 bb s = do
  let sender = listedFirst :: Member sender '[sender, receiver]
  let receiver = listedSecond :: Member receiver '[sender, receiver]

  keys <- locally receiver  \un -> liftIO $ genKeys $ un singleton s
  pks <- (receiver, \un -> let (pk1, pk2, _) = un singleton keys
                           in return (pk1, pk2)) ~~> sender @@ nobody
  encrypted <- (sender, \un -> let (b1, b2) = un singleton bb
                               in liftIO $ encryptS (un singleton pks) b1 b2) ~~> receiver @@ nobody
  locally receiver  \un -> liftIO $ decryptS (un singleton keys)
                                                           (un singleton s)
                                                           (un singleton encrypted)

reveal :: forall ps m. (KnownSymbols ps) => Faceted ps '[] Bool -> Choreo ps m Bool
reveal shares = xor <$> (gather ps ps shares >>= naked ps)
  where ps = allOf @ps
\end{minted}
    \end{minipage}
    \caption{Various helpers for the GMW protocol.
      \inlinecode{fAnd} computes the result of an AND gate on secret-shared inputs using pairwise oblivious transfer. The choreography works for an arbitrary number of parties, and leverages the 1 out of 2 OT defined earlier.
      \inlinecode{xor} computes the result of an OR gate as a standard non-choreographic function.
      \inlinecode{secretShare} handles Input gate secret sharing \inlinecode{p}'s secret value among \inlinecode{parties}
         and for revealing a secret-shared value.
      \inlinecode{ot} performs 1 out of 2 oblivious transfer (OT) using RSA public-key encryption. The choreography involves exactly two parties, \inlinecode{sender} and \inlinecode{receiver}.
      \inlinecode{genShares} uses \inlinecode{Quire} to map each member \inlinecode{p} in \inlinecode{ps} to a generated secret share \inlinecode{Bool}.
      \inlinecode{encryptS} \inlinecode{decryptS} which are omitted for brevity use the cryptonite library for encryption and decryption.
    }
    \label{fig:gmw-helpers-multichor-example}
  \end{mdframed}
    \Description{Fifty nine lines of Haskell code using the MultiChor library.
	This code defines functions "secretShare", "genShares", "xor", "fAnd", "ot2", and "reveal".}
\end{figure*}

\paragraph{Additive secret sharing}
We begin by describing additive secret sharing, a common building block in MPC protocols. A secret bit $x$ can be \emph{secret shared} by generating $n$ random \emph{shares} $s_1, \dots, s_n$ such that $x = \sum_{i=1}^n s_i$. If $n-1$ of the shares are generated uniformly and independently randomly, and the final share is chosen to satisfy the property above, then the shares can be safely distributed to the $n$ parties without revealing $x$---recovering $x$ requires access to all $n$ shares. Importantly, secret shares are \emph{additively homomorphic}---adding together shares of secrets $x$ and $y$ produces a share of $x+y$.

\MultiChor choreograpies for performing secret sharing in the arithmetic field of booleans appear in Figure~\ref{fig:gmw-helpers-multichor-example}. The function \inlinecode{secretShare} takes a single secret bit located at party \inlinecode{p}, generates shares \inlinecode{shares :: Quire parties Bool} which maps each member in \inlinecode{parties} to a share, then uses \inlinecode{scatter} to send a share to each member.
However \inlinecode{scatter} would return a \inlinecode{Faceted parties '[p] Bool} since by default it includes the sender. Since the share should be secret we should instead have a \inlinecode{Faceted parties '[] Bool} we accomplish this by deconstructing and reconstructing via \inlinecode{PIndexed} and using \inlinecode{othersForget (First @@ nobody)}. This \inlinecode{Faceted} bit is located at all parties, such that the bits held by the parties sum up to the original secret.
\inlinecode{reveal} takes such a shared value and broadcasts
all the shares so everyone can reconstruct the plain-text.

\paragraph{Oblivious transfer}
The other important building block of the GMW protocol is oblivious transfer (OT)~\cite{naor2001efficient}. OT is a 2-party protocol between a \emph{sender} and a \emph{receiver}. In the simplest variant (\emph{1 out of 2} OT, used in GMW), the sender inputs two secret bits $b_1$ and $b_2$, and the receiver inputs a single secret \emph{select bit} $s$. If $s=0$, then the receiver receives $b_1$. If $s=1$, then the receiver receives $b_2$. Importantly, the sender does \emph{not} learn which of $b_1$ or $b_2$ has been selected, and the receiver does \emph{not} learn the non-selected value.

Importantly, oblivious transfer is a \emph{two-party protocol}, it would be a type-error for any third-parties to be involved in the implementation. \MultiChor's \inlinecode{Faceted} values and utilities for type-safe embedding of conclaved sub-protocols within arbitrarily large censuses make it possible to embed the use of pairwise oblivious transfer between parties in a general version of multi-party GMW.

\paragraph{The GMW protocol}
The complete GMW protocol operates as summarized earlier, by secret sharing input values and then evaluating the circuit gate-by-gate. Our implementation as a \MultiChor choreography appears in Figure~\ref{fig:gmw-multichor-example}, defined as a recursive function over the structure of the circuit. The choreography returns a \inlinecode{Faceted} value, representing the secret-shared output of the circuit. For ``input'' gates (lines~4--6), the choreography runs the secret sharing protocol in Figure~\ref{fig:gmw-helpers-multichor-example} to distribute shares of the secret value. For XOR gates (lines~16--18) Figure ~\ref{fig:gmw-multichor-example}, the parties recursively run the GMW protocol to compute the two inputs to the gate and then each party computes one share of the gate's output by XORing their shares of the inputs. This approach leverages the additive homomorphism of additive secret shares. For AND gates (lines~13--15) Figure ~\ref{fig:gmw-multichor-example}, the parties compute shares of the gate's inputs, then use the \inlinecode{fAnd} protocol to perform multiplication of the two inputs. This implements the protocol as described in  Section 3.2.1 of \cite{evans2018pragmatic} namely the \emph{Generalization to more than two parties} case. Since additive secret shares are not multiplicatively homomorphic, this operation leverages the oblivious transfer protocol to perform the multiplication.

\paragraph{Computing secret-shared AND via OT}
To compute the result of an AND gate, the parties compute \emph{pair-wise} ANDs using their respective shares of the input values, then use the results to derive shares of the gate's output. The \inlinecode{fAnd} choreography (Figure~\ref{fig:gmw-helpers-multichor-example} lines~17--39) takes \inlinecode{Faceted} values holding the parties' shares of the input values, and returns a \inlinecode{Faceted} value representing each party's share of the output. On line~25, the parties perform a \inlinecode{fanOut} to begin the pairwise computation; the \inlinecode{fanIn} on line~27 completes the pairing for each computation, and uses \inlinecode{conclaveTo} (line 34) to embed pairwise OTs (via \inlinecode{ot2}) in the larger set of parties.

Our implementation of GMW leverages \MultiChor's \inlinecode{Faceted} values and utilities for type-safe parallel, conclaved, and pairwise choreographies to build a fully-general implementation of the protocol that works for an arbitrary number of parties.

\clearpage
\section{Census Polymorphic Server Backups in \Chorus}
\label{sec:chorus-kvs}

\Cref{fig:kvs-chorus} shows a \Chorus choreography in which a client \inlinecode[rust]{Client} submits a query to a server \inlinecode[rust]{Server},
and \inlinecode[rust]{Server} replicates its store across a parametric list of backup servers.
\begin{figure*}[bp]
  \begin{mdframed}
    \begin{minipage}[b]{13cm}
\begin{minted}{Rust}
struct HandleRequest<Backups, BackupsPresent, BSpine>
{   request: Located<Request, Server>,
    _phantoms: PhantomData<(Backups, BackupsPresent, BSpine)>,
}
impl<Backups: LocationSet, BackupsPresent, BSpine>
  Choreography<Located<Response, Server>> for HandleRequest<Backups, BackupsPresent, BSpine>
where Backups: Subset<HCons<Server, Backups>, BackupsPresent>,
      Backups: LocationSetFoldable<HCons<Server, Backups>, Backups, BSpine>
{   type L = HCons<Server, Backups>;
    fn run(self, op: &impl ChoreoOp<Self::L>) -> Located<Response, Server> {
        match op.broadcast(Server, self.request) {
            Request::Put(key, value) => {
                let oks = op.parallel(Backups::new(), ||{ handle_put(key.clone(), value) });
                let gathered = op.fanin::<Response, Backups, HCons<Server, HNil>, _, _, _, _>(
                        Backups::new(),
                        Gather{values: &oks, phantom: PhantomData}
                    );
                op.locally(Server, |un| {
                  let ok = un.unwrap::<Quire<Response, Backups>, _, _>(&gathered)
                               .get_map().into_values().all(|response|{response == 0});
                  if ok {
                    return handle_put(key.clone(), value)
                  } else {
                    return -1
                  }
                })
            }
            Request::Get(key) => op.locally(Server, |_| { handle_get(key.clone()) })
        }
    }
}

struct KVS<Backups: LocationSet,
           BackupsPresent,
           BackupsAreServers,
           BSpine>
{   request: Located<Request, Client>,
    _phantoms: PhantomData<(Backups, BackupsPresent, BackupsAreServers, BSpine)>,
}
impl<Backups: LocationSet, BackupsPresent, BackupsAreServers, BSpine>
  Choreography<Located<Response, Client>> for KVS<Backups, BackupsPresent, BackupsAreServers, BSpine>
where Backups: Subset<HCons<Client, HCons<Server, Backups>>, BackupsPresent>,
      Backups: Subset<HCons<Server, Backups>, BackupsAreServers>,
      Backups: LocationSetFoldable<HCons<Server, Backups>, Backups, BSpine>
{   type L = HCons<Client, HCons<Server, Backups>>;
    fn run(self, op: &impl ChoreoOp<Self::L>) -> Located<Response, Client> {
        let request = op.comm(Client, Server, &self.request);
        let response = op.conclave(HandleRequest::<Backups, _, _>{
            request: request,
            _phantoms: PhantomData
        }).flatten();
        op.comm(Server, Client, &response)
    }
}
\end{minted}
    \end{minipage}
    \caption{A census-polymorphic key-value store choreography in \Chorus. }
    \label{fig:kvs-chorus}
    \Description{Fifty four lines of Rust code using the ChoRus library.
	  This code defines two structs, "KVS" and "HandleRequest", both implementing "Choreography".}
  \end{mdframed}
\end{figure*}
The entry point is the \inlinecode[rust]{run} method of \inlinecode[rust]{KVS} (lines~46--53),
but nearly all the action happens inside the \inlinecode[rust]{conclave} (lines~48--51) to the sub-choreography \inlinecode[rust]{HandleRequest}.
Notably, \inlinecode[rust]{HandleRequest} does not include \inlinecode[rust]{Client} in its census (\inlinecode[rust]{L}, on line~9).
The body of \inlinecode[rust]{HandleRequest} (the \inlinecode[rust]{run} method on lines~10--30)
immediately broadcasts the request so that \inlinecode[rust]{Server} and all the backups can branch on its structure together.
In the event of a \inlinecode[rust]{Get} request, \inlinecode[rust]{Server} handles it unilaterally (line~28).
In the event of a \inlinecode[rust]{Put} request, first the backups all handle it in parallel (lines~13).
Then the backups all send their results (either \inlinecode[rust]{0} for "success", or an error code)
back to \inlinecode[rust]{Server}; these values are stored in the \inlinecode[rust]{Quire} value \inlinecode[rust]{gathered} (line~14).
If all of the backups' responses are "OK", then \inlinecode[rust]{Server} handles the request as well (lines~19-20);
their result with be sent to the client in the main choreography (line~22).
If any of the backups reported an error, then \inlinecode[rust]{Server} instead responds with an error code (\inlinecode[rust]{-1})
to indicate that the system has lost synchronization.
The entire system is generic over the number of (and identities of) the backup servers.
The version of \inlinecode[rust]{Gather} used in \inlinecode[rust]{HandleRequest} is shown in \Cref{fig:gather-chorus}.

\begin{figure*}[tbp]
  \begin{mdframed}
    \begin{minipage}[b]{13cm}
\begin{minted}{Rust}
struct Gather<'a,
              V,
              Senders: LocationSet + Subset<Census, SendersPresent>,
              Receivers: LocationSet + Subset<Census, ReceiversPresent>,
              Census: LocationSet,
              SendersPresent,
              ReceiversPresent>
{   values: &'a Faceted<V, Senders>,
    phantom: PhantomData<(Census, SendersPresent, Receivers, ReceiversPresent)>,
}
impl<'a,
     V: Portable + Copy,
     Senders: LocationSet + Subset<Census, SendersPresent>,
     Receivers: LocationSet + Subset<Census, ReceiversPresent>,
     Census: LocationSet,
     SendersPresent,
     ReceiversPresent>
  FanInChoreography<V> for Gather<'a, V, Senders, Receivers, Census, SendersPresent, ReceiversPresent>
{   type L = Census;
    type QS = Senders;
    type RS = Receivers;
    fn run<Sender: ChoreographyLocation,
           _SendersPresent, _ReceiversPresent, SenderPresent, SenderInSenders>(
        &self,
        op: &impl ChoreoOp<Self::L>,
    ) -> MultiplyLocated<V, Self::RS>
    where Self::QS: Subset<Self::L, SendersPresent>,
          Self::RS: Subset<Self::L, ReceiversPresent>,
          Sender: Member<Self::L, SenderPresent>,
          Sender: Member<Self::QS, SenderInSenders>,
    {   let x = op.locally(Sender::new(), |un| *un.unwrap(self.values));
        let xx = op.multicast::<Sender, V, Self::RS, SenderPresent, ReceiversPresent>(
                     Sender::new(),
                     <Self::RS>::new(),
                     &x);
        xx
    }
}
\end{minted}
    \end{minipage}
      \caption{A census-polymorphic "gather" function in \Chorus,
               specialized for use in the \inlinecode[rust]{KVS} choreography in \Cref{fig:kvs-chorus}.
    }
    \label{fig:gather-chorus}
    \Description{Thirty eight lines of Rust code using the ChoRus library.
	  This code defines a struct "Gather" that implements "FanInChoreography".}
  \end{mdframed}
\end{figure*}

\clearpage
\section{A Cryptographically Fair Lottery in \ChoreographyTS}
\label{sec:lottery}

DPrio~\cite{dprio2023} is a recent variation on a secure aggregation
protocol called Prio~\cite{corrigan2017prio}.
DPrio supports all of the same security guarantees as Prio and adds a layer of
differential privacy (DP)
so that client inputs cannot be inferred by the receiving analyst.
To avoid summarizing the polynomial-based zero-knowlege-proofs used in Prio,
we excerpt the novel part of DPrio as a stand-alone protocol for this case study.
The key mechanism of DPrio is that every client generates noise for the DP layer, and
the servers randomly choose whose noise to actually use.
The choreography in Figure~\ref{fig:lottery-choreots-example} and Figure ~\ref{fig:lottery-choreots-example-2}  implements this process,
except that, since we are not doing the rest of the secure aggregation,
the servers are just selecting which one of the values submitted by the clients will be revealed to the analyst.

Lines 14--24 the clients choose their inputs and secret-share them to the servers.
This process is basically the same as in GMW, except the parties generating and receiving shares are distinct subsets of the census.
Lines 35--79 implement the lottery itself.
Finally, the analyst receives shares of the chosen client's secret and sums them together to
learn the final result on line 88--90.

A more casual lottery could be implemented by having one server choose a client-index at random
and inform the other servers of the choice,
but then everyone would have to trust that the choice truly was random.
DPrio ensures the fairness of the lottery in two steps:
First, \emph{all} the servers independently generate random values up to some multiple
of the number of clients.
Second, before the clients reveal their randomness to each other,
they \emph{commit} to their randomness by broadcasting salted hashes.
The actual client-index used is the modulo of the sum of these random values.
The commitment process just prevents any server from waiting until the end to select their value;
without it the last server would be able to calculate their "random" value to result in
any index they wanted.
If there is a failed commitment check a \inlinecode{Error("Commitment failed")} is thrown
at any parties that detect it, which will prevent the choreography from completing.

The three \inlinecode{parallel} blocks on lines 35, 41, and 46 of Figure~\ref{fig:lottery-choreots-example}
could be combined into one without changing the semantics of the choreography with the type
\mintinline[breaklines]{TypeScript}|Faceted<{rho: FiniteField; psi: FiniteField; alpha: string;}, SL>|.
Instead we chose this presentation to more closely match the order of communication expressed by the protocol in DPrio \cite{dprio2023}.
In contrast, it would not be safe to combine the \inlinecode{fanIn} on line 57
with the ones on lines 61 and 65; it's precisely this sequential separation that ensures no server sends their \inlinecode{rho}
until they've received all the \inlinecode{alpha_}.

\begin{figure*}
  \begin{mdframed}
    \begin{minipage}[b]{13cm}
\begin{minted}{Typescript}
const field = new Field(999983);
const maxSalt = 2 ^ 18;
const minSalt = 2 ^ 20;

export const lottery = <SL extends Location, CL extends Location>(
  serverLocations: SL[], clientLocations: CL[], askQuestion: (query: string) => Promise<string>
): Choreography<"analyst" | SL | CL, undefined, MultiplyLocated<FiniteField, "analyst">> =>
  async ({ locally, fanin, fanout, parallel }) => {
    const secret = await parallel(clientLocations, async () => {
      const secretStr = await askQuestion("Secret: ");
      return parseInt(secretStr);
    });

    const clientShares = await parallel(clientLocations, async (_, unwrap) => {
      const freeShares = Array.from({ length: serverLocations.length - 1 }, () => field.rand());
      const lastShare = field.sub(unwrap(secret),
        freeShares.reduce((a, b) => field.add(a, b), field.zero));
      const shares = freeShares.concat(lastShare);
      const serverToShares: Record<string, FiniteField> = {};
      for (const [index, server] of serverLocations.entries()) {
        serverToShares[server] = shares[index] as FiniteField
      }
      return serverToShares;
    });

    const serverShares = await fanout(serverLocations, (server) =>
      async ({ fanin }) => fanin(clientLocations, [server], (client) =>
        async ({ locally, comm }) => {
          const share = await locally(client, (unwrap) => unwrap(clientShares)[server]);
          return comm(client, server, share);
        })
    );

    // 1) Each server selects a random number; τ is some multiple of the number of clients.
    const rho = await parallel(serverLocations, async () => {
      const tauStr = await askQuestion("Pick a number from 1 to tau:");
      return parseInt(tauStr);
    });

    // Salt value
    const psi = await parallel(serverLocations, async () =>
      Math.floor(Math.random() * (maxSalt - minSalt + 1)) + minSalt
    );

    // 2) Each server computes and publishes the hash α = H(ρ, ψ) to serve as a commitment
    const alpha = await parallel(serverLocations, async (_, unwrap) => {
      const rhoValue = unwrap(rho);
      const psiValue = unwrap(psi);
      return hash(rhoValue, psiValue);
    });

    const alpha_ = await fanin(serverLocations, serverLocations, (server) =>
      async ({ multicast }) => multicast(server, serverLocations, alpha)
    );

    // 3) Every server opens their commitments by publishing their ψ and ρ to each other
    const psi_ = await fanin(serverLocations, serverLocations, (server) =>
      async ({ multicast }) => multicast(server, serverLocations, psi)
    );

    const rho_ = await fanin(serverLocations, serverLocations, (server) =>
      async ({ multicast }) => multicast(server, serverLocations, rho)
    );
\end{minted}
    \end{minipage}
    \caption{A federated-lottery protocol (part 1 of 2). See Figure~\ref{fig:lottery-choreots-example} for part 2 of 2.}
    \Description{63 lines of Typescript code using \ChoreographyTS defining the first part of a choreography named "lottery".}
    \label{fig:lottery-choreots-example}
    \Description{Sixty three lines of TypeScript code using the ChoreTS library.
	  This code defines a choreographic function "lottery".
	  The code continues in Figure 15.}
  \end{mdframed}
\end{figure*}

\begin{figure*}[tbp]
  \begin{mdframed}
    \begin{minipage}[b]{13cm}
\begin{minted}{Typescript}
    // 4) All servers verify each other's commitment by checking α = H(ρ, ψ)
    await parallel(serverLocations, async (server, unwrap) => {
      if (unwrap(alpha_)[server] != hash(unwrap(rho_)[server], unwrap(psi_)[server])) {
        throw new Error("Commitment failed");
      }
    });

    // 5) If all the checks are successful, then sum random values to get the random index.
    const omega = await parallel(serverLocations, async (_, unwrap) => {
      const randomValues = Object.values(unwrap(rho_)) as [FiniteField];
      return randomValues.reduce((a, b) => field.add(a, b), field.zero) % clientLocations.length;
    });

    const chosenShares = await parallel(serverLocations, async (_, unwrap) =>
      Object.values(unwrap(serverShares))[unwrap(omega)] as FiniteField
    );

    // Server sends the chosen shares to the analyst
    const allShares = await fanin(serverLocations, ["analyst"], (server) =>
      async ({ locally, comm }) => {
        const chosenShares_ = await locally(server, (unwrap) => unwrap(chosenShares));
        return comm(server, "analyst", chosenShares_);
      });

    return await locally("analyst", (unwrap) =>
      Object.values(unwrap(allShares) as Record<string, FiniteField>)
        .reduce((a, b) => field.add(a, b), field.zero));
  };
\end{minted}
    \end{minipage}
    \caption{A federated-lottery protocol (part 2 of 2).
      One of the secret values chosen by the clients is revealed to the analyst;
      as long as at least one server acts honestly
      (randomly generates their \inlinecode{rho} on line 29),
      the choice of which value to reveal will be random.
      Only the analyst learns any of the clients' secrets;
      they only learn the one secret, and they do not learn which one it was.
      The algorithm-step numbers and the unicode variable names align with
      the instructions in Section 6.2 of Keller~\textit{et~al}~\cite{dprio2023}.
      Client secrets are chosen at-will from a finite field (the type \inlinecode{FiniteField});
      we used the finite field of size 999983.}
    \Description{27 lines of Typescript code using \ChoreographyTS defining the second part of a choreography named "lottery".}
    \label{fig:lottery-choreots-example-2}
    \Description{The continuation of the code from Figure 14. The total number of lines in both figures is 91.}
  \end{mdframed}
\end{figure*}

\clearpage
\section{A Formal Conclaves-\&-MLVs Language}\label{sec:more-formalism}

We present the \HLSCentral CP system.
The syntax of \HLSCentral and our overarching computational model and proof-approach are loosely based on
Chorλ~\cite{chor-lambda}.
\HLSCentral is a higher-order choreographic lambda calculus;
we omit recursion and polymorphism because they are orthogonal to our goals here.
Specifically, we will show that multiply-located values
and conclavinging operations are sufficient for a sound
CP language without further KoC management.
In Sections~\ref{sec:syntax} through~\ref{sec:semantics}
we describe the syntax, type system, and semantics of \HLSCentral.
As in other choreographic languages, the primary semantics describes the intended \emph{meaning} of choreographies
and can be used to reason about their behavior,
but is not the "ground truth" of concurrent execution.
Sections~\ref{sec:local-lang} through~\ref{sec:networks} describe the languages of distributed processes,
\HLSLocal and \HLSNet,
and define endpoint projection for \HLSCentral.
In Section~\ref{sec:deadlock-freedom}, we prove that the behavior of a choreography's projection in \HLSNet
matches that of the original \HLSCentral choreography, and that \HLSCentral's type system ensures deadlock-freedom.

\subsection{Syntax}\label{sec:syntax}
The syntax of \HLSCentral is in Figure~\ref{fig:syntax}.
Location information sufficient for typing, semantics, and EPP is explicit in
the expression forms.
We distinguish between "pairs"
($\PAIR V_1 V_2$, of type $(d_1 × d_2)@\nonempty{p}$)
and "tuples"
($(V_1, V_2)$, of type $(T_1, T_2)$)
so that we can have a distinguishable concept of "data" as "stuff that can be sent";
we do not believe this to have any theoretic significance.

The superscript-marked identifier $\nonempty{p}$ is a semantic token representing a set of parties;
an unmarked $p$ is a completely distinct token representing a single party.
Note the use of a superscript "$+$" to denote sets of parties
instead of a hat or boldface;
this denotes that these lists may never be empty.\footnote{
Later, we'll use an "$\ast$" to denote a possibly-empty set or list,
and a "$?$" to denote "zero or one".
}
The typing and semantic rules will enforce this invariant as needed.
When a set of parties should be understood as "context"
rather than "attribute" (\eg in the typing rules),
we write Θ rather than $\nonempty{p}$;
this is entirely to clarify intent and the distinction has no formal meaning.

\begin{figure}[tbp]
\footnotesize
    \begin{mdframed}
\begin{align*}
M  \BNF   &  V                       && \text{Values.}          \\
   \BNFOR &  M M                     && \text{Function application.}          \\
   \BNFOR &  \CASE{\nonempty{p}}{M}{x}{M}{x}{M}    \quad&& \text{Branching on a disjoint-sum value.}          \\
                                            \\
V  \BNF   &  x                       && \text{Variables.}          \\
   \BNFOR &  (λ x:T \DOT M)@\nonempty{p}            && \text{Function literals annotated with participants.}          \\
   \BNFOR &  ()@\nonempty{p}                      && \text{Multiply-located unit.}          \\
   \BNFOR &  \INL V                  && \text{Injection to a disjoint-sum.}           \\
   \BNFOR &  \INR V                  && \text{}           \\
    \BNFOR &  \PAIR V V               && \text{Construction of data pairs (products).}           \\
   \BNFOR &  (V, \dots, V)           && \text{Construction of heterogeneous tuples.}           \\
   \BNFOR &  \FST{\nonempty{p}}      && \text{Projection of data pairs.}           \\
   \BNFOR &  \SND{\nonempty{p}}      && \text{}           \\
   \BNFOR &  \LOOKUP{n}{\nonempty{p}}   && \text{Projection of tuples.}           \\
   \BNFOR &  \COMM{p}{\nonempty{p}}     && \text{Send to one or more recipients.}            \\
                                            \\
d  \BNF   &  ()         && \text{We provide a simple algebra of "data" types,}   \\
   \BNFOR &  d + d                   && \text{which can encode booleans or other finite types}           \\
   \BNFOR &  d × d                   && \text{and could be extended with natural numbers if desired.}   \\
                                            \\
T  \BNF   &  d@\nonempty{p}          && \text{A complete multiply-located data type.}             \\
    \BNFOR &  (T → T)@\nonempty{p}          && \text{Functions are located at their participants.}             \\
   \BNFOR &  (T, \dots, T)           && \text{A fixed-length heterogeneous tuple.}  \\
\end{align*}
    \caption{The complete syntax of the \HLSCentral language.}
    \label{fig:syntax}
    \Description[A BNF syntax for a choreographic lambda calculus.]
                {A BNF syntax for a choreographic lambda calculus.
                There are three expression-forms M,
                including the V form for values,
                of which there are eleven sub-forms.
                There are also forms for types.
                Both types and expressions have party-annotations.}
    \end{mdframed}
\end{figure}

\subsection{The Mask Operator}\label{sec:masking}
Here we introduce the \mask operator,
the purpose of which is to allow Theorem~\ref{theorem:preservation}
(semantic stepping preserves types, Section~\ref{sec:preservation-proof})
to hold
without adding sub-typing or polymorphism to \HLSCentral.
\mask is a partial function defined in Figure~\ref{fig:masking};
the left-hand argument is either a type (in which case it returns a type)
or a value (in which case it returns a value).
The effect of \mask is very similar to EPP,
except that it projects to a set of parties instead of just one,
and instead of introducing a ⊥ symbol it is simply undefined in some cases.
Because it is used during type-checking, errors related to it are caught at that time.

Consider an expression using a "masking identity" function:
$(λ x:()@\set{p} \DOT x)@\set{p} ()@\set{p,q}$,
where the lambda is an identity function \emph{application of which}
turns a multiply-located unit value into one located at just $p$.
Clearly, the lambda should type as $(()@\set{p} → ()@\set{p})@\set{p}$;
and so the whole application expression should type as $()@\set{p}$.
Masking in the typing rules lets this work as expected,
and similar masking in the semantic rules ensures type preservation.

\begin{figure}[tbp]
\footnotesize
    \begin{mdframed}
\begin{gather*}
\myference{MTData}
          {\nonempty{p} ∩ Θ ≠ ∅}
          {d@\nonempty{p} \mask Θ \DEF d@(\nonempty{p} ∩ Θ)}
          \quad
\myference{MTFunction}
          {\nonempty{p} \subseteq Θ}
          {(T → T')@\nonempty{p} \mask Θ \DEF (T → T')@\nonempty{p}}
          \vdbl
\myference{MTVector}
          {T_1' = T_1 \mask Θ, \quad \dots \quad T_n' = T_n \mask Θ}
          {(T_1, \dots, T_n) \mask Θ \DEF (T_1', \dots, T_n')}
          \vdbl
\myference{MVLambda}
          {\nonempty{p} \subseteq Θ}
          {(λ x:T \DOT M)@\nonempty{p} \mask Θ \DEF (λ x:T \DOT M)@\nonempty{p}}
          \quad
\myference{MVUnit}
          {\nonempty{p} ∩ Θ ≠ ∅}
          {()@\nonempty{p} \mask Θ \DEF ()@(\nonempty{p} ∩ Θ)}
          \vdbl
\myference{MVInL}
          {V' = V \mask Θ}
          {\INL V \mask Θ \DEF \INL V'}
          \quad
\myference{MVInR}
          {\dots}
          {\dots}
          \quad
\myference{MVProj1}
          {\nonempty{p} \subseteq Θ}
          {\FST{\nonempty{p}} \mask Θ \DEF \FST{\nonempty{p}}}
          \quad
\myference{MVProj2}
          {\dots}
          {\dots}
          \vdbl
\myference{MVPair}
          {V_1' = V_1 \mask Θ \quad V_2' = V_2 \mask Θ}
          {\PAIR V_1 V_2 \mask Θ \DEF \PAIR V_1' V_2'}
          \quad
\myference{MVVector}
          {V_1' = V_1 \mask Θ \quad \dots \quad V_n' = V_n \mask Θ}
          {(V_1, \dots, V_n) \mask Θ \DEF (V_1', \dots, V_n')}
          \vdbl
\myference{MVProjN}
          {\nonempty{p} \subseteq Θ}
          {\LOOKUP{n}{\nonempty{p}} \mask Θ \DEF \LOOKUP{n}{\nonempty{p}}}
          \quad
\myference{MVCom}
          {s \in Θ \quad \nonempty{r} \subseteq Θ}
          {\COMM{s}{\nonempty{r}} \mask Θ \DEF \COMM{s}{\nonempty{r}}}
          \quad
\myference{MVVar}
          {}
          {x \mask Θ \DEF x}
\end{gather*}
    \caption{Definition of the \mask operator.}
    \label{fig:masking}
    \Description[Inference rules defining a partial function "mask".]
                {Inference rules defining a partial function "mask"
                denoted by a rightward triangle.
                The right-hand argument is a non-empty set of parties,
                and the left-hand argument is either a type or a value.
                It's defined as the left-hand argument re-located
                to the right-hand argument, provided the new locations
                are a subset of the original locations and that the new value
                is still semantically useable.}
    \end{mdframed}
\end{figure}

\subsection{Typing Rules}\label{sec:typing}
The typing rules for \HLSCentral are in Figure~\ref{fig:typing}.
A judgment $Θ;Γ ⊢ M : T$ says that $M$ has type $T$ in the context
of a non-empty set of participating parties Θ
and a (possibly empty) list of variable bindings $Γ=(x_1:T_1), \dots (x_n:T_n)$.
In \textsc{TLambda} and \textsc{TProjN} we write preconditions
$\noop{\nonempty{p}}{T}$ meaning $T = T \mask \nonempty{p}$,
\ie masking to those parties is a "no-op".

\begin{figure}[tbp]
\footnotesize
    \begin{mdframed}
\begin{gather*}
\myference{TLambda}
          {\nonempty{p};Γ,(x:T) ⊢ M : T' \quad
           \nonempty{p} \subseteq Θ \quad
           \noop{\nonempty{p}}{T}}
          {Θ;Γ ⊢ (λ x:T \DOT M)@\nonempty{p} : (T → T')@\nonempty{p}}
          \quad
\myference{TVar}
          {x : T \in Γ \quad T' = T \mask Θ}
          {Θ;Γ ⊢ x : T' }
          \vdbl
\myference{TApp}
          {Θ;Γ ⊢ M : (T_a → T_r)@\nonempty{p} \quad
           Θ;Γ ⊢ N : T_a' \quad
           T_a' \mask \nonempty{p} = T_a}
          {Θ;Γ ⊢ M N : T_r}
          \vdbl
\myference{TCase}
          {Θ;Γ ⊢ N : T_N \quad
           (d_l + d_r)@\nonempty{p} = T_N \mask \nonempty{p} \\
           \nonempty{p};Γ,(x_l:d_l@\nonempty{p}) ⊢ M_l : T \quad
           \nonempty{p};Γ,(x_r:d_r@\nonempty{p}) ⊢ M_r : T \quad
           \nonempty{p} \subseteq Θ}
          {Θ;Γ ⊢ \CASE{\nonempty{p}}{N}{x_l}{M_l}{x_r}{M_r} : T}
          \vdbl
\myference{TUnit}
          {\nonempty{p} \subseteq Θ}
          {Θ;Γ ⊢ ()@\nonempty{p} : ()@\nonempty{p}}
          \quad
\myference{TPair}
          {Θ;Γ ⊢ V_1 : d_1@\nonempty{p_1} \quad
           Θ;Γ ⊢ V_2 : d_2@\nonempty{p_2} \quad
           \nonempty{p_1} ∩ \nonempty{p_2} ≠ ∅}
          {Θ;Γ ⊢ \PAIR V_1 V_2 : (d_1 × d_2)@(\nonempty{p_1} ∩ \nonempty{p_2})}
          \vdbl
\myference{TVec}
          {Θ;Γ ⊢ V_1 : T_1 \quad \dots \quad Θ;Γ ⊢ V_n : T_n}
          {Θ;Γ ⊢ (V_1, \dots, V_n) : (T_1, \dots T_n)}
          \quad
\myference{TInl}
          {Θ;Γ ⊢ V : d@\nonempty{p}}
          {Θ;Γ ⊢ \INL V : (d + d')@\nonempty{p}}
          \quad
\myference{TInr}{\dots}{\dots}
          \vdbl
\myference{TProjN}
          {\nonempty{p} \subseteq Θ \quad
           \noop{\nonempty{p}}{(T_1, \dots, T_n)}}
          {Θ;Γ ⊢ \LOOKUP{i}{\nonempty{p}} : ((T_1, \dots, T_i, \dots, T_n) → T_i)@\nonempty{p}}
          \quad
\myference{TProj2}{\dots}{\dots}
          \vdbl
\myference{TProj1}
          {\nonempty{p} \subseteq Θ}
          {Θ;Γ ⊢ \FST{\nonempty{p}} : ((d_1 × d_2)@\nonempty{p} → d_1@\nonempty{p})@\nonempty{p}}
          \vdbl
\myference{TCom}
          {s \in \nonempty{s} \quad
           \nonempty{s}\cup\nonempty{r} \subseteq Θ}
          {Θ;Γ ⊢ \COMM{s}{\nonempty{r}} : (d@\nonempty{s} → d@\nonempty{r})@(\set{s}\cup\nonempty{r})}
\end{gather*}
    \caption{\HLSCentral typing rules.}
    \label{fig:typing}
    \Description[Inference rules for he-lambda-small's type system.]
                {Inference rules for he-lambda-small's type system.
                There are thirteen of them, corresponding to the thirteen
                total expression forms.}
    \end{mdframed}
\end{figure}

Examine \textsc{TCase} as the most involved example.
The actual judgment says that in the context of Θ and Γ,
the case expression types as $T$.
The first two preconditions say that
the guard expression $N$ must type in the parent context
as some type $T_N$, which masks to the explicit party-set $\nonempty{p}$
as a sum-type $(d_l + d_r)@\nonempty{p}$.
The only rule by which it can do that is \textsc{MTData},
so we can deduce that $T_N = (d_l + d_r)@\nonempty{q}$,
where $\nonempty{q}$ is some unspecified superset of $\nonempty{p}$.
The third and forth preconditions say that $M_l$ and $M_r$
must both type as $T$ in the context of $\nonempty{p}$ instead of Θ
and with the respective $x_l$ and $x_r$ bound to the right and left
data types at $\nonempty{p}$.
The final precondition says that $\nonempty{p}$ is a subset of Θ,
\ie everyone who's supposed to be branching is actually present to do so.

The other rules are mostly normal, with similar masking of types and narrowing
of participant sets as needed.
In \textsc{TVar}, the Θ context overrides (masks) the type bindings in Γ.
In isolation, some expressions such as $\INR ()@\set{p}$
or the projection operators
are flexible about their exact types;
additional annotations could give them monomorphic typing,
if that was desirable.

\subsection{Masked Substitution}\label{sec:substitution}

For \mask to fulfil its purpose during semantic evaluation,
it may need to be applied arbitrarily many times with different party-sets
inside the new expressions, and it may not even be defined for all such
party-sets.
Conceptually, this just recapitulates the masking performed in \textsc{TVar}.
To formalize these subtleties, in Figure~\ref{fig:substitution} we specialize the normal variable-substitution
notation $M[x:=V]$ to perform location-aware substitution.
In Section~\ref{sec:substitution-proof} we prove Theorem~\ref{theorem:substitution},
which shows that this specialized substitution operation
satisfies the usual concept of substitution.

\begin{figure}[tbp]
\footnotesize
    \begin{mdframed}
\begin{align*}
M[x:=V] \DEF \text{by pattern matching on $M$:}& \\
y            \DEFCASE & \begin{cases}
                                        y ≡ x & \DEFCASE  V  \\
                                        y ≢ x & \DEFCASE  y
                                        \end {cases} \\
N_1 N_2     \DEFCASE & N_1[x:=V] N_2[x:=V] \\
(λ y:T \DOT N)@\nonempty{p}  \DEFCASE & \begin{cases}
                                        V \mask \nonempty{p} = V'
                                            & \DEFCASE (λ y:T \DOT N[x:=V'])@\nonempty{p} \\
                                        \text{otherwise} & \DEFCASE M
                                        \end{cases} \\
\CASEm{\nonempty{p}}{N}{x_l}{M_l}{x_r}{M_r} \DEFCASE & \begin{cases}
                                        V \mask \nonempty{p} = V'
                                            & \DEFCASE \CASEm{\nonempty{p}}
                                                            {N[x:=V]}{x_l}{M_l[x:=V']}
                                                            {x_r}{M_r[x:=V']} \\
                                        \text{otherwise}
                                            & \DEFCASE \CASEm{\nonempty{p}}
                                                            {N[x:=V]}{x_l}{M_l}{x_r}{M_r}
                                        \end{cases} \\
\INL V_1    \DEFCASE & \INL V_1[x:=V] \\
\INR V_2    \DEFCASE & \INR V_2[x:=V] \\
\PAIR V_1 V_2  \DEFCASE & \PAIR V_1[x:=V] V_2[x:=V] \\
(V_1, \dots, V_n) \DEFCASE & (V_1[x:=V], \dots, V_n[x:=V]) \\
\begin{rcases}
    ()@\nonempty{p}
    \qquad \FST{\nonempty{p}}
    \qquad \SND{\nonempty{p}} \\
    \qquad \LOOKUP{\nonempty{p}}{i}
    \qquad \COMM{s}{\nonempty{r}}
\end{rcases}\DEFCASE & M
\end{align*}
    \caption{The customised substitution used in \HLSCentral's semantics.}
    \label{fig:substitution}
    \Description[A case-wise definition of variable substitution.]
                {A case-wise definition of variable substitution.
                 Most cases are normal; some involve masking
                 and when the masking is undefined they revert to a no-op.}
    \end{mdframed}
\end{figure}

\subsection{Centralized Semantics}\label{sec:semantics}

The semantic stepping rules for evaluating \HLSCentral expressions
are in Figure~\ref{fig:semantics}.
In Sections~\ref{sec:local-lang}, \ref{sec:projection}, and \ref{sec:networks}
we will develop the "ground truth" of the distributed process semantics and show that
the \HLSCentral's semantics correctly capture distributed behavior.

\HLSCentral is equipped with a substitution-based semantics that,
after accounting for the \mask operator and the specialized implementation of
substitution, is quite standard among lambda-calculi.
In particular, we make no effort here to support the out-of-order execution
supported by some choreography languages.
Because the language and corresponding computational model are parsimonious,
no step-annotations are needed for the centralized semantics.

\begin{figure}[tbp]
    \begin{mdframed}
\begin{gather*}
\myference{AppAbs}
          {V' = V \mask \nonempty{p}}
          {((λ x:T \DOT M)@\nonempty{p}) V \step M[x := V']}
          \quad
\myference{App1}
          {N \step N'}
          {V N \step V N'}
          \quad
\myference{App2}
          {M \step M'}
          {M N \step M' N}
          \vdbl
\myference{Case}
          {N \step N'}
          {\CASE{\nonempty{p}}{N}{x_l}{M_l}{x_r}{M_r}
            \step \CASE{\nonempty{p}}{N'}{x_l}{M_l}{x_r}{M_r}}
          \vdbl
\myference{CaseL}
          {V' = V \mask \nonempty{p}}
          {\CASE{\nonempty{p}}{\INL V}{x_l}{M_l}{x_r}{M_r} \step M_l[x_l := V']}
          \quad
\myference{CaseR}
          {\dots}
          {\dots}
          \vdbl
\myference{Proj1}
          {V' = V_1 \mask \nonempty{p}}
          {\FST{\nonempty{p}} (\PAIR V_1 V_2) \step V'}
          \quad
\myference{Proj2}
          {\dots}
          {\dots}
          \quad
\myference{ProjN}
          {V' = V_i \mask \nonempty{p}}
          {\LOOKUP{i}{\nonempty{p}} (V_1, \dots, V_i, \dots, V_n) \step V'}
          \vdbl
\myference{Com1}
          {()@\nonempty{p} \mask \set{s} = ()@s}
          {\COMM{s}{\nonempty{r}} ()@\nonempty{p} \step ()@\nonempty{r}}
          \quad
\myference{ComPair}
          {\COMM{s}{\nonempty{r}} V_1 \step V_1' \quad \COMM{s}{\nonempty{r}} V_2 \step V_2'}
          {\COMM{s}{\nonempty{r}} (\PAIR V_1 V_2) \step \PAIR V_1' V_2'}
          \vdbl
\myference{ComInl}
          {\COMM{s}{\nonempty{r}} V \step V'}
          {\COMM{s}{\nonempty{r}} (\INL V) \step \INL V'}
          \quad
\myference{ComInr}
          {\dots}
          {\dots}
\end{gather*}
    \caption{\HLSCentral's semantics.}
    \label{fig:semantics}
    \Description[Infernce rules for the central language.]
                {Thirteen inference rules defining the semantics of
                 choreographies.
                Most of them are entirely normal lambda-calculus rules
                except that they use a mask operator and a specialized
                notion of substitution.
                The exceptions are the COM rules, which use each other
                as recursive preconditions to replace the location annotations
                on unit values.}
    \end{mdframed}
\end{figure}

The \textsc{Com1} rule simply replaces one location-annotation with another.
\textsc{ComPair}, \textsc{ComInl}, and \textsc{ComInr} are defined recursively
amongst each other and \textsc{Com1};
the effect of this is that "data" values can be sent but other values
(functions and variables) cannot.

As is typical for a typed lambda calculus, \HLSCentral enjoys preservation and progress.
We prove these properties in Sections~\ref{sec:preservation-proof} and~~\ref{sec:progress-proof} respectively.

\subsection{The Local Process Language}\label{sec:local-lang}

In order to define EPP and a "ground truth" for \HLSCentral computation,
we need a locally-computable language, \HLSLocal, into which it can project.
\HLSLocal is very similar to \HLSCentral;
to avoid ambiguity we denote \HLSLocal expressions $B$ (for "behavior")
instead of $M$ (which denotes a \HLSCentral expression)
and \HLSLocal values $L$ instead of $V$.
The syntax is presented in Figure~\ref{fig:local-syntax}.

\begin{figure}[tbp]
    \begin{mdframed}
    \begin{align*}
        B \BNF   & L \BNFOR B B && \text{\small Process expressions.} \\
          \BNFOR & \CASE{}{B}{x}{B}{x}{B} \\[0.5em]
        L \BNF   & x \BNFOR () \BNFOR   λ x \DOT B
                     && \text{\small Process values.} \\
          \BNFOR & \INL L \BNFOR \INR L \BNFOR  \PAIR L L \\
          \BNFOR & \FST{} \BNFOR \SND{} \\
          \BNFOR & (L, \dots, L) \BNFOR \LOOKUP{n}{} && \text{} \\
          \BNFOR & \RECV{p} \BNFOR \SEND{p^{\ast}}
                     && \text{\small Receive from one party. Send to many.} \\
          \BNFOR & \SEND{p^{\ast}}^{\ast}
                    && \text{\small Send to many \emph{and} keep for oneself.} \\
          \BNFOR & ⊥                  && \text{\small "Missing" (located someplace else).}
    \end{align*}
    \caption{Syntax for the \HLSLocal language.}
    \label{fig:local-syntax}
    \Description[A BNF for a simple lambda calculus with "send" and "receive" operators.]
                {A BNF for a simple lambda calculus with "send" and "receive" operators.
                By design, it's very similar to the central language, just without party annotations
                and with "com" replaced by "send", "send*", and "recv".}
    \end{mdframed}
\end{figure}

\HLSLocal differs from \HLSCentral in a few ways.
It's untyped, and the party-set annotations are mostly missing.
\HLSCentral's $\COMM{p}{\nonempty{q}}$ operator is replaced by $\SEND{\nonempty{q}}$ and $\RECV{p}$,
as well as a $\SEND{\nonempty{q}}^{\ast}$, which differs from $\SEND{\nonempty{q}}$ only in that
the process which calls it keeps a copy of the sent value for itself.
Syntactically, the recipient lists of $\SEND{}$ and $\SEND{}^{\ast}$ may be empty;
this keeps semantics consistent in the edge case implied by
a \HLSCentral expression like $\COMM{s}{\set{s}}$ (which is useless but legal).
Finally, the value-form ⊥ ("bottom") is a stand-in for parts of the choreography that do not involve the target party.
In the context of choreographic languages, ⊥ does not denote an error but should instead be read as "unknown"
or "somebody else's problem".

The behavior of ⊥ during semantic evaluation can be handled a few different ways,
the pros-and-cons of which are not important in this work.
We use a ⊥-normalizing "floor" function, defined in Figure~\ref{fig:floor},
during EPP and semantic stepping to avoid ever handling
⊥-equivalent expressions like $\PAIR ⊥ ⊥$ or $⊥ ()$.

\begin{figure}[tbp]
\footnotesize
    \begin{mdframed}
\begin{align*}
\FLR{B}                        \DEF      \text{by pattern matching on $B$:}
  & \qquad\qquad\qquad\qquad  \text{\small{(Observe that floor is idempotent.)}} \\
B_1 B_2                      \DEFCASE &
  \begin{cases}
    \FLR{B_1} = ⊥, \FLR{B_2} = L \DEFCASE & ⊥  \\
    \text{else}              \DEFCASE & \FLR{B_1} \FLR{B_2}
  \end{cases}  \\
\CASE{}{B_G}{x_l}{B_l}{x_r}{B_r} \DEFCASE &
  \begin{cases}
    \FLR{B_G} = ⊥                \DEFCASE & ⊥ \\
    \text{else}      \DEFCASE & \CASE{}{\FLR{B_G}}{x_l}{\FLR{B_l}}{x_r}{\FLR{B_r}}
  \end{cases}  \\
λ x \DOT B'                  \DEFCASE & λ x \DOT \FLR{B'} \\
\INL L                       \DEFCASE & \begin{cases}
  \FLR{L} = ⊥                \DEFCASE & ⊥ \\
  \text{else}                \DEFCASE & \INL \FLR{L}
  \end{cases} &&      \\
\INR L                       \DEFCASE & \begin{cases}
  \FLR{L} = ⊥                \DEFCASE & ⊥ \\
  \text{else}                \DEFCASE & \INR \FLR{L}
  \end{cases} &&      \\
\PAIR L_1 L_2                \DEFCASE & \begin{cases}
  \FLR{L_1} = ⊥ = \FLR{L_2}  \DEFCASE & ⊥ \\
  \text{else}                \DEFCASE & \PAIR \FLR{L_1} \FLR{L_2}
  \end{cases} &&         \\
(L_1, \dots, L_n)            \DEFCASE & \begin{cases}
  \forall_{i\in[1,n]} \FLR{L_i} = ⊥ \DEFCASE & ⊥ \\
  \text{else}                \DEFCASE & (\FLR{L_1}, \dots, \FLR{L_n})
  \end{cases} && \\
\begin{rcases}
  x \\
  () \\
  \FST{} \\
  \SND{} \\
  \LOOKUP{i}{} \\
  \SEND{p^{\ast}} \\
  \SEND{p^{\ast}}^{\ast} \\
  \RECV{p} \\
  ⊥
\end{rcases}                 \DEFCASE &  B
\end{align*}
    \caption{The "floor" function, which reduces ⊥-based expressions.}
    \label{fig:floor}
    \Description[A casewise definition of a function using the bottom-brackets associated with a real-number "floor" function.]
                {A casewise definition of a function, using the bottom-brackets associated with a real-number "floor" function,
                that takes a local-language expression and returns it with "bottom" values simplified.}
    \end{mdframed}
\end{figure}

\HLSLocal's semantic stepping rules are given in Figure~\ref{fig:local-semantics}.
Local steps are labeled with send ($⊕$) and receive ($⊖$) sets, like so:
$B \prcstep{\set{(p,L_1), (q,L_2)}}{\set{(r, L_3), (s, L_4)}} B'$,
or $B \prcstep{μ}{η} B'$ when we don't need to inspect the contents of the annotations.
The floor function is used to keep expressions normalized during evaluation.
Otherwise, most of the rules are analogous to the corresponding \HLSCentral rules from Figure~\ref{fig:semantics}.
The \textsc{LSend-} rules are defined recursively, similar to the \textsc{Com-} rules.
\textsc{LSendSelf} shows that $\SEND{}^{\ast}$ is exactly like $\SEND{}$
except it locally acts like \inlinecode{id} instead of returning ⊥.
\textsc{LRecv} shows that the $\RECV{}$ operator ignores its argument and can return
\emph{anything}, with the only restriction being that the return value must be reflected in the receive-set step-annotation.

\begin{figure}[tbp]
\footnotesize
    \begin{mdframed}
\begin{gather*}
\myference{LAbsApp}
          {}
          {(λ x \DOT B) L \prcstep{∅}{∅} \FLR{B[x:=L]}}
          \quad
\myference{LApp1}
          {B \prcstep{μ}{η} B'}
          {L B \prcstep{μ}{η} \FLR{L B'}}
          \quad
\myference{LApp2}
          {B \prcstep{μ}{η} B'}
          {B B_2 \prcstep{μ}{η} \FLR{B' B_2}}
          \vdbl
\myference{LCase}
          {B \prcstep{μ}{η} B'}
          {\CASE{}{B}{x_l}{B_l}{x_r}{B_r} \prcstep{μ}{η}
           \FLR{\CASE{}{B'}{x_l}{B_l}{x_r}{B_r}}}
          \vdbl
\myference{LCaseL}
          {}
          {\CASE{}{\INL L}{x_l}{B_l}{x_r}{B_r} \prcstep{∅}{∅} \FLR{B_l[x_l := L]}}
          \quad
\myference{LCaseR}
          {\dots}
          {\dots}
          \vdbl
\myference{LProj1}
          {}
          {\FST{} (\PAIR L_1 L_2) \prcstep{∅}{∅} L_1}
          \quad
\myference{LProj2}
          {\dots}
          {\dots}
          \quad
\myference{LProjN}
          {}
          {\LOOKUP{i}{} (L_1, \dots, L_i, \dots, L_n) \prcstep{∅}{∅} L_i}
          \vdbl
\myference{LSend1}
          {}
          {\SEND{p^{\ast}} () \prcstep{\set{(p, ()) \mid p \in p^{\ast}}}{∅} ⊥}
          \quad
\myference{LSendPair}
          {\SEND{p^{\ast}} L_1 \prcstep{μ_1}{∅} ⊥ \quad
           \SEND{p^{\ast}} L_2 \prcstep{μ_2}{∅} ⊥}
          {\SEND{p^{\ast}} (\PAIR L_1 L_2)
           \prcstep{\set{(p, \PAIR L_1 L_2) \mid p \in p^{\ast}}}{∅}
           ⊥}
          \vdbl
\myference{LSendInL}
          {\SEND{p^{\ast}} L \prcstep{μ}{∅} ⊥}
          {\SEND{p^{\ast}} (\INL L)
           \prcstep{\set{(p, \INL L) \mid p \in p^{\ast}}}{∅}
           ⊥}
          \quad
\myference{LSendInR}
          {\dots}
          {\dots}
          \quad
\myference{LSendSelf}
          {\SEND{p^{\ast}} L \prcstep{μ}{∅} ⊥}
          {\SEND{p^{\ast}}^\ast L \prcstep{μ}{∅} L}
          \vdbl
\myference{LRecv}
          {}
          {\RECV{p} L_0 \prcstep{∅}{\set{(p, L)}} L}
          \quad
\end{gather*}
    \caption{The semantics of \HLSLocal.}
    \label{fig:local-semantics}
    \Description[Fifteen inference rules defining the semantics of the local process language.]
                {Inference rules defining the substitution-based semantic stepping of the local process language.
                There are fifteen rules, roughly corresponding to the similar rules from the choreographic semantics.}
    \end{mdframed}
\end{figure}

\subsection{Endpoint Projection}\label{sec:projection}
Endpoint projection (EPP) is the translation between the choreographic language \HLSCentral
and the local process language \HLSLocal;
necessarily it's parameterized by the specific
local process you're projecting \emph{to}.
$⟦M⟧_p$ is the projection of $M$ to $p$, as defined in Figure~\ref{fig:eep}.
It does a few things:
Most location annotations are removed, some expressions become ⊥,
⊥-based expressions are normalized by the floor function,
and $\COMM{s}{\nonempty{r}}$ becomes $\SEND{\nonempty{r}}$, $\SEND{\nonempty{r}}^{\ast}$, or $\RECV{s}$,
keeping only the identities of the peer parties and not the local party.

\begin{figure}[tbp]
\footnotesize
    \begin{mdframed}
\begin{align*}
⟦M⟧_p                        \DEF      \text{by pattern matching on $M$:}& \\
N_1 N_2                      \DEFCASE & \FLR{⟦N_1⟧_p ⟦N_2⟧_p} \\
\CASEm{\nonempty{p}}{N}{x_l}{M_l}{x_r}{M_r} \DEFCASE &
  \begin{cases}
    p \in \nonempty{p}       \DEFCASE & \FLR{
      \CASE{}{⟦N⟧_p}{x_l}{⟦M_l⟧_p}{x_r}{⟦M_r⟧_p} } \\
    \text{else}              \DEFCASE & \FLR{
      \CASE{}{⟦N⟧_p}{x_l}{⊥}{x_r}{⊥} }
  \end{cases}  \\
x                            \DEFCASE &  x        \\
(λ x:T \DOT N)@\nonempty{p}  \DEFCASE &
  \begin{cases}
    p \in \nonempty{p}       \DEFCASE & λ x \DOT ⟦N⟧_p \\
    \text{else}              \DEFCASE & ⊥
  \end{cases}  \\
()@\nonempty{p}              \DEFCASE &
  \begin{cases}
    p \in \nonempty{p}       \DEFCASE & () \\
    \text{else}              \DEFCASE & ⊥
  \end{cases}  \\
\INL V                       \DEFCASE & \FLR{\INL ⟦V⟧_p}  &&      \\
\INR V                       \DEFCASE & \FLR{\INR ⟦V⟧_p}  &&      \\
\PAIR V_1 V_2                \DEFCASE & \FLR{\PAIR ⟦V_1⟧_p ⟦V_2⟧_p} &&         \\
(V_1, \dots, V_n)            \DEFCASE & \FLR{(⟦V_1⟧_p, \dots, ⟦V_n⟧_p)} &&       \\
\FST{\nonempty{p}}           \DEFCASE &
  \begin{cases}
    p \in \nonempty{p}       \DEFCASE & \FST{} \\
    \text{else}              \DEFCASE & ⊥
  \end{cases}  \\
\SND{\nonempty{p}}           \DEFCASE &
  \begin{cases}
    p \in \nonempty{p}       \DEFCASE & \SND{} \\
    \text{else}              \DEFCASE & ⊥
  \end{cases}  \\
\LOOKUP{i}{\nonempty{p}}     \DEFCASE &
  \begin{cases}
    p \in \nonempty{p}       \DEFCASE & \LOOKUP{i}{} \\
    \text{else}              \DEFCASE & ⊥
  \end{cases}  \\
\COMM{s}{\nonempty{r}}       \DEFCASE &
  \begin{cases}
    p = s, p \in \nonempty{r}      \DEFCASE & \SEND{\nonempty{r} ∖ \set{p}}^\ast \\
    p = s, p \not\in \nonempty{r}  \DEFCASE & \SEND{\nonempty{r}} \\
    p \not = s, p \in \nonempty{r} \DEFCASE & \RECV{s} \\
    \text{else}              \DEFCASE & ⊥
  \end{cases}
\end{align*}
    \caption{EPP from \HLSCentral to \HLSLocal.}
    \label{fig:eep}
    \Description[A casewise definition of a function denoted by double-square-brackets and parameterized by a party-name subscript.]
                {A casewise definition of a function, denoted by enclosing the argument in double-square-brackets,
                parameterized by a party name in subscript,
                that takes a He-Lambda-Small expression and returns the party's view of it in the local process language.}
    \end{mdframed}
\end{figure}

\subsection{Process Networks}\label{sec:networks}
A single party evaluating local code can hardly be considered the ground truth of choreographic computation;
for a message to be sent it must be received \emph{by} someone (and \textit{visa-versa}).
Our third "language", \HLSNet, is just concurrent asynchronous threads of \HLSLocal.
An \HLSNet "network" $\mathcal{N}$
is a dictionary mapping each party in its domain to a \HLSLocal program representing that party's current place in the execution.
We express party-lookup as $\mathcal{N}(p) = B$.
A singleton network, written $\mathcal{N} = p[B]$, has the one party $p$ in its domain and assigns the expression $B$ to it.
Parallel composition of networks is expressed as $\mathcal{N} \mid \mathcal{N}'$
(the order doesn't matter).
Thus, the following statements are basically equivalent:
\begin{itemize}
  \item $\mathcal{N}(p) = B$
  \item $\mathcal{N} = p[B] \mid \mathcal{N}'$
  \item $p[B] \in \mathcal{N}$
\end{itemize}
When many compositions need to be expressed at once, we can write
$\mathcal{N} = Π_{p \in \nonempty{p}} p[B_p]$.
Parallel projection of all participants in $M$ is expressed as
$⟦M⟧ = Π_{p \in \roles{M}} p[⟦M⟧_p]$.
For example, if $p$ and $q$ are the only parties in $M$, then
$⟦M⟧ = p[⟦M⟧_p] \mid q[⟦M⟧_q]$.

The rules for \HLSNet semantics are in Figure~\ref{fig:networks}.
\HLSNet semantic steps are annotated with \emph{incomplete} send actions;
$\mathcal{N} \netstep{p}{\set{\dots,(q_i, L_i),\dots}} \mathcal{N}'$
indicates a step in which $p$ sent a respective $L_i$ to each of the listed $q_i$
and the $q_i$s have \emph{not} been noted as receiving.
When there are no such incomplete sends and the $p$ doesn't matter,
it may be omitted
(\eg $\mathcal{N} \netstep{}{∅} \mathcal{N}'$
instead of $\mathcal{N} \netstep{p}{∅} \mathcal{N}'$).
\textbf{Only $∅$-annotated steps are "real";}
other steps are conceptual justifications used in the semantics's derivation trees.
In other words, \HLSLocal semantics only elevate to \HLSNet semantics
when the message-annotations cancel out.
Rule \textsc{NCom} allows annotations to cancel out.
For example the network
$⟦\COMM{s}{\set{p,q}} ()@\set{s}⟧$
gets to $⟦()@\set{p,q}⟧$
by a \emph{single} \textsc{NCom} step.
The derivation tree for that step starts at the top with \textsc{NPro}:
$s[\SEND{\set{p,q}} ()] \netstep{s}{\set{(p,()),(q,())}} s[⊥]$;
this justifies two nestings of \textsc{NCom} in which the $p$ step and $q$ step
(in either order)
compose with the $s$ step and remove the respective party from the step-annotation.

\begin{figure}[tbp]
\footnotesize
    \begin{mdframed}
\begin{gather*}
\myference{NPro}
          {B \prcstep{μ}{∅} B'}
          {p[B] \netstep{p}{μ} p[B']}
          \quad
\myference{NCom}
          {\mathcal{N} \netstep{s}{μ∪\set{(r,L)}} \mathcal{N}'
           \quad B \prcstep{∅}{\set{(s, L)}} B'}
          {\mathcal{N} \mid r[B] \netstep{s}{μ} \mathcal{N}' \mid r[B']}
          \quad
\myference{NPar}
          {\mathcal{N} \netstep{}{∅} \mathcal{N}'}
          {\mathcal{N} \mid \mathcal{N}^{+} \netstep{}{∅} \mathcal{N}' \mid \mathcal{N}^{+}}
\end{gather*}
    \caption{Semantic rules for \HLSNet.}
    \label{fig:networks}
    \Description[Inference rules for networks of processes, showing when and how the local semantics can actually be applied.]
                {Three inference rules for networks of processes,
                showing when and how the local semantic stepping rules can be applied in the "real world" of communicating processes.}
    \end{mdframed}
\end{figure}

\subsection{Deadlock Freedom}\label{sec:deadlock-freedom}
Above we introduced the necessary machinery of EPP and evaluation of a network of communicating processes.
In Section~\ref{sec:soundness-proof} we prove that EPP is \emph{sound}
(Theorem~\ref{theorem:soundness}, any behavior possible for the \HLSNet projection of a choreography is also possible in the original \HLSCentral).
In Section~\ref{sec:completeness-proof} we prove that EPP is \emph{complete}
(Theorem~\ref{theorem:completeness}, any behavior possible in \HLSCentral is also possible in the \HLSNet projection).

The central promise of choreographic programming is that participants in well-formed choreographies
will never get stuck waiting for messages they never receive.
This important property, \textit{"deadlock freedom by design"}, is trivial once our previous theorems are in place.

\begin{corollary}[Deadlock Freedom]\label{theorem:deadlock}
  If $Θ;∅ ⊢ M : T$ and $⟦M⟧ \netstep{}{∅}^{\ast} \mathcal{N}$,
    then either $\mathcal{N} \netstep{}{∅}^{\ast} \mathcal{N}'$
    or for every $p\in\roles{M}$, $\mathcal{N}(p)$ is a value.

    This follows from Theorem~\ref{theorem:soundness}, Theorem~\ref{theorem:preservation},
    Theorem~\ref{theorem:progress}, and Theorem~\ref{theorem:completeness}.
\end{corollary}

\section{Proof of The Substitution Theorem}\label{sec:substitution-proof}

Theorem~\ref{theorem:substitution} says that
if $Θ;Γ,(x:T_x) ⊢ M : T$ and $Θ;Γ ⊢ V : T_x$,
then $Θ;Γ ⊢ M[x := V] : T$.
We first prove a few lemmas.

\begin{lemma}[Conclave]\label{theorem:conclave}
    If $Θ;Γ ⊢ V : T$ and $Θ' \subseteq Θ$
    and $T' = T \mask Θ'$ is defined
    then $V' = V \mask Θ'$ is defined,
    and $Θ';Γ ⊢ V' : T'$.
\end{lemma}

\subsection{Proof of Lemma~\ref{theorem:conclave}}
This is vacuous if $T'$ doesn't exist, so assume it does.
Do induction on the definition of masking for $T$:

\begin{itemize}
\item \textsc{MTData}: $Θ;Γ ⊢ V : d@\nonempty{p}$ and $\nonempty{p} ∩ Θ' ≠ ∅$
  so $T' = d@(\nonempty{p} ∩ Θ')$.
  Consider cases for typing of $V$:
  \begin{itemize}
    \item \textsc{TVar}: $V' = V$ by \textsc{MVVar} and it types by \textsc{TVar} b.c. $T'$ exists.
    \item \textsc{TUnit}: We've already assumed the preconditions for \textsc{MVUnit}, and it types.
    \item \textsc{TPair}: $V = \PAIR V_1 V_2$,
      and $Θ;Γ ⊢ V_1 : d_1@(\nonempty{p_1} \supseteq \nonempty{p})$
      and $Θ;Γ ⊢ V_2 : d_2@(\nonempty{p_2} \supseteq \nonempty{p})$.
      By \textsc{MTData}, these larger-owernership types will still mask with $Θ'$,
      so this case come by induction.
    \item \textsc{TInL}, \textsc{TInR}: Follows by simple induction.
  \end{itemize}
\item \textsc{MTFunction}: $T' = T$ and $\nonempty{p} \subseteq Θ'$,
  so lambdas and function-keywords all project unchanged, and the respective typings hold.
\item \textsc{MTVector}: Simple induction.
\end{itemize}

\begin{lemma}[Quorum]\label{theorem:quorum}
    \textbf{A)} If $Θ;Γ,(x:T_x) ⊢ M : T$ and $T_x' = T_x \mask Θ$, then $Θ;Γ,(x:T_x') ⊢ M : T$.

    \textbf{B)} If $Θ;Γ,(x:T_x) ⊢ M : T$ and $T_x \mask Θ$ is not defined, then $Θ;Γ ⊢ M : T$.
\end{lemma}

\subsection{Proof of Lemma~\ref{theorem:quorum}}
By induction on the typing of M.
The only case that's not recursive or trivial is \textsc{TVar},
for which we just need to observe that masking on a given party-set is idempotent.

\begin{lemma}[Unused]\label{theorem:unused}
  If $Θ;Γ ⊢ M : T$ and $x \not \in Γ$, then $M[x := V] = M$.
\end{lemma}
\subsection{Proof of Lemma~\ref{theorem:unused}}
By induction on the typing of $M$.
There are no non-trivial cases.

\subsection{Theorem~\ref{theorem:substitution}}

\begin{theorem}[Substitution]\label{theorem:substitution}
  If $Θ;Γ,(x:T_x) ⊢ M : T$ and $Θ;Γ ⊢ V : T_x$,
  then $Θ;Γ ⊢ M[x := V] : T$.
\end{theorem}

The proof is in 13 cases.
\textsc{TProjN}, \textsc{TProj1}, \textsc{TProj2}, \textsc{TCom}, and \textsc{TUnit}
are trivial base cases.
\textsc{TInL}, \textsc{TInR}, \textsc{TVec}, and \textsc{TPair}
are trivial recursive cases.

\begin{itemize}
  \item \textsc{TLambda} where $T_x' = T_x \mask \nonempty{p}$:
  $M = (λ y : T_y \DOT N)@\nonempty{p}$ and $T = (T_y → T')@\nonempty{p}$.
  \begin{enumerate}
      \item $Θ;Γ,(x:T_x) ⊢ (λ y : T_y \DOT N)@\nonempty{p} : (T_y → T')@\nonempty{p}$ by assumption.
      \item $Θ;Γ ⊢ V : T_x$ by assumption.
      \item $\nonempty{p};Γ,(x:T_x),(y:T_y) ⊢ N : T'$ per preconditions of \textsc{TLambda}.
      \item $Θ;Γ,(y:T_y) ⊢ V : T_x$ by weakening (or strengthening?) \#2.
      \item $V' = V \mask \nonempty{p}$ and $\nonempty{p}; Γ,(y:T_y) ⊢ V' : T_x'$ by Lemma~\ref{theorem:conclave}.
      \item $\nonempty{p};Γ,(x:T_x'),(y:T_y) ⊢ N : T'$ by applying Lemma~\ref{theorem:quorum} to \#3.
      \item $\nonempty{p};Γ,(y:T_y) ⊢ N[x:=V'] : T'$ by induction on \#6 and \#5.
      \item $M[x:=V] = (λ y : T_y \DOT N[x:=V'])@\nonempty{p}$ by definition,
     which typechecks by \#7 and \textsc{TLambda}. \textbf{QED.}
  \end{enumerate}
  \item \textsc{TLambda} where $T_x \mask \nonempty{p}$ is undefined:
  $M = (λ y : T_y \DOT N)@\nonempty{p}$.
  \begin{enumerate}
      \item $\nonempty{p};Γ,(x:T_x),(y:T_y) ⊢ N : T'$ per preconditions of \textsc{TLambda}.
      \item $\nonempty{p};Γ,(y:T_y) ⊢ N : T'$ by Lemma~\ref{theorem:quorum} B.
      \item $N[x:=V] = N$ by Lemma~\ref{theorem:unused},
     so regardless of the existence of $V \mask \nonempty{p}$ the substitution is a noop,
     and it typechecks by \#2 and \textsc{TLambda}.
  \end{enumerate}
  \item \textsc{TVar}: Follows from the relevant definitions, whether $x ≡ y$ or not.
  \item \textsc{TApp}: This is also a simple recursive case;
  the masking of $T_a$ doesn't affect anything.
  \item \textsc{TCase}: Follows the same logic as \textsc{TLambda},
  just duplicated for $M_l$ and $M_r$.
\end{itemize}

\section{Proof of The Preservation Theorem}\label{sec:preservation-proof}
Theorem~\ref{theorem:preservation} says that
if $Θ;∅ ⊢ M : T$ and $M \step M'$,
then $Θ;∅ ⊢ M' : T$.
We'll need a few lemmas first.

\begin{lemma}[Sub-Mask]\label{theorem:sub-mask}
  If $Θ;Γ ⊢ V : d@\nonempty{p}$ and $∅ ≠ \nonempty{q} \subseteq \nonempty{p}$,
    then \textbf{A:} $d@\nonempty{p} \mask \nonempty{q} = d@\nonempty{q}$ is defined
    and \textbf{B:} $V \mask \nonempty{q}$ is also defined and types as $d@\nonempty{q}$.
\end{lemma}
\subsection{Proof of Lemma~\ref{theorem:sub-mask}}
Part A is obvious by \textsc{MTData}.
Part B follows by induction on the definition of masking for values.
\begin{itemize}
\item \textsc{MVLambda}: Base case; can't happen because it wouldn't allow a data type.
\item \textsc{MVUnit}: Base case; passes definition and typing.
\item \textsc{MVInL}, \textsc{MVInR}: Recursive cases.
\item \textsc{MVPair}: Recursive case.
\item \textsc{MVVector}: Can't happen because it wouldn't allow a data type.
\item \textsc{MVProj1}, \textsc{MVProj2}, \textsc{MVProjN}, and \textsc{MVCom}:
  Base cases, can't happen because they wouldn't allow a data type.
\item \textsc{MVVar}: Base case, trivial.
\end{itemize}

\begin{lemma}[Maskable]\label{theorem:maskable}
  If $Θ;Γ ⊢ V : T$ and $T \mask \nonempty{p} = T'$,
  then \textbf{A:} $V \mask \nonempty{p} = V'$ is defined
    and \textbf{B:} $Θ;Γ ⊢ V' : T'$.
\end{lemma}
\subsection{Proof of Lemma~\ref{theorem:maskable}}
By induction on the definition of masking for values.
\begin{itemize}
\item \textsc{MVLambda}: Base case. From the type-masking assumption, \textsc{MTFunction},
  $\nonempty{p}$ is a superset of the owners,
  so $T' = T$, so $V' = V$.
\item \textsc{MVUnit}: Base case; passes definition and typing.
\item \textsc{MVInL}, \textsc{MVInR}: Recursive cases.
\item \textsc{MVPair}: Recursive case.
\item \textsc{MVVector}: Recursive case.
\item \textsc{MVProj1}, \textsc{MVProj2}, \textsc{MVProjN}, and \textsc{MVCom}:
  From the typing assumption, $\nonempty{p}$ is a superset of the owners,
  so $T' = T$ and $V' = V$.
\item \textsc{MVVar}: Base case, trivial.
\end{itemize}

\begin{lemma}[Exclave]\label{theorem:exclave}
  If $Θ;∅ ⊢ M : T$ and $Θ \subseteq Θ'$
  then $Θ';∅ ⊢ M : T$.
\end{lemma}
\subsection{Proof of Lemma~\ref{theorem:exclave}}
By induction on the typing of $M$.
\begin{itemize}
\item \textsc{TLambda}: The recursive typing is unaffected,
  and the other tests are fine with a larger set.
\item \textsc{TVar}: Can't apply with an empty type context.
\item All other cases are unaffected by the larger party-set.
\end{itemize}

\subsection{Theorem~\ref{theorem:preservation}}

\begin{theorem}[Preservation]\label{theorem:preservation}
  If $Θ;∅ ⊢ M : T$ and $M \step M'$, then $Θ;∅ ⊢ M' : T$.
\end{theorem}

We prove this by induction on typing rules for $M$.
The eleven base cases (values) fail the assumption that $M$ can step,
so we consider the recursive cases:

\begin{itemize}
\item \textsc{TCase}: $M$ is of form $\CASE{\nonempty{p}}{N}{x_l}{M_l}{x_r}{M_r}$.
  There are three ways it might step:
  \begin{itemize}
  \item \textsc{CaseL}: $N$ is of form $\INL V$, $V'$ exists, and $M' = M_l[x_l := V']$.
    \begin{enumerate}
    \item $\nonempty{p};(x_l:d_l@\nonempty{p}) ⊢ M_l : T$ by the preconditions of \textsc{TCase}.
    \item $Θ;∅ ⊢ V : d_l@\nonempty{p}$ because $N$ must type by \textsc{TInL}.
    \item $\nonempty{p};∅ ⊢ V' : d_l@\nonempty{p}$ by Lemma~\ref{theorem:conclave} and \textsc{MTData}.
    \item $\nonempty{p};∅ ⊢ M_l[x_l := V'] : T$ by Lemma~\ref{theorem:substitution}.
    \item $Θ;∅ ⊢ M_l[x_l := V'] : T$ by Lemma~\ref{theorem:exclave}. \textbf{QED.}
    \end{enumerate}
  \item \textsc{CaseR}: Same as \textsc{CaseL}.
  \item \textsc{Case}: $N \step N'$, and by induction and \textsc{TCase},
    $Θ;Γ⊢ N' : T_N$,
    so the original typing judgment will still apply.
  \end{itemize}
\item \textsc{TApp}: $M$ is of form $F A$, and $F$ is of a function type and $A$ also types
  (both in the empty typing context).
  If the step is by \textsc{App2}or \textsc{App1}, then recursion is easy.
  There are eight other ways the step could happen:
  \begin{itemize}
  \item \textsc{AppAbs}: $F$ must type by \textsc{TLambda}.
    $M = ((λ x : T_x \DOT B)@\nonempty{p}) A$.
    We need to show that $A' = A \mask \nonempty{p}$ exists and $Θ;∅ ⊢ B[x := A'] : T$.
    \begin{enumerate}
    \item $\nonempty{p};(x:T_x) ⊢ B : T$ by the preconditions of \textsc{TLambda}.
    \item $Θ;∅ ⊢ A : T_a'$ such that $T_x = T_a' \mask \nonempty{p}$,
       by the preconditions of \textsc{TApp}.
    \item $A'$ exists and $\nonempty{p};∅ ⊢ A' : T_x$ by Lemma~\ref{theorem:conclave} on \#2.
    \item $\nonempty{p};∅ ⊢ B[x := A'] : T$ by Lemma~\ref{theorem:substitution}.
    \item \textbf{QED.} by Lemma~\ref{theorem:exclave}.
    \end{enumerate}
  \item \textsc{Proj1}: $F = \FST{\nonempty{p}}$ and $A = \PAIR V_1 V_2$ and
    $M' = V_1 \mask \nonempty{p}$.
    Necessarily, by \textsc{TPair} $Θ;∅ ⊢ V_1 : d_1@\nonempty{p_1}$
    where $\nonempty{p} \subseteq \nonempty{p_1}$.
    By Lemma~\ref{theorem:sub-mask}, $Θ;∅ ⊢ M' : T$.
  \item \textsc{Proj2}: same as \textsc{Proj1}.
  \item \textsc{ProjN}: $F = \LOOKUP{i}{\nonempty{p}}$ and $A = (\dots, V_i, \dots)$
    and $M' = V_i \mask \nonempty{p}$.
    Necessarily, by \textsc{TVec} $Θ;∅ ⊢ V_i : T_i$ and $Θ;∅ ⊢ A : (\dots, T_i, \dots)$.
    By \textsc{TApp}, $(\dots, T_i, \dots) \mask \nonempty{p} = T_a$,
    so by \textsc{MTVector} $T_i \mask \nonempty{p}$ exists
    and (again by \textsc{TApp} and \textsc{TProjN}) it must equal $T$.
    \textbf{QED.} by Lemma~\ref{theorem:maskable}.
  \item \textsc{Com1}: By \textsc{TCom} and \textsc{TUnit}.
  \item \textsc{ComPair}: Recusion among the \textsc{Com*} cases.
  \item \textsc{ComInl}:  Recusion among the \textsc{Com*} cases.
  \item \textsc{ComInr}:  Recusion among the \textsc{Com*} cases.
  \end{itemize}
\end{itemize}

\section{Proof of The Progress Theorem}\label{sec:progress-proof}

\begin{theorem}[Progress]\label{theorem:progress}
  If $Θ;∅ ⊢ M : T$, then either M is of form $V$ (which cannot step)
  or their exists $M'$ s.t. $M \step M'$.
\end{theorem}

The proof is by induction of typing rules.
There are eleven base cases and two recursive cases.
Base cases:
\begin{itemize}
\item \textsc{TLambda}
\item \textsc{TVar} (can't happen, by assumption)
\item \textsc{TUnit}
\item \textsc{TCom}
\item \textsc{TPair}
\item \textsc{TVec}
\item \textsc{TProj1}
\item \textsc{TProj2}
\item \textsc{TProjN}
\item \textsc{TInl}
\item \textsc{TInr}
\end{itemize}

Recursive cases:
\begin{itemize}
\item \textsc{TCase}: $M$ is of form $\CASE{\nonempty{p}}{N}{x_l}{M_l}{x_r}{M_r}$
  and ${Θ;∅ ⊢ N : (d_l + d_r)@\nonempty{p}}$.
  By induction, either $N$ can step, in which case M can step by \textsc{Case},
  or $N$ is a value.
  The only typing rules that would give an $N$ of form $V$ the required type are
  \textsc{TVar} (which isn't compatible with the assumed empty $Γ$),
  and \textsc{TInl} and \textsc{TInr}, which respectively force $N$ to have the required forms
  for $M$ to step by \textsc{CaseL} or \textsc{CaseR}.
  From the typing rules, \textsc{MTData}, and the first part of Lemma~\ref{theorem:conclave},
  the masking required by the step rules is possible.
\item \textsc{TApp}: $M$ is of form $F A$, and $F$ is of a function type and $A$ also types
  (both in the same empty $Γ$).
  By induction, either $F$ can step (so $M$ can step by \textsc{App2}),
  or $A$ can step (so $M$ can step by \textsc{App1}),
  or $F$ and $A$ are both values.
  Ignoring the impossible \textsc{TVar} cases,
  there are five ways an $F$ of form $V$ could type as a function;
  in each case we get to make some assumption about the type of $A$.
  Furthermore, by \textsc{TApp} and Lemma~\ref{theorem:conclave},
  we know that $A$ can mask to the owners of $F$.
  \begin{itemize}
  \item \textsc{TProj1}: $A$ must be a value of type $(d_1×d_2)@\nonempty{q}$,
    and must type by \textsc{TPair}, so it must have form $\PAIR V_1 V_2$,
    so $M$ must step by \textsc{Proj1}.
    We know $V_1$ can mask by \textsc{MVPair}.
  \item \textsc{TProj2}: (same as \textsc{TProj1})
  \item \textsc{TProjN}: $A$ must be a value of type $(T_1,\dots,T_n)$ with $i ≤ n$
    and must type by \textsc{TVec}, so it must have from $(V_1,\dots,V_n)$.
    $M$ must step by \textsc{ProjN}.
    We known $V_i$ can step by \textsc{MVVector}.
  \item \textsc{TCom}: $A$ must be a value of type $d@\nonempty{q}$,
      such that $d@\nonempty{q} \mask \nonempty{s} = d@\nonempty{s}$.
          For that to be true, \textsc{MTData} requires that $\nonempty{s} \subseteq \nonempty{q}$.
    $A$ can type that way under \textsc{TUnit}, \textsc{TPair}, \textsc{TInl}, or \textsc{TInr},
    which respectively force forms $()@\nonempty{q}$, $\PAIR V_1 V_2$, $\INL V$, and $\INR V$,
    which respectively require that $M$ reduce by
    \textsc{Com1}, \textsc{ComPair}, \textsc{ComInl}, and \textsc{ComInr}.
          In the case of $()$, this follows from Lemma~\ref{theorem:sub-mask},
          since $\set{s} \subseteq \nonempty{s} \subseteq \nonempty{q}$;
    the other three are recursive among each other.
  \item \textsc{TLambda}: $M$ must reduce by \textsc{AppAbs}.
      By the assumption of \textsc{TApp} and Lemma~\ref{theorem:maskable}, it can.
  \end{itemize}
\end{itemize}

\section{Proof of The Soundness Theorem}\label{sec:soundness-proof}
Theorem~\ref{theorem:soundness} says that
if $Θ;∅ ⊢ M : T$ and $⟦M⟧ \netstep{}{∅}^{\ast} \mathcal{N}_n$,
then there exists $M'$ such that
$M \step^{\ast} M'$ and $\mathcal{N}_n \netstep{}{∅}^{\ast} ⟦M'⟧$.
We'll need a few lemmas first.

\begin{lemma}[Values]\label{theorem:values}
  \textbf{A):} $⟦V⟧_p = L$.
  \textbf{B):} If $⟦M⟧_p = L \neq ⊥$ then $M$ is a value $V$.

  Proof is by inspection of the definition of projection.
\end{lemma}
\begin{corollary}\label{theorem:values-cor}
  If $N$ is well-typed and $⟦N⟧$ can step at all,
    then \textbf{(A)} $N$ can step to some $N'$
    and \textbf{(B)} $⟦N⟧$ can multi-step to $⟦N'⟧$ with empty annotation.

    \textbf{A} follows from Lemma~\ref{theorem:values} and Theorem~\ref{theorem:progress}.
    \textbf{B} is just Theorem~\ref{theorem:completeness}.
\end{corollary}

\begin{lemma}[Determinism]\label{theorem:determinism}
  If
  $\mathcal{N}_a \mid \mathcal{N}_0 \netstep{}{∅} \mathcal{N}_a \mid \mathcal{N}_1$
  s.t. for every $p[B_0] \in \mathcal{N}_0$, $\mathcal{N}_1(p) \neq B_0$, \\
    \emph{and}
  $\mathcal{N}_b \mid \mathcal{N}_0 \netstep{}{∅} \mathcal{N}_c \mid \mathcal{N}_2$
  s.t. the domain of $\mathcal{N}_2$ equals the domain of $\mathcal{N}_0$,  
    then \emph{either}
    \begin{itemize}
        \item $\mathcal{N}_2 = \mathcal{N}_0$, \emph{or}
        \item $\mathcal{N}_2 = \mathcal{N}_1$ and $\mathcal{N}_b = \mathcal{N}_c$.
    \end{itemize}
\end{lemma}

\subsection{Proof of Lemma~\ref{theorem:determinism}} First, observe that for every non-value expression in the process language,
there is at most one rule in the process semantics by which it can step.
(For values, there are zero.)
Furthermore, the only way for
the step annotation and resulting expression to \emph{not} be fully determined
by the initial expression
is if the justification is based on a \textsc{LRecv} step,
in which case the send-annotation will be empty
and the resulting expression will match the (single) item in the receive-annotation.

$\mathcal{N}_a \mid \mathcal{N}_0 \netstep{}{∅} \mathcal{N}_a \mid \mathcal{N}_1$
must happen by \textsc{NPar}, so consider the $\mathcal{N}_0$ step that enables it;
call that step \stepname{S}.
\stepname{S} can't be by \textsc{NPar};
that would imply parties in $\mathcal{N}_0$ who don't step.
\begin{itemize}
    \item If \stepname{S} is by \textsc{NPro}, then $\mathcal{N}_0 = p[B_0]$ is a singleton
  and \stepname{S} is justified by a process step with empty annotation.
  As noted above, that process step is the only step $B_0$ can take,
  so the
  $\mathcal{N}_b \mid \mathcal{N}_0 \netstep{}{∅} \mathcal{N}_c \mid \mathcal{N}_2$
  step must either be a \textsc{NPar} composing some other party(ies) step
  with $\mathcal{N}_0$ (satisfying the first choice),
  or a \textsc{NPar} composing \stepname{S} with $\mathcal{N}_b$
  (satisfying the second).
\item If \stepname{S} is by \textsc{NCom}, then there must be both
  a singleton \textsc{NPro} step justified by a process step
  (by some party $s$)
  with nonempty send-annotation
  and a nonempty sequence of other party steps
  (covering the rest of $\mathcal{N}_0$'s domain)
  that it gets matched with
  each with a corresponding receive-annotation.
  The send-annotated \textsc{NPro} step is deterministic in the same way as
  an empty-annotated \textsc{NPro} step.
  In order for the parties to cancel out, it can only compose by \textsc{NCom}
  with (a permutation of) the same sequence of peers.
  Considered in isolation, the peers are non-deterministic,
  but their process-steps can only be used in the network semantics by composing
  with $s$ via \textsc{NCom},
  and their resulting expressions are determined by the matched process annotation,
  which is determined by $s$'s step. \\
  Thus, for any $p[B_2] \in \mathcal{N}_2$,
  $B_2 \neq \mathcal{N}_0(p)$ implies that
  for all $q[B_2'] \in \mathcal{N}_2$, $B_2' = \mathcal{N}_1(p)$.
  In the case where $\mathcal{N}_2 = \mathcal{N}_1$,
  the step from $\mathcal{N}_0$ could only have composed with
  $\mathcal{N}_b$ by \textsc{NPar},
  so $\mathcal{N}_b = \mathcal{N}_c$, Q.E.D.
\end{itemize}

\begin{lemma}[Parallelism]\label{theorem:parallelism}
  \textbf{A):} If $\mathcal{N}_1 \netstep{}{∅}^{\ast} \mathcal{N}_1'$
  and $\mathcal{N}_2 \netstep{}{∅}^{\ast} \mathcal{N}_2'$
  then $\mathcal{N}_1 \mid \mathcal{N}_2 \netstep{}{∅}^{\ast}
  \mathcal{N}_1' \mid \mathcal{N}_2 \netstep{}{∅}^{\ast}
  \mathcal{N}_1' \mid \mathcal{N}_2'$. \\
  \textbf{B):} If $\mathcal{N}_1 \mid \mathcal{N}_2 \netstep{}{∅}^{\ast}
  \mathcal{N}_1' \mid \mathcal{N}_2 \netstep{}{∅}^{\ast}
  \mathcal{N}_1' \mid \mathcal{N}_2'$,
  then $\mathcal{N}_1 \netstep{}{∅}^{\ast} \mathcal{N}_1'$
  and $\mathcal{N}_2 \netstep{}{∅}^{\ast} \mathcal{N}_2'$.
\end{lemma}

\subsection{Proof of Lemma~\ref{theorem:parallelism}}
\textbf{A} is just repeated application of \textsc{NPar}. \\
For \textbf{B}, observer that in the derivation tree of ever step of the sequence, some (possibly different)
minimal sub-network will step by \textsc{NPro} or {NCom} as a precondition
to some number of layers of \textsc{NPar}.
The domains of these minimal sub-networks will be subsets of the domains of $\mathcal{N}_1$
and $\mathcal{N}_2$ respectively,
so they can just combine via \textsc{NPar} to get the needed step in the respective sequences for
$\mathcal{N}_1$ and $\mathcal{N}_2$.

\subsection{Theorem~\ref{theorem:soundness}}
\begin{theorem}[Soundness]\label{theorem:soundness}
  If $Θ;∅ ⊢ M : T$ and $⟦M⟧ \netstep{}{∅}^{\ast} \mathcal{N}_n$,
  then there exists $M'$ such that
  $M \step^{\ast} M'$ and $\mathcal{N}_n \netstep{}{∅}^{\ast} ⟦M'⟧$.
\end{theorem}

Declare the predicate $\mathsf{sound}(\mathcal{N})$ to mean that
there exists some $M_{\mathcal{N}}$ such that
$M \step^{\ast} M_{\mathcal{N}}$
and $\mathcal{N} \netstep{}{∅}^{\ast} ⟦M_{\mathcal{N}}⟧$.

Consider the sequence of network steps
$⟦M⟧ = \mathcal{N}_0 \netstep{}{∅} \dots \netstep{}{∅} \mathcal{N}_n$.
By Corollary~\ref{theorem:values-cor}, $\mathsf{sound}(\mathcal{N}_0)$.
Select the largest $i$ s.t. $\mathsf{sound}(\mathcal{N}_i)$.
We will derive a contradiction from an assumption that
$\mathcal{N}_{i+1}$ is part of the sequence;
this will prove that $i=n$, which completes the proof of the Theorem.

Choose a sequence of network steps (of the possibly many such options)
$\mathcal{N}_i = \mathcal{N}^a_i \netstep{}{∅} \dots \netstep{}{∅}
\mathcal{N}^a_m = ⟦M^a⟧$
where $M \step^{\ast} M^a$.

Assume $\mathcal{N}_{i+1}$ is part of the original sequence.
Decompose the step to it as
$\mathcal{N}_i = \mathcal{N}^0_i \mid \mathcal{N}^1_i \netstep{}{∅}
\mathcal{N}^0_i \mid \mathcal{N}^1_{i+1} = \mathcal{N}_{i+1}$
where $\mathcal{N}^1_i$'s domain is as large as possible.
We will examine two cases:
either the parties in $\mathcal{N}^1_i$ make steps in the sequence to
$\mathcal{N}^a_m$, or they do not.
Specifically, consider the largest $j$ s.t.
$\mathcal{N}^a_j = \mathcal{N}^b_j \mid \mathcal{N}^1_i$.

\begin{itemize}
\item Suppose $j < m$. \\
  By Lemma~\ref{theorem:determinism} and our decision that $j$ is as large as possible,
  $\mathcal{N}^a_{j+1} = \mathcal{N}^b_j \mid \mathcal{N}^1_{i+1}$.
  Thus we have
  $\mathcal{N}^0_i \mid \mathcal{N}^1_i \netstep{}{∅}^{\ast}
   \mathcal{N}^b_j \mid \mathcal{N}^1_i \netstep{}{∅}
   \mathcal{N}^b_j \mid \mathcal{N}^1_{i+1}$.
  By Lemma~\ref{theorem:parallelism}, we can reorganize that into an alternative sequence where
  $\mathcal{N}^0_i \mid \mathcal{N}^1_i \netstep{}{∅}
   \mathcal{N}^0_i \mid \mathcal{N}^1_{i+1} \netstep{}{∅}^{\ast}
   \mathcal{N}^b_j \mid \mathcal{N}^1_{i+1}$.
  Since $\mathcal{N}^0_i \mid \mathcal{N}^1_{i+1} = \mathcal{N}_{i+1}$
  and $\mathcal{N}^a_{j+1} \netstep{}{∅}^{\ast} ⟦M^a⟧$,
  this contradicts our choice that $i$ be as large as possible.
\item Suppose $j = m$, so $⟦M^a⟧ = \mathcal{N}^b_m \mid \mathcal{N}^1_i$.\\
  By Lemma~\ref{theorem:parallelism}, $⟦M^a⟧$ can step (because $\mathcal{N}^1_i$ can step)
  so by Corollary~\ref{theorem:values-cor}, $M^a \step M^{a+1}$.
  We can repeat our steps from our choice of
  $\mathcal{N}^a_i \netstep{}{∅}^{\ast} \mathcal{N}^a_m = ⟦M^a⟧$,
  but using $M^{a+1}$ instead of $M^a$.
        Since \HLSCentral doesn't have recursion, eventually we'll arrive at a $M^{a++}$
  that can't step, and then-or-sooner we'll be in the first case above.
  Q.E.D.
\end{itemize}

\section{Proof of The Completeness Theorem}\label{sec:completeness-proof}
Theorem~\ref{theorem:completeness} says that
if $Θ;∅ ⊢ M : T$ and $M \step M'$,
then $⟦M⟧ \netstep{}{∅}^{\ast} ⟦M'⟧$.
We'll need a few lemmas first.

\begin{lemma}[Cruft]\label{theorem:cruft}
  If $Θ;∅ ⊢ M : T$ and $p \not\in Θ$,
  then $⟦M⟧_p = ⊥$.
\end{lemma}
\subsection{Proof of Lemma~\ref{theorem:cruft}}
By induction on the typing of $M$:
\begin{itemize}
\item \textsc{TLambda}:
  $\nonempty{p} \subseteq Θ$, therefore $p \not\in \nonempty{p}$,
  therefore $⟦M⟧_p = ⊥$.
\item \textsc{TVar}: Can't happen because $M$ types with empty $Γ$.
\item \textsc{TUnit}, \textsc{TCom}, \textsc{TProj1}, \textsc{TProj2},
  and \textsc{TProjN}:
  Same as \textsc{TLambda}.
\item \textsc{TPair}, \textsc{TVec}, \textsc{TInl}, and \textsc{TInr}:
  In each of these cases we have some number of recursive typing judgments
  to which we can apply the inductive hypothesis.
  This enables the respective cases of the definition of floor
  (as used in the respective cases of the definition of projection)
  to map to $⊥$.
\item \textsc{TApp}: $M = N_1 N_2$.
  By induction, $⟦N_1⟧_p = ⊥$ and $⟦N_2⟧_p = ⊥$,
  so $⟦M⟧_p = ⊥$
\item \textsc{TCase}: Similar to \textsc{TLambda},
  by induction the guard projects to $⊥$ and therefore the whole thing does too.
\end{itemize}

\begin{lemma}[Existence]\label{theorem:existence}
  If $Θ;Γ ⊢ V : d@\nonempty{p}$ and $p,q \in \nonempty{p}$,
  then $⟦V⟧_p = ⟦V⟧_q \neq ⊥$.
\end{lemma}
\subsection{Proof of Lemma~\ref{theorem:existence}}
By induction on possible typings of $V$:
\begin{itemize}
\item \textsc{TVar}: Projection is a no-op on variables.
\item \textsc{TUnit}: $⟦V⟧_p = ⟦V⟧_q = ()$.
\item \textsc{TPair}: $p,q \in \nonempty{p_1} ∩ \nonempty{p_2}$,
  so both are in each of them, so we can recurse on $V_1$ and $V_2$.
\item \textsc{TInl} and \textsc{TInr}: simple induction.
\end{itemize}

\begin{lemma}[Bottom]\label{theorem:bottom}
  If $Θ;∅ ⊢ M : T$ and $⟦M⟧_p = ⊥$ and $M \step M'$
  then $⟦M'⟧_p = ⊥$.
\end{lemma}
\subsection{Proof of Lemma~\ref{theorem:bottom}}
By induction on the step $M \step M'$.
\begin{itemize}
\item \textsc{AppAbs}: $M = (λ x:T_x \DOT N)@\nonempty{p} V$,
  and necessarily $⟦(λ x:T_x \DOT N)@\nonempty{p}⟧_p = ⊥$.
  Since the lambda doesn't project to a lambda, $p\not\in\nonempty{p}$.
  $M' = N[x:=V\mask\nonempty{p}]$.
        By \textsc{TLambda}, Lemma~\ref{theorem:substitution}, and Lemma~\ref{theorem:cruft},
  $⟦N[x:=V\mask\nonempty{p}]⟧_p = ⊥$.
\item \textsc{App1}: $M = V N$
  and necessarily $⟦V⟧_p = ⟦N⟧_p = ⊥$.
  By induction on $N \step N'$, $⟦N'⟧_p = ⊥$.
\item \textsc{App2}: Same as \textsc{App1}.
\item \textsc{Case}: The guard must project to $⊥$, so this follows from induction.
\item \textsc{CaseL} (and \textsc{CaseR} by mirror image):
  $M = \CASE{\nonempty{p}}{\INL V}{x_l}{M_l}{x_r}{M_r}$
  and $M' = M_l[x_l := V\mask\nonempty{p}]$.
  Necessarily, $⟦V⟧_p = ⊥$.
  By \textsc{TCase} and \textsc{MTData}, $\INL V$ types as data,
        so by Lemma~\ref{theorem:existence} $p \not\in \nonempty{p}$.
        By \textsc{TCase}, Lemma~\ref{theorem:substitution}, and Lemma~\ref{theorem:cruft},
  $⟦M'⟧_p = ⟦M_l[x_l := V\mask\nonempty{p}]⟧_p = ⊥$.
\item \textsc{Proj1}: $M = \FST{\nonempty{p}}(\PAIR V_1 V_2)$,
  and $p \not \in \nonempty{p}$.
  $M' = V_1 \mask \nonempty{p}$.
  Since $Θ;∅ ⊢ V_1 : T'$ (by \textsc{TPair})
  and $T' \mask \nonempty{p} = T''$ is defined
  (by \textsc{TApp} and the indifference of \textsc{MTData} to the data's structure),
        by Lemma~\ref{theorem:conclave} $\nonempty{p};∅ ⊢ V_1 \mask \nonempty{p} : T''$.
        By Lemma~\ref{theorem:cruft} this projects to $⊥$.
\item \textsc{Proj2}, \textsc{ProjN}, and \textsc{Com1} are each pretty similar to
  \textsc{Proj1}.
\item \textsc{Com1}, \textsc{ComPair}, \textsc{ComInl}, and \textsc{ComInr}:
    For $M$ to project to ⊥, $p$ must be neither a sender nor a recipient.
    By induction among these cases (with \textsc{Com1} as the base case),
        $M'$ will be some structure of $()@\nonempty{r}$;
        since $p\not\in\nonempty{r}$ and projection uses floor,
        this will project to ⊥.
\end{itemize}

\begin{lemma}[Masked]\label{theorem:masked}
  If $p \in \nonempty{p}$ and $V' = V \mask \nonempty{p}$
  then $⟦V⟧_p = ⟦V'⟧_p$.
\end{lemma}
\subsection{Proof of Lemma~\ref{theorem:masked}}
By (inductive) case analysis of endpoint projection:
\begin{itemize}
\item $⟦x⟧_p = x$. By \textsc{MVVar} the mask does nothing.
\item $⟦(λ x:T \DOT M)@\nonempty{q}⟧_p$:
  Since $V \mask \nonempty{p}$ is defined, by \textsc{MVLambda} it does nothing.
\item $⟦()@\nonempty{q}⟧_p$: By \textsc{MVUnit} $V' = ()@(\nonempty{p} ∩ \nonempty{q})$.
  $p$ is in that intersection iff $p \in \nonempty{q}$,
  so the projections will both be $()$ or $⊥$ correctly.
\item $\INL V_l$, $\INR V_r$, $\PAIR V_1 V_2$, $(V_1, \dots, V_n)$: simple recursion.
\item $\FST{\nonempty{q}}$, $\SND{\nonempty{q}}$,
  $\LOOKUP{i}{\nonempty{q}}$, $\COMM{q}{\nonempty{q}}$:
  Since the masking is defined, it does nothing.
\end{itemize}

\begin{lemma}[Floor Zero]\label{theorem:floor-zero}
  $⟦M⟧_p = \FLR{⟦M⟧_p}$
\end{lemma}
\subsection{Proof of Lemma~\ref{theorem:floor-zero}}
There are thirteen forms.
Six of them (application, case, injection-r/l, pair and vector)
apply floor directly in the definition of projection.
Six of them (variable, unit, the three lookups, and $\langword{com}$)
can only project to values such that floor is a no-op.
For a lambda $(λ x:T_x \DOT N)@\nonempty{p}$, the proof is by induction on the body $N$.

\begin{lemma}[Distributive Substitution]\label{theorem:distributive-substitution}
  If $Θ;(x : T_x) ⊢ M : T$ and $p \in Θ$, \\
  then $⟦M[x:=V]⟧_p = \FLR{⟦M⟧_p[x := ⟦V⟧_p]}$.
    (Because $⟦V⟧_p$ may be ⊥, this isn't really distribution; an extra flooring operation is necessary.)
\end{lemma}
\subsection{Proof of Lemma~\ref{theorem:distributive-substitution}}
It'd be more elegant if substitution really did distribute over projection,
but this weaker statement is what we really need anyway.
The proof is by inductive case analysis on the form of $M$:
\begin{itemize}
\item $\PAIR V_1 V_2$: $⟦M[x:=V]⟧_p = ⟦\PAIR V_1[x:=V] V_2[x:=V]⟧_p$\\
  $= \FLR{\PAIR ⟦V_1[x:=V⟧_p ⟦V_2[x:=V]⟧_p}$ \\
  and $⟦M⟧_p[x := ⟦V⟧_p] = \FLR{\PAIR ⟦V_1⟧_p ⟦V_2⟧_p}[x := ⟦V⟧_p]$.
  \begin{itemize}
  \item Suppose one of $⟦V_1⟧_p$, $⟦V_2⟧_p$ is not $⊥$.
    Then \\
    $⟦M⟧_p[x := ⟦V⟧_p] = (\PAIR \FLR{⟦V_1⟧_p} \FLR{⟦V_2⟧_p})[x := ⟦V⟧_p]$ \\
          which by Lemma~\ref{theorem:floor-zero}
    $= (\PAIR ⟦V_1⟧_p ⟦V_2⟧_p)[x := ⟦V⟧_p]$ \\
    $= \PAIR (⟦V_1⟧_p[x := ⟦V⟧_p]) (⟦V_2⟧_p[x := ⟦V⟧_p])$. \\
    Thus $\FLR{⟦M⟧_p[x := ⟦V⟧_p]}
     = \FLR{\PAIR (⟦V_1⟧_p[x := ⟦V⟧_p]) (⟦V_2⟧_p[x := ⟦V⟧_p])}$. \\
    By induction,
    $⟦V_1[x:=V]⟧_p = \FLR{⟦V_1⟧_p[x := ⟦V⟧_p]}$ \\
          and
    $⟦V_2[x:=V]⟧_p = \FLR{⟦V_2⟧_p[x := ⟦V⟧_p]}$;
    with that in mind,
    \begin{itemize}
    \item Suppose one of $⟦V_1[x:=V]⟧_p$, $⟦V_1[x:=V]⟧_p$ is not $⊥$. \\
      $\FLR{⟦M⟧_p[x := ⟦V⟧_p]}
       = \PAIR \FLR{⟦V_1⟧_p[x := ⟦V⟧_p]} \FLR{⟦V_2⟧_p[x := ⟦V⟧_p]}$, \\
      and $⟦M[x:=V]⟧_p = \PAIR \FLR{⟦V_1[x:=V⟧_p} \FLR{⟦V_2[x:=V]⟧_p}$ \\
       $= \PAIR ⟦V_1[x:=V⟧_p ⟦V_2[x:=V]⟧_p$
      Q.E.D.
    \item Otherwise, $\FLR{⟦M⟧_p[x := ⟦V⟧_p]} = ⊥ = ⟦M[x:=V]⟧_p$.
    \end{itemize}
  \item Otherwise, $⟦M⟧_p[x := ⟦V⟧_p] = \FLR{\PAIR ⊥ ⊥}[x := ⟦V⟧_p] = ⊥$. \\
      Note that, by induction \textit{etc},
    $⟦V_1⟧_p = ⊥ = ⟦V_1⟧_p[x := ⟦V⟧_p] = \FLR{⟦V_1⟧_p[x := ⟦V⟧_p]}
     = ⟦V_1[x:=V]⟧_p$,
    and the same for $V_2$, so
    $⟦M[x:=V]⟧_p = ⊥$, Q.E.D.
  \end{itemize}
\item $\INL V_l$, $\INR V_r$, $(V_1, \dots, V_n)$:
  Follow the same inductive pattern as $\PAIR$.
\item $N_1 N_2$:
  $⟦M[x:=V]⟧_p = ⟦N_1[x:=V] N_2[x:=V]⟧_p = \FLR{⟦N_1[x:=V]⟧_p ⟦N_2[x:=V]⟧_p}$ \\
  $= \begin{cases}
    \FLR{⟦N_1[x:=V]⟧_p} = ⊥, \FLR{⟦N_2[x:=V]⟧_p} = L :& ⊥ \\
    \text{else} :& \FLR{⟦N_1[x:=V]⟧_p} \FLR{⟦N_2[x:=V]⟧_p}
  \end{cases}$ \\
  $= \begin{cases}
    ⟦N_1[x:=V]⟧_p = ⊥, ⟦N_2[x:=V]⟧_p = L :& ⊥ \\
    \text{else} :& ⟦N_1[x:=V]⟧_p ⟦N_2[x:=V]⟧_p
  \end{cases}$ \\
  and $\FLR{⟦M⟧_p[x:=⟦V⟧_p]} = \FLR{\FLR{⟦N_1⟧_p ⟦N_2⟧_p}[x:=⟦V⟧_p]}$ \\
  $= \begin{cases}
    \FLR{⟦N_1⟧_p} = ⊥, \FLR{⟦N_2⟧_p} = L :& \FLR{⊥[x:=⟦V⟧_p]} = ⊥ \\
    \text{else} :& \FLR{ (\FLR{⟦N_1⟧_p} \FLR{⟦N_2⟧_p})[x:=⟦V⟧_p] } \\
                 & \quad= \FLR{ (⟦N_1⟧_p[x:=⟦V⟧_p]) (⟦N_2⟧_p[x:=⟦V⟧_p]) }
  \end{cases}$ \\
  $= \begin{cases}
    \FLR{⟦N_1⟧_p[x:=⟦V⟧_p]} = ⊥, \FLR{⟦N_2⟧_p[x:=⟦V⟧_p]} = L : ⊥ \\
    \text{else} : \FLR{⟦N_1⟧_p[x:=⟦V⟧_p]} \FLR{⟦N_2⟧_p[x:=⟦V⟧_p]}
  \end{cases}$ \\
  (Note that we collapsed the $\FLR{⟦N_1⟧_p} = ⊥,\dots$ case.
  We can do that because if $⟦N_1⟧_p = ⊥$ then so does $\FLR{⟦N_1⟧_p[x:=⟦V⟧_p]}$
  and if $⟦N_2⟧_p = L$ then $\FLR{⟦N_2⟧_p[x:=⟦V⟧_p]}$ is also a value.) \\
  By induction, $⟦N_1[x:=V]⟧_p = \FLR{⟦N_1⟧_p[x := ⟦V⟧_p]}$ \\
  and $⟦N_2[x:=V]⟧_p = \FLR{⟦N_2⟧_p[x := ⟦V⟧_p]}$.
\item $y$: trivial because EPP and floor are both no-ops.
\item $(λ y:T_y \DOT N)@\nonempty{p}$:
  \begin{itemize}
  \item If $p \not\in \nonempty{p}$, both sides of the equality are $⊥$.
  \item If $V' = V \mask \nonempty{p}$ is defined, then \\
    $⟦(λ y:T_y \DOT N)@\nonempty{p}[x:=V]⟧_p
    =⟦(λ y:T_y \DOT N[x:=V'])@\nonempty{p}⟧_p
    =  λ y \DOT ⟦N[x:=V']⟧_p$ \\
    and
    $\FLR{⟦(λ y:T_y \DOT N)@\nonempty{p}⟧_p[x := ⟦V⟧_p]}$ \\
    $= \FLR{(λ y \DOT ⟦N⟧_p)[x := ⟦V⟧_p]  }$ \\
    $= \FLR{ λ y \DOT (⟦N⟧_p[x := ⟦V⟧_p]) }$ \\
          $= \FLR{ λ y \DOT (⟦N⟧_p[x := ⟦V'⟧_p])}$ (by Lemma~\ref{theorem:masked}) \\
    $=  λ y \DOT \FLR{(⟦N⟧_p[x := ⟦V'⟧_p])}$ \\
    Then we do induction on $N$ and $V'$.
  \item Otherwise, substitution in the central program is a no-op.  
    \begin{itemize}
    \item $⟦(λ y:T_y \DOT N)@\nonempty{p}[x:=V]⟧_p = ⟦(λ y:T_y \DOT N)@\nonempty{p}⟧_p
      = λ y \DOT ⟦N⟧_p$ \\
      and \\ $\FLR{⟦(λ y:T_y \DOT N)@\nonempty{p}⟧_p[x := ⟦V⟧_p]}
      = \FLR{(λ y \DOT ⟦N⟧_p)[x := ⟦V⟧_p]} \\
      = \FLR{λ y \DOT (⟦N⟧_p[x := ⟦V⟧_p])}$ \\
      $= λ y \DOT \FLR{⟦N⟧_p[x := ⟦V⟧_p]}$.
    \item Since we already known
      $(λ y:T_y \DOT N)@\nonempty{p}[x:=V] = (λ y:T_y \DOT N)@\nonempty{p}$,
            we can apply Lemma~\ref{theorem:substitution} to $M$ and unpack the typing of
      $M[x:=V] = M$
      to get $\nonempty{p};(y:T_y) ⊢ N : T'$.
  \item By Lemma~\ref{theorem:unused}, we get $N[x:=V] = N$.
    \item By induction on $N$ and $V$, we get
      $\FLR{⟦N⟧_p[x := ⟦V⟧_p]} = ⟦N[x:=V]⟧_p =  ⟦N⟧_p$,
      QED.
    \end{itemize}
  \end{itemize}
\item $\CASE{\nonempty{p}}{N}{x_l}{N_l}{x_r}{N_r}$: 
  \begin{itemize}
  \item If $⟦N⟧_p = ⊥$ then $\FLR{⟦N⟧_p[x:=⟦V⟧_p]} = ⊥ = ⟦N[x:=V]⟧_p$ (by induction),
    so both halfs of the equality are $⊥$.
  \item Else if $p \not \in \nonempty{p}$, then we get \\
    $⟦\CASE{\nonempty{p}}{N[x:=V]}{x_l}{N_l'}{x_r}{N_r'}⟧_p
    = \CASE{\nonempty{p}}{⟦N[x:=V]⟧_p}{x_l}{⊥}{x_r}{⊥}$ \\
    and \\
    $\FLR{⟦\CASE{\nonempty{p}}{N}{x_l}{N_l}{x_r}{N_r}⟧_p[x := ⟦V⟧_p]} \\
    = \FLR{(\CASE{\nonempty{p}}{⟦N⟧_p}{x_l}{⊥}{x_r}{⊥})[x := ⟦V⟧_p]} \\
    = \FLR{\CASE{\nonempty{p}}{⟦N⟧_p[x := ⟦V⟧_p]}{x_l}{⊥}{x_r}{⊥}}$. \\
    Since we've assumed $\FLR{⟦N⟧_p[x:=⟦V⟧_p]} \neq ⊥$,
    these are equal by induction.
  \item Else if $V' = V \mask \nonempty{p}$ is defined then we can do induction similar
    similar to how we did for the respective lambda case, except the induction is
    three-way.
  \item Otherwise, it's similar to the respective lambda case, just more verbose.
  \end{itemize}
\item $()@\nonempty{p}$, $\FST{\nonempty{p}}$, $\SND{\nonempty{p}}$,
  $\LOOKUP{i}{\nonempty{p}}$, and $\COMM{s}{\nonempty{r}}$:
  trivial because substitution and floor are no-ops.
\end{itemize}

\begin{lemma}[Weak Completeness]\label{theorem:weak-completeness}
  If $Θ;∅ ⊢ M : T$ and $M \step M'$
  then $⟦M⟧_p \prcstep{μ}{η}^{?} ⟦M'⟧_p$.  
  (\ie it takes zero or one steps to get there.)
\end{lemma}
\subsection{Proof of Lemma~\ref{theorem:weak-completeness}}
If $⟦M⟧_p = ⊥$ then this is follows trivially from Lemma~\ref{theorem:bottom},
so assume it doesn't.
We proceed with induction on the form of $M \step M'$:
\begin{itemize}
\item \textsc{AppAbs}: $M = (λ x:T_x \DOT N)@\nonempty{p} V$,
  and $M' = N[x:=V\mask\nonempty{p}]$.
  By assumption, the lambda doesn't project to $⊥$, so $p \in \nonempty{p}$
  and $⟦M⟧_p \prcstep{∅}{∅} \FLR{⟦N⟧_p[x:=⟦V⟧_p]}$ by \textsc{LAbsApp}. \\
        By Lemma~\ref{theorem:masked} and Lemma~\ref{theorem:distributive-substitution}
  $\FLR{⟦N⟧_p[x:=⟦V⟧_p]} = \FLR{⟦N⟧_p[x:=⟦V\mask\nonempty{p}⟧_p]}
  = ⟦N[x:=V\mask\nonempty{p}]⟧_p = ⟦M'⟧_p$.
\item \textsc{App1}: $M = V N \step V N' = M'$.
  By induction, $⟦N⟧_p \prcstep{μ}{η}^{?} ⟦N'⟧_p$.
  \begin{itemize}
  \item Assume $⟦V⟧_p = ⊥$.
    By our earlier assumption, $⟦N⟧_p \neq ⊥$.
    Since $⟦N⟧_p$ can step; that step justifies a \textsc{LApp1} step
    with the same annotations.
          If $⟦N'⟧_p$ is a value then
    that'll be handled by the floor built into \textsc{LApp1}.
  \item Otherwise, the induction is even simpler,
    we just don't have to worry about possibly collapsing the whole thing to $⊥$.
  \end{itemize}
\item \textsc{App2}:
  $M = N_1 N_2 \step N_1' N_2 = M'$.
  By induction, $⟦N_1⟧_p \prcstep{μ}{η}^{?} ⟦N_1'⟧_p$.
  \begin{itemize}
  \item Assume $⟦N_2⟧_p = L$.
    By our earlier assumption, $⟦N_1⟧_p \neq ⊥$.
    Since $⟦N_1⟧_p$ steps, that step justifies a \textsc{LApp2} step
    with the same annotations.
         If $⟦N_1'⟧_p$ is a value then
    that'll be handled by the floor built into \textsc{LApp2}.
  \item Otherwise, the induction is even simpler.
  \end{itemize}
\item \textsc{Case}: By our assumptions, the guard can't project to $⊥$;
  we just do induction on the guard to satisfy \textsc{LCase}.
\item \textsc{CaseL} (\textsc{CaseR} mirrors):
  $M = \CASE{\nonempty{p}}{\INL V}{x_l}{M_l}{x_r}{M_r}$,
  and $⟦M⟧_p = \CASE{}{\INL ⟦V⟧_p}{x_l}{B_l}{x_r}{B_r}$.
  $⟦M⟧_p \prcstep{∅}{∅} \FLR{B_l[x_l := ⟦V⟧_p]}$ by \textsc{LCaseL}.
  $M' = M_l[x_l := V\mask\nonempty{p}]$.
  If $p \in \nonempty{p}$
  then $B_l = ⟦M_l⟧_p$
        and by Lemma~\ref{theorem:masked} and Lemma~\ref{theorem:distributive-substitution}
  $\FLR{B_l[x_l := ⟦V⟧_p]} = \FLR{⟦M_l⟧_p[x_l := ⟦V⟧_p]}
  = \FLR{⟦M_l⟧_p[x_l := ⟦V\mask\nonempty{p}⟧_p]}$ \\
  $= ⟦M_l[x_l := V\mask\nonempty{p}]⟧_p
  = ⟦M'⟧_p$. \\
  Otherwise, $B_l[x_l := ⟦V⟧_p] = ⊥$
        and by \textsc{TCase}, Lemma~\ref{theorem:substitution}, and Lemma~\ref{theorem:cruft},
  $⟦M'⟧_p = ⊥$.
\item \textsc{Proj1}: $M = \FST{\nonempty{p}} (\PAIR V_1 V_2)$
  and $M' = V_1 \mask \nonempty{p}$.
  Since we assumed $⟦M⟧_p \neq ⊥$, $p \in \nonempty{p}$. \\
  $⟦M⟧_p = \FST{} \FLR{\PAIR ⟦V_1⟧_p ⟦V_2⟧_p} = \FST{} (\PAIR ⟦V_1⟧_p ⟦V_2⟧_p)$
        by Lemma~\ref{theorem:existence} and \textsc{TPair}.
  This steps by \textsc{LProj1} to $⟦V_1⟧_p$,
        which equals $⟦M'⟧_p$ by Lemma~\ref{theorem:masked}.
\item \textsc{Proj2}, \textsc{ProjN}: Same as \textsc{Proj1}.
\item \textsc{Com1}: $M = \COMM{s}{\nonempty{r}} ()@\nonempty{p}$
  and $M' = ()@\nonempty{r}$.
  \begin{itemize}
  \item $s = p$ and $p \in \nonempty{r}$:
    By \textsc{MVUnit}, $p \in \nonempty{p}$,
    so $⟦M⟧_p = \SEND{\nonempty{r} ∖ \set{p}}^{\ast} ()$,
    which steps by \textsc{LSendSelf} (using \textsc{LSend1}) to $()$.
    $⟦M'⟧_p = ()$.
  \item $s = p$ and $p \not\in \nonempty{r}$:
    By \textsc{MVUnit}, $p \in \nonempty{p}$,
    so $⟦M⟧_p = \SEND{\nonempty{r}} ()$,
    which steps by \textsc{LSend1} to $⊥$.
    $⟦M'⟧_p = ⊥$.
  \item $s \neq p$ and $p \in \nonempty{r}$:
    $⟦M⟧_p = \RECV{s} ⟦()@\nonempty{p}⟧_p$,
    which can step
    (arbitrarily, but with respective annotation)
    by \textsc{LRecv} to $⟦M'⟧_p$.
  \item Otherwise, we violate our earlier assumption.
  \end{itemize}
\item \textsc{ComPair}, \textsc{ComInl}, and \textsc{ComInr}:
  Each uses the same structure of proof as \text{Com1},
  using induction between the cases
  to support the respective process-semantics step.
\end{itemize}

\subsection{Theorem~\ref{theorem:completeness}}
\begin{theorem}[Completeness]\label{theorem:completeness}
  If $Θ;∅ ⊢ M : T$ and $M \step M'$,
  then $⟦M⟧ \netstep{}{∅}^{\ast} ⟦M'⟧$.
\end{theorem}

The proof is by case analysis on the semantic step $M \step M'$:
\begin{itemize}
\item \textsc{AppAbs},
  \textsc{CaseL},
  \textsc{CaseR},
  \textsc{Proj1},
  \textsc{Proj2},
  and \textsc{ProjN}:
  Necessarily, the set of parties $\nonempty{p}$ for whom
  $⟦M⟧_{p\in\nonempty{p}} \neq ⊥$ is not empty.
  For every $p \in \nonempty{p}$,
        by Lemma~\ref{theorem:weak-completeness} $⟦M⟧_p \prcstep{∅}{∅}^{?} ⟦M'⟧_p$
  (checking the cases to see that the annotations are really empty!).
  By \textsc{NPro}, each of those is also a
  network step,
        which by Lemma~\ref{theorem:parallelism} can be composed in any order to get
  $⟦M⟧ \netstep{}{∅}^{\ast} \mathcal{N}$.
  For every $p \in \nonempty{p}$,
  $\mathcal{N}(p) = ⟦M'⟧_p$,
        and (by Lemma~\ref{theorem:bottom}) for every $q \not\in \nonempty{p}$,
  $\mathcal{N}(q) = ⊥ = ⟦M'⟧_q$,
  Q.E.D.
\item \textsc{Com1},
  \textsc{ComPair},
  \textsc{ComInl},
  and \textsc{ComInr}:
  $M = \COMM{s}{\nonempty{r}} V$.
  By the recursive structure of \textsc{Com1}, \textsc{ComPair}, \textsc{ComInl},
  and \textsc{ComInr}, $M'$ is some structure of
  $\set{\PAIR, \INL{}, \INR{}, ()@\nonempty{r}}$,
  and $⟦M'⟧_{r\in\nonempty{r}} = ⟦V⟧_s$.
  For every $q \not\in \nonempty{r} ∪ \set{s}$, $⟦M⟧_q = ⊥ = ⟦M'⟧_q$
        by Lemma~\ref{theorem:bottom}.
  Consider two cases:
  \begin{itemize}
  \item $s \not\in \nonempty{r}$: \\
      By Lemma~\ref{theorem:weak-completeness}
    $⟦M⟧_s = \SEND{\nonempty{r}} ⟦V⟧_s
    \prcstep{\set{(r, ⟦V⟧_s) \mid r \in \nonempty{r}}}{∅} ⊥$.\\
    By the previously mentioned structure of $M'$, $⟦M'⟧_s = ⊥$. \\
    For every $r \in \nonempty{r}$,
    by Lemma~\ref{theorem:weak-completeness}
    $⟦M⟧_r = \RECV{s} ⟦V⟧_r
    \prcstep{∅}{\set{(s,⟦V⟧_s)}} ⟦V⟧_s = ⟦M'⟧_{r}$. \\
    By \textsc{NPro},
    $s[⟦M⟧_s] \netstep{s}{\set{(r, ⟦V⟧_s) \mid r \in \nonempty{r}}} s[⊥=⟦M'⟧_s]$.\\
    This composes in parallel with each of the $r_{\in\nonempty{r}}[⟦M⟧_r]$
    by \textsc{NCom} in any order until the unmactched send is empty.
    Everyone in and not-in $\nonempty{r} ∪ \set{s}$ has stepped, if needed,
    to the respective projection of $M'$.
  \item $s \in \nonempty{r}$: Let $\nonempty{r_0} = \nonempty{r} ∖ \set{s}$. \\
    By Lemma~\ref{theorem:weak-completeness}
    $⟦M⟧_s = \SEND{\nonempty{r_0}}^{\ast} ⟦V⟧_s
    \prcstep{\set{(r, ⟦V⟧_s) \mid r \in \nonempty{r_0}}}{∅} ⟦V⟧_s
    = ⟦M'⟧_{s\in \nonempty{r}}$. \\
    For every $r \in \nonempty{r_0}$,
    by Lemma~\ref{theorem:weak-completeness}
    $⟦M⟧_r = \RECV{s} ⟦V⟧_r
    \prcstep{∅}{\set{(s,⟦V⟧_s)}} ⟦V⟧_s = ⟦M'⟧_{r}$. \\
    We proceed as in the previous case.
  \end{itemize}
\item \textsc{App1} (\textsc{App2} and \textsc{Case} are similar):
  $M = V N$.
  By induction, $⟦N⟧ \netstep{}{∅}^{\ast} ⟦N'⟧$.
  Every $N$ step in that process in which a single party advances by \textsc{NPro}
  can justify a corresponding $M$ step by \textsc{LApp1}.
  \textsc{NCom} steps are basically the same: each of the participating parties will
  justify a \textsc{LApp1} $M$ step with a $N$ step;
  since this doesn't change the send \& receive annotations,
  the cancellation will still work.
\end{itemize}

\end{document}